\documentclass{jfm}
\usepackage{graphicx}
\usepackage{natbib}
\usepackage{color}
\usepackage{latexsym}
\usepackage{subfig}
\usepackage{natbib}
\usepackage{upgreek}
\usepackage{mathrsfs}
\usepackage{float}
\usepackage{caption}
\usepackage{soul}
\usepackage{textcomp}
\usepackage{stmaryrd}
\usepackage{amsmath}
\usepackage{mathtools}
\usepackage{epstopdf, epsfig}
\usepackage{amsmath,etoolbox}
\usepackage{tabularx}
\usepackage{bigints}
\usepackage{lipsum}
\usepackage{physics}
\usepackage[normalem]{ulem}
\usepackage{xargs}                      % Use more than one optional parameter in a new commands
\usepackage[pdftex,dvipsnames]{xcolor}  % Coloured text etc.
\usepackage[colorinlistoftodos,prependcaption,textsize=small]{todonotes}

\AtBeginEnvironment{pmatrix}{\setlength{\arraycolsep}{2pt}}
\def\ee{{\rm e}}
\def\ii{{\rm i}}
\shortauthor{S. Gururaj \& A. Guha}
\title[Internal wave triads in a {vertically} bounded domain with mild-slope bathymetry]{Resonant and near-resonant internal wave triads for non-uniform stratifications. Part 2: {Vertically} bounded domain with mild-slope bathymetry}
\author{Saranraj Gururaj\aff{1}\corresp{\email{gmsaranraj@gmail.com}}  \and Anirban Guha\aff{1} }

\affiliation{
\aff{1} School of Science and Engineering, University of Dundee, DD1 4HN, U.K.
}

\begin{document}

\maketitle
\begin{abstract}
Weakly nonlinear internal wave-wave interaction is a key mechanism that cascades energy from large to small scales, leading to ocean turbulence and mixing. Oceans typically have a non-uniform density stratification profile; moreover, submarine topography leads to a spatially varying {bathymetry} ($h$). { Under these conditions and assuming mild-slope bathymetry, we employ multiple-scale analysis to derive the wave amplitude equations for weakly nonlinear wave-wave interactions.} The waves are assumed to have a slowly (rapidly) varying amplitude (phase)  in space and time. For uniform stratifications,  the horizontal wavenumber ($k$) condition for waves ($1$,$2$,$3$), given by ${k}_{(1,a)}+{k}_{(2,b)}+{k}_{(3,c)}=0$, is unaffected as $h$ is varied, where $(a,b,c)$ denote the modenumber. Moreover, the nonlinear coupling coefficients (NLC) are proportional to $1/h^2$, implying that triadic waves grow faster while travelling up a seamount.
For non-uniform stratifications, triads that do not satisfy the condition $a=b=c$ may not satisfy the horizontal wavenumber condition as $h$ is varied, and unlike uniform stratification, the NLC may not decrease (increase) monotonically with increasing (decreasing) $h$. {NLC, and hence wave growth rates for weakly nonlinear wave-wave interactions, can also vary rapidly with $h$.} The most unstable daughter wave combination of a triad with a mode-1 parent wave can also change for relatively small changes in $h$. We also investigate higher-order self-interactions in the presence of a monochromatic, small amplitude {topography}; here the {topography} behaves as a zero frequency wave. We derive the amplitude evolution equations and show that higher-order self-interactions might be a viable mechanism of energy cascade.

\end{abstract}

\section{Introduction}
Low-mode, long wavelength
internal gravity waves in oceans can travel thousands of kilometers from their generation site without dissipation \citep{zhao}. The energy in these {long} waves can cascade to small scales through a variety of mechanisms, such as  nonlinear interactions among the waves \citep{winter,mac_2013}, scattering through interaction with the seafloor {topography}  \citep{sonya_legg}, and scattering through interaction with turbulent quasigeostrophic flows \citep{vanneste_2019}. This transfer of energy to small  scales will eventually lead to turbulence and mixing, which is  essential for maintaining the meridional overturning circulation \citep{MUNK}. {In weakly nonlinear wave-wave interactions, an internal gravity wave can become unstable via resonant triad interactions if it has the largest frequency in the triad \citep{hasselmann_1967};
through this mechanism, energy is irreversibly transferred from a high frequency and low wavenumber primary wave to lower frequency and higher wavenumber secondary waves.} {In a resonant internal wave triad, a wave of angular frequency $\omega_3$ and wavevector $\mathbf{k}_{3}$ can resonantly transfer its energy to two ‘daughter’ waves when both the conditions $\mathbf{k}_{3} = \mathbf{k}_{1} + \mathbf{k}_{2}$ and $\omega_3
= \omega_1 + \omega_2$ are met \citep{davis_acrivos_1967,hasselmann_1967,phillips1967dynamics}. The former condition is a consequence of the quadratic nonlinearity
of the Navier-Stokes equations.}

Since ocean's density stratification is non-uniform, recent efforts have been directed towards understanding energy transfer in non-uniformly stratified fluids. {In \cite{varma}, the conditions for the existence of resonant weakly nonlinear wave-wave interactions in a non-uniform stratification was studied. They proved that resonant triads and self-interactions can exist if (i) they satisfy the horizontal wavenumber condition, and (ii) {each} wave's {functional form in $z-$direction} is  non-orthogonal to the nonlinear forcing terms.} 
\cite{wunsch} studied self-interaction of an internal wave mode in the presence of  a non-uniform stratification, the latter was simplified using a $3$-layer model. {It was shown that amplitude of the superharmonic wave, which is forced by the self-interaction of a parent wave, can be highly sensitive to changes in the stratification profile characteristics such as pycnocline depth and strength.} Moreover, \cite{liang_zareei_alam_2017} showed that self-interaction also occurs in the presence of uniform stratification, provided the nonlinear terms in the free surface boundary condition are taken into account. {Self-interaction was also numerically studied by \cite{sutherland}, and it was observed that, in the presence of non-uniform stratification,  self-interaction of an internal wave mode was more dominant than triadic interactions for low Coriolis frequency.} Furthermore, \cite{baker_2020} studied self-interaction of a mode under angular frequency mismatch, and  found that  the daughter wave (superharmonic wave) can return its energy to the parent wave.

Apart from the weakly nonlinear wave-wave interactions, wave-{topography} interactions (where the {topography} is of small amplitude) have also been extensively studied. In \cite{buhler_2011}, the decay of a mode--$1$ internal tide due to its interaction with a small amplitude sea floor {topography} was studied using ray-tracing. {It was shown that if the bottom bathymetry is `resonant' (see \S \ref{Section_6} for more detail),  the internal mode--$1$  interacts with the bathymetry and resonantly gives its energy to the higher modes.} The {topography} in this case acts like a stationary wave with zero angular frequency. {In \cite{couston_2017}, this scattering process was explored in a 3--dimensional setting, where the mode--$1$ internal wave was obliquely incident on a small amplitude bottom topography.} { \cite{buhler_2011} and \cite{li_mei_2014} also focused on scattering of an internal gravity wave (IGW) by a small amplitude, stationary, zero mean random topography. \cite{li_mei_2014} consider topographies that vary in zonal and meridional  directions.  Both studies, under realistic parameters,  estimate a decay length scale of about $500-1000$kms for the mode-1 wave.}

{In \cite{Mathur_14}, Green's function approach along with numerical simulations was used to study IGW scattering under the linear, inviscid limit in the presence of constant and non-constant buoyancy frequency in a 2-dimensional setting for large amplitude topographies. Height of the topography and criticality ($C_r$) were the two main factors that influence IGW scattering. In general, subcritical ($C_r < 1$) topographies were found to scatter the incoming wave lesser than supercritical topographies ($C_r > 1$). Critical topographies ($C_r \approx 1$) were the most proficient in scattering the incoming wave. 
Scattering of large amplitude waves, breaking, and the ensuing kinetic energy dissipation is a very important quantity to study since it provides an estimate for local diffusivity. 
Internal wave breaking due to different types of topographies was focused in \cite{sonya_2014}. In particular, a condition for internal wave breaking was given using the incoming wave's Froude number. The Froude number for a mode-1 wave is defined as:
\begin{equation}
    Fr = \frac{U\pi\alpha}{H\omega}
    \label{eqn:Froude_number}
\end{equation}
where $U$ and $\omega$ are respectively the peak horizontal velocity and frequency of the wave, and $H$ is the depth of the domain. $\alpha$ is the slope of the wave. As the mode-1 wave shoals up a large amplitude topography its $Fr$ increases.  It is empirically determined that if wave's $Fr$ reaches a range of $(0.3-1)$ due to shoaling, then the wave is prone to breaking. Wave's local Froude number can also increase due to reflection from a topography. Highly nonlinear features such as bores were observed in regions of the topography where the local Froude number was greater than $1$ \citep{Legg_2004}. Scattering and dissipation due to large amplitude highly supercritical topographies were focused in \cite{Klymak_2013}. Interestingly, it was observed that a mode-1 wave loses a maximum $\sim20\%$ of its energy at an isolated tall supercritical topography.}

%\textcolor{blue}{In \cite{gururaj_guha_2020}, the effect of slowly varying stratification on internal gravity waves triad in an unbounded domain was studied. In particular, the effect of stratification's variation on nonlinear coupling coefficients, detuning, group speed of the wave packets was studied. It was found that different triads undergo different amount of detuning for the same change in background stratification.}

\begin{figure}
 \centering{\includegraphics[width=0.8\textwidth]{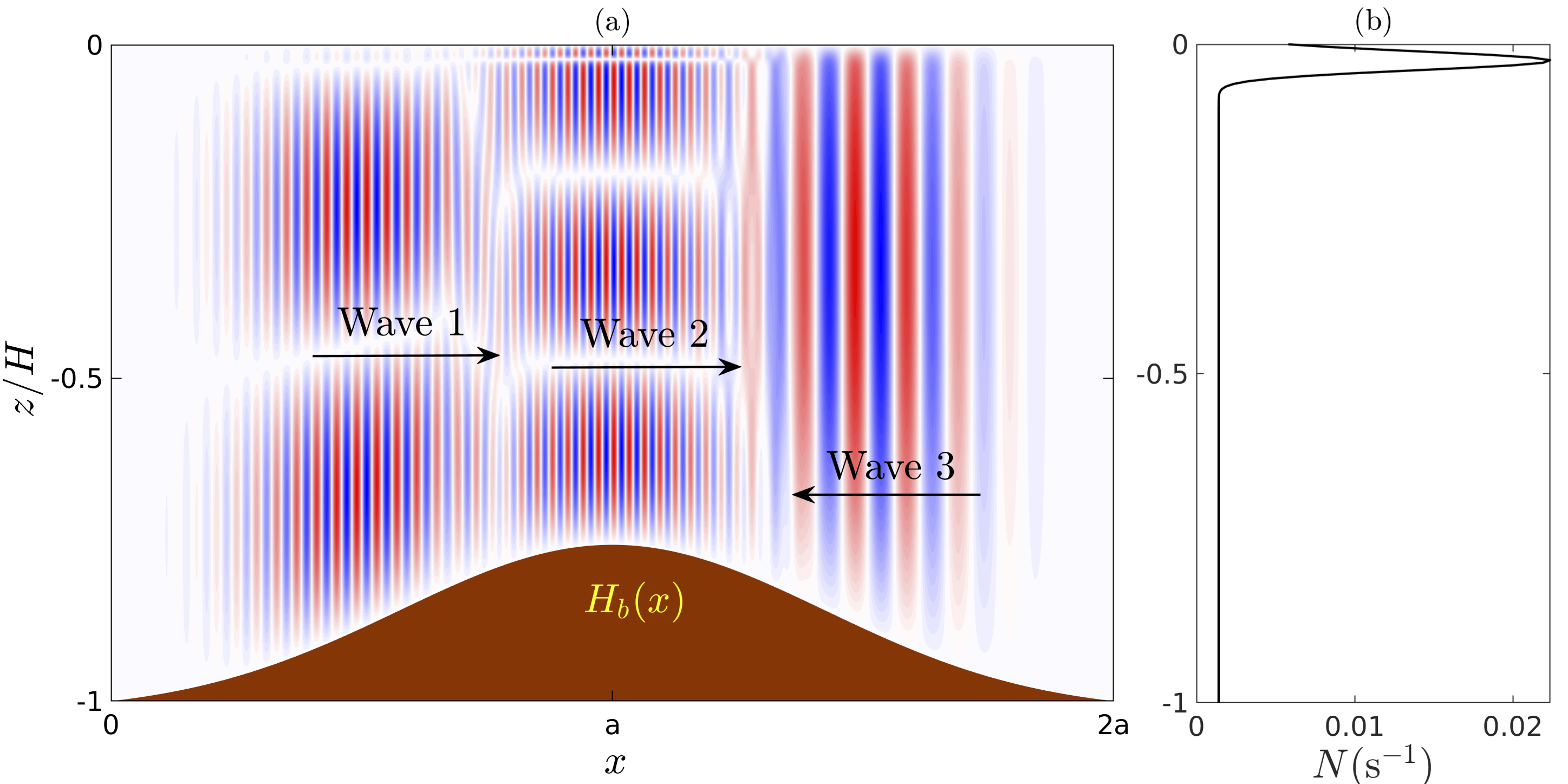}} 
  \caption{A general schematic of the problem to be studied. {The figure shows the streamfunction field of three wave packets interacting in the presence of a varying {bathymetry} $h(x)$. $H$ is the mean depth (equivalent to the depth in flat bathymetry situation), while $H_b(x)$ denotes the submarine topography shape.} (b) The stratification profile used in constructing the modes. The same non-uniform stratification model is used throughout the paper.}
  \label{fig:sc}
\end{figure}

This paper is the `Part-2' of \cite{gururaj_guha_2020}, in which the effect of non-uniform (albeit slowly varying) stratification on internal wave triads in an \emph{unbounded domain} was theoretically and numerically studied. It was shown that the variation in stratification profile may significantly affect the nonlinear coupling coefficients, (vertical wavenumber) detuning, and group speed of the wave packets constituting a triad, and hence the ensuing energy transfer. Different triads were also observed to undergo different amounts of detuning for the same change in the background stratification. The present paper extends the paradigm explored in \cite{gururaj_guha_2020} to a {\emph{vertically bounded domain}} with a {\emph{mild slope bathymetry}}. To the best of our knowledge, this is the first work that considers the effect of the variation of the ocean depth on internal wave triads.
% \textcolor{blue}{Wave-wave interactions in nonuniformly stratified vertically bounded domains have been previously considered in various studies \citep{baker_2020,VARMA_2020,young_g}. In this paper, reduced order model for wave-wave interactions in a region of varying $h$ are derived, and hence spatially varying nonlinear coupling coefficients, group speed and detuning are all involved. Moreover, the equations can model wave or finite width wave packets in a region of varying $h$. In \S \ref{Section_6}, we have also derived (and validated numerically) reduced order equations which model higher order self interactions in the presence of a small amplitude topography. Equations in \S \ref{Section_6} can also be used to model standard resonant self interactions in the presence of slowly varying large amplitude topographies.}
%To the best of our knowledge, previous studies on internal gravity wave triads have focused on unbounded domains or domains which have a constant depth 
 A simplified schematic of the setup is given in figure \ref{fig:sc}. 
 %\textcolor{red}{IN FIG-1, MAKE THE SET-UP LIKE YOUR TALK IN CIVIL. ALSO WHY USE N-hat?}  
 The motivation behind this study stems from the simple fact that ocean depth varies spatially (hence consideration of constant depth might be an over-simplification), hence waves can move from one {depth} to another while they are interacting in a {medium of varying stratification}. While the effect of change in {fluid depth} on resonant and near-resonant interactions between three distinct waves is the  primary focus of this paper, we have also studied higher order self-interactions among the internal waves in the presence of a small amplitude {topography}. As an analogy, such higher order interactions have been studied in \cite{alam_2009} for surface waves. 
{Wave-wave interactions in nonuniformly stratified vertically bounded domains have been previously considered in {various} studies Refs. \citep{baker_2020,VARMA_2020,young_g}. In this paper, reduced order model for wave-wave interactions in a region of varying $h$ are derived, and hence spatially varying nonlinear coupling coefficients, group speed and detuning are all involved. Moreover, the equations can model wave-wave interactions of wave trains or finite width wave packets in a region of varying $h$. In \S \ref{Section_6}, we have also derived (and validated numerically) reduced order equations which model higher order self interactions in the presence of a small amplitude topography. Equations in \S \ref{Section_6} can also be used to model standard resonant self interactions in the presence of slowly varying large amplitude topographies.}
 
%To study this we assume that the constituent waves of the triad have wavenumbers which is at least an order of magnitude higher than the bottom bathymetry's wavenumbers. In such scenarios, the horizontal wavenumber of the internal waves undergo a variation in space.

%In addition to perfect triads, we have also focused on near-resonant triads, that is, waves which \emph{almost} satisfy the triad condition. Such triads, in the context of internal gravity waves, was studied by \cite{mcewan}. Near resonant triads have also have previously been studied by \cite{lambnear}; it was shown that near-resonant triads can occur when internal gravity waves generated via tide-bathymetry interactions interact among themselves. The interaction strength was also found to be comparable to that of an exact triad. \\

% The paper is organized as follows. In \S \ref{Section:2}, we derive the amplitude evolution equations of the constituent waves of a triad in the presence of a slowly varying  bathymetry using the inviscid, incompressible, two-dimensional (2D) Navier-Stokes equations under Boussinesq approximation on the $f$-plane.
The paper is organized as follows. In \S \ref{Section:2}, we derive the amplitude evolution equations of the constituent waves of a triad in the presence of a slowly varying  bathymetry using the {Boussinesq Navier-Stokes equations in the $f-$plane}. {To derive these equations, the streamfunction, buoyancy perturbation, and meridional velocity due to each wave are assumed to be a product of a slowly varying amplitude and a rapidly varying phase that are functions of space and time.} In \S \ref{Section_3} and \S \ref{section:4}, the effect on the horizontal wavenumber condition when waves interact in a region of varying ocean depth in the presence of uniform and non-uniform stratification are respectively studied.
%In \S \ref{section:4}, the effect on the horizontal wavenumber condition when waves interact in a region of varying ocean depth in the presence of non-uniform stratification is studied. 
In \S \ref{section_5}, we have studied the effect of ocean depth variation on the rate of energy transfer in triadic interactions and self-interactions in the presence of a non-uniform stratification.
In \S \ref{Section_6}, we analyze higher order self-interactions of a wave in the presence of small-amplitude monochromatic {topography}. In \S \ref{Section:7}, the reduced order equations derived in this paper are validated by solving the full {Boussinesq equations} using an open source code Dedalus \citep{Dedalus}. The paper has been summarized in \S \ref{Section:8}.

\vspace{-0.2cm}
\section{Derivation of the governing equations in terrain-following coordinates \label{Section:2}}
 
The incompressible, inviscid, 2D (in the $x$--$z$ plane) { Navier-Stokes equations on the $f$--plane under the Boussinesq approximation,  hereafter referred to as the Boussinesq equations,} can be expressed in terms of the perturbation streamfunction $\psi$, meridional velocity $v$ (along $y$-direction), and the perturbation buoyancy $b$ as follows:
\begin{subequations}
\begin{align}
\frac{\partial }{\partial t}\left(\nabla^{2}\psi\right)  + \frac{\partial b}{\partial x} - f\frac{\partial v}{\partial z} & = -\{\nabla^{2}\psi,\psi\}, \label{eqn:NS_stream}\\
\frac{\partial v}{\partial t} + f\frac{\partial \psi}{\partial z}  & = -\{v,\psi\}, \label{eqn:corilios}\\
\frac{\partial b}{\partial t} - N^{2}\frac{\partial \psi}{\partial x} & = -\{b,\psi\}. \label{eqn:material_cons}
\end{align}
\end{subequations}
% \begin{equation}
% \frac{\partial }{\partial t}\left(\nabla^{2}\psi\right)  + \frac{\partial b}{\partial x} - f\frac{\partial v}{\partial z} = -\{\nabla^{2}\psi,\psi\}, \label{eqn:NS_stream}
% \end{equation}
% \addtocounter{equation}{-1}
% \renewcommand{\theequation}{\arabic{section}.\arabic{equation}b}
% \begin{equation}
% \frac{\partial v}{\partial t} + f\frac{\partial \psi}{\partial z}   = -\{v,\psi\}, \label{eqn:corilios}
% \end{equation}
% \addtocounter{equation}{-1}
% \renewcommand{\theequation}{\arabic{section}.\arabic{equation}c}
% \begin{equation}
% \frac{\partial b}{\partial t} - N^{2}\frac{\partial \psi}{\partial x}  = -\{b,\psi\}. \label{eqn:material_cons}
% \end{equation}
%\renewcommand{\theequation}{\arabic{section}.\arabic{equation}}

\noindent Here $N^{2}(z) \equiv -\left(g/\rho^{*}\right)\left(d \bar{\rho}/d z\right)$  is the squared buoyancy frequency,  $\bar{\rho}$ is the base density profile,  $\rho^{*}$ is the reference density, and $g$ is the acceleration due to gravity (directed along $-z$). The perturbation buoyancy is defined as $b \equiv -g\rho/\rho^{*}$, where $\rho$ is the perturbation density. 
 
%with respect to the wavenumbers of the waves involved in triad interactions. 
%weakly  vary with respect to variation of the waves.
%$N^{2}$, is defined in terms of the base density profile, $\bar{\rho}$, as 
%(in which $g$ is the gravity and $\rho^{*}$ is the characteristic density value).
% Horizontal velocity (x-direction), $u$, and the vertical velocity, $w$, are given by  $u = \partial \psi/\partial z$, $w = -\partial \psi/\partial x$.  The buoyancy frequency in general is function of $z$, which here is assumed to weakly vary with respect to variation of the waves. $b = -g\rho'/\rho^{*}$ is the buoyancy perturbation, where $\rho'$ is the perturbation density from the mean density $\bar{\rho}$. This problem is done without considering the Coriolis effects. 
The operator $\{G_1,G_2\} \equiv  (\partial G_1/\partial x)(\partial G_2/\partial z) - (\partial G_1/\partial z)(\partial G_2/\partial x)$ denotes the Poisson bracket, and $f$ is the Coriolis frequency.
%o{TThe physics is expected to be qualitatively similar if we consider the problem in 3D\todhis is a problematic statement}, hence for simplification we have restricted our analyses to 2D. 
% The Coriolis effect has also been neglected since it would not change the results qualitatively. 
Viscous effects have been neglected owing to the fact that we consider waves with long wavelengths.
%The rationale for neglecting viscous effects is because  our main focus is on waves with long wavelengths.
%Furthermore, our focus is on wave triads whose wavelengths are long enough, hence viscous effects are of little consequence. 

% The fluid domain is defined by $\mathcal{B}=\{(x,z)\in \mathbb{R}^2: h( x)\!<\!z\!<0\}$; the free surface ($z=0$) is modelled as a rigid-lid (i.e. zero vertical velocity, which leads to $\psi(x,0) = 0$) while the bottom boundary at  $z = h(x)$ satisfies the impenetrable boundary condition,  yielding $\psi(x,h(x)) = 0$. 

The fluid domain is bounded at the top ($z=0$) by  a rigid-lid (i.e., zero vertical velocity,  leading to the boundary condition $\psi(x,0) = 0$). The bottom boundary at $z = h(x)$ satisfies the impenetrable boundary condition  $\psi(x,h(x)) = 0$. 
 
%The fluid domain is bounded by the free surface at $z=0$, which is modelled as a rigid-lid (i.e. zero vertical velocity,  leading to the boundary condition $\psi(x,0) = 0$). Moreover, fluid domain is enclosed by the bottom boundary at $z = h(x)$ where the impenetrable boundary condition,  given by $\psi(x,h(x)) = 0$, is satisfied. 
 
%Here $\epsilon_k$ is a parameter denoting the variation of the bottom bathymetry with respect to the horizontal wavenumber of the internal gravity waves that we discuss later in the paper.
%When $\epsilon_k \ll \mathcal{O}(1)$, the bottom bathymetry  varies slowly with $x$. However $\epsilon_k$ may also be an $\mathcal{O}(1)$ quantity, implying topographic variations occuring at the leading order.

%is assumed to be bounded in vertical direction between free surface ($z=0$) and the bottom bathymetry at $z=h(x)$. At $z = 0$, we use the rigid-lid approximation -- the vertical velocity is assumed to be zero, leading to the boundary condition $\psi(x,0) = 0$. We use the impenetrable boundary condition at the bottom bathymetry, $z = h(\epsilon_k x)$, yielding $\psi(x,h) = 0$. Here $\epsilon_k$ is a parameter which denotes the variation of the bottom bathymetry with respect to the horizontal wavenumber of the waves forming the triads. \\

% We define $\epsilon_k$ as:
% \begin{equation}
%     \epsilon_k = \textnormal{max} \left(\abs{\frac{1}{k_{\textnormal{min}}}\frac{1}{h}\frac{\partial h}{\partial x}}  \right),
%     \label{eqn:eps_k}
% \end{equation}
Instead of solving the fully nonlinear equations \eqref{eqn:NS_stream}--\eqref{eqn:material_cons} numerically, we combine (\ref{eqn:NS_stream})--(\ref{eqn:material_cons}) into a single equation and employ a multiple-scale analysis. To this end, we perform $\partial$\eqref{eqn:NS_stream}\!$/\partial t+$    $f\partial$\eqref{eqn:corilios}\!$/\partial z$  $-\partial$\eqref{eqn:material_cons}\!$/\partial x$, which results in
\begin{equation}
\frac{\partial^{2} }{\partial t^{2}}\left(\nabla^{2}\psi\right) + N^{2}\frac{\partial^{2} \psi}{\partial x^{2}} + f^{2}\frac{\partial^{2} \psi}{\partial z^{2}}  = -\frac{\partial}{\partial t} \left(\{\nabla^{2}\psi,\psi\}\right) + \frac{\partial }{\partial x}\left(\{b,\psi\}\right) - f\frac{\partial }{\partial z}\left(\{v,\psi\}\right) .
\label{eqn:combined}
\end{equation} 

\noindent Following the approach of \cite{mauge}, we now change the governing equations to terrain following coordinates, where a new variable ($\eta$) is defined as:
\begin{equation}
    \eta \equiv - \frac{z}{h(x)}.
\label{eqn:eta_def}
\end{equation} 
According to the definition (\ref{eqn:eta_def}), the bottom boundary condition at $z=h(x)$ would now be enforced at $\eta = -1$, while the surface boundary condition at $z=0$ remains unaltered, except that it is now at $\eta=0$. The governing equations, which are in the $x$--$z$ coordinates, need to be transformed into the $x$--$\eta$ coordinates. The correspondence between the variables in the $x$--$z$ and $x$--$\eta$ coordinate systems are as follows:
%We define the following functions:
\begin{equation}
\psi(x,z,t) \Rightarrow \Psi(x,\eta,t), \hspace{1cm} b(x,z,t) \Rightarrow B(x,\eta,t), \hspace{1cm} v(x,z,t) \Rightarrow \mathcal{V}(x,\eta,t).
\label{eqn:change_var}
\end{equation}
%On substituting the new variables in \eqref{eqn:combined},  
On transforming the differential operators from $x$--$z$ coordinates to $x$--$\eta$ coordinates  and substituting the transformed variables in \eqref{eqn:combined}, we arrive at
% \begin{align}
% \frac{\partial^{2} }{\partial t^{2}}\left[({L}_{xx}+{L}_{\eta\eta})\Psi\right] + N^{2}(- h(x)\eta){L}_{xx} (\Psi) +  f^2{L}_{\eta\eta}(\Psi) & = -\frac{\partial}{\partial t} \left[\mathcal{J}\{({L}_{xx}+{L}_{\eta\eta})\Psi,\Psi\}\right] \nonumber \\ & + {L}_{x}\left(\mathcal{J}\{B,\Psi\}\right) - f{L}_{\eta}\left(\mathcal{J}\{v,\Psi\}\right),
% \label{eqn:combined_eta}
% \end{align}
\begin{align}
\left[\frac{\partial^{2} }{\partial t^{2}}({L}_{xx}+{L}_{\eta\eta}) + N^{2}(-h(x)\eta){L}_{xx}  +  f^2{L}_{\eta\eta}\right]\Psi & = -\frac{\partial}{\partial t} \left[\mathcal{J}\{({L}_{xx}+{L}_{\eta\eta})\Psi,\Psi\}\right] \nonumber \\ & + {L}_{x}\left(\mathcal{J}\{B,\Psi\}\right) - f{L}_{\eta}\left(\mathcal{J}\{\mathcal{V},\Psi\}\right),
\label{eqn:combined_eta}
\end{align}
where the operators $ L_{x}, L_{\eta}, L_{xx},  L_{\eta\eta}$, and $\mathcal{J}\{G_1,G_2\}$ have the following definitions:
 \begin{subequations}
\begin{equation}
L_{x} \equiv  \frac{\partial}{\partial x} + \frac{\partial \eta}{\partial x}\frac{\partial}{\partial \eta}, \hspace{1cm} L_{\eta} \equiv -\frac{1}{h}\frac{\partial}{\partial \eta}, \hspace{1cm} L_{\eta\eta} \equiv  \frac{1}{h^{2}}\frac{\partial^{2}}{\partial \eta^{2}},
\label{eqn:operator_def_1}
\end{equation}
\begin{equation}
L_{xx} \equiv  \frac{\partial^{2}}{\partial x^{2}} + \frac{\eta^{2}}{h^{2}}\left(\frac{\partial h}{\partial x}\right)^{2}\frac{\partial^{2}}{\partial \eta^{2}} - 2\frac{\eta}{h}\left(\frac{\partial h}{\partial x}\right)\frac{\partial^{2}}{\partial \eta \partial x} + \frac{\eta}{h}\left[\frac{2}{h}\left(\frac{\partial h}{\partial x}\right)^{2} - \frac{\partial^{2} h}{\partial x^{2}} \right]\frac{\partial}{\partial \eta},
\label{eqn:operator_def_2}
\end{equation}
\begin{equation}
\mathcal{J}\{G_1,G_2\} \equiv  {L}_{x} (G_1){L}_{\eta} (G_2) - {L}_{\eta} (G_1){L}_{x} (G_2).
\label{eqn:operator_def_3}
\end{equation}
\end{subequations}

For performing multiple-scale analysis, we assume wavelike perturbations, and the streamfunction due to the $j$-th  wave ($j = 1,2,3$) is given according to the following ansatz:
\begin{equation}
\Psi_{j} =  a_{j}(\epsilon_{x} x,\epsilon_{t} t)\Xi_{j}(x,\eta,t)  + \mathrm{c.c.}, 
\label{eqn:stream_ans}
\end{equation}
where `c.c.' denotes the complex conjugate, $a_{j}$ is the slowly varying complex amplitude, and {$\Xi_j(x,\eta,t)$} is the rapidly varying phase part of the $j$-th wave.  The small parameters $\epsilon_{t}$ and $\epsilon_{x}$ are respectively used to denote the weak variation of the amplitude function with time and streamwise ($x$) direction. The amplitude is assumed to be an $\mathcal{O}(\epsilon_{a})$ quantity, where $\epsilon_{a}$ is a small parameter. The bathymetry ($h$), { which is simply the negative of the fluid depth},  is assumed to be of the form:
\begin{equation}
    h = -H + \epsilon_h H_b(k_b x),
\label{eqn:topo}
\end{equation}
where $H$ represents the mean depth of the fluid domain, $H_b$ denotes the submarine topography shape, $\epsilon_h$ is its amplitude, 
%\textcolor{blue}{and $k_b$ represents a nondimensional wavenumber associated with the bathymetry.}
and $k_b^{-1}$ represents the length scale of the bathymetry. %\textcolor{purple}{Hence bathymetry $(h(x))$  is simply the negative of the fluid depth}.
We always assume the bathymetry to have a `mild slope'; for this  we use an analog condition of that used for surface gravity waves \citep{meyer,kirby_1986}:
%$\epsilon_h$ and $\epsilon_k$ are parameters, where $H_b$ is same order of magnitude as $H$. $H$ represents the mean depth of the fluid domain.  The bathymetry ($h$) is always assumed to be slowly varying, that is:
% \begin{equation}
%     \textcolor{blue}{\frac{1}{\mathcal{K}_{j}}}\frac{1}{k_{j}h}\frac{\partial h}{\partial x}  =\mathcal{O}(\epsilon_h \epsilon_k) \ll \mathcal{O}(1),
%     \label{eqn:wav_no_assumption}
% \end{equation}
\begin{equation}
   {\frac{1}{\mathcal{K}_{j}}}\frac{\partial h}{\partial x}  =\mathcal{O}(\epsilon_h \epsilon_k) \ll \mathcal{O}(1),
    \label{eqn:wav_no_assumption}
\end{equation}
% \begin{equation}
%     \frac{1}{h}\frac{\partial h}{\partial x} \sim \mathcal{O} (\epsilon_h k_b) = \mathcal{O}(\epsilon_h \epsilon_k k_{\textnormal{IW}}) \ll \mathcal{O}(1),
%     \label{eqn:wav_no_assumption}
% \end{equation} 
where $\mathcal{K}_j\equiv k_j h$ is the nondimensional horizontal
wavenumber ($k_j$ being the horizontal wavenumber) of the $j-$th internal wave. 
%$k_j$ is the horizontal wavenumber of the $j-$th internal wave.
%\textcolor{blue}{ Moreover, the relation $k_{j}^{-1} = \epsilon_k k_b^{-1}$ }
Moreover, the relation $k_{j}^{-1} = \epsilon_k k_b^{-1}$ is used in \eqref{eqn:wav_no_assumption},
% where $k_j$ is the horizontal wavenumber of the $j-$th wave. Moreover, the relation $k_{\textnormal{j}}^{-1} = \epsilon_k k_b^{-1}$ is the horizontal length scale of the internal gravity waves considered, \textcolor{red}{ \sout{ with $k_b =\mathcal{O}(1)$.}} 
%In this paper, we \emph{only} consider the situation $\epsilon_h \epsilon_k \ll \mathcal{O}(1)$, which means that
which implies that either of the parameters, $\epsilon_h$ or $\epsilon_k$, could be a small quantity while the other could potentially be an $\mathcal{O}(1)$ quantity.
%In other words, if the amplitude of the bathymetry is comparable to the mean depth, the topographic variation is mild; whereas if the amplitude of the bathymetry is small, the topographic variation could be comparable to the scale of the internal waves.
%The assumption regarding the bathymetry slope made in \eqref{eqn:wav_no_assumption} is similar to the mild slope approximation used in studying surface gravity waves in the presence of slowly varying mean depth (For example in \cite{meyer,kirby_1986}).
We note in passing that the mild slope condition in our case can \emph{still} lead to internal gravity wave scattering \footnote{Internal wave scattering is largely dependent on the slope of the wave, which is almost constant (wave's slope is dependent on $N$, which is nearly constant away from the pycnocline)  even for higher modes whose horizontal wavenumber is much larger.}.  Scaling analysis to find the relations between these small parameters is given in appendix \ref{app:A}.

\subsection{Leading order analysis} 
\label{subsec:los}
\begin{figure}
 \centering{\includegraphics[width=0.8\linewidth]{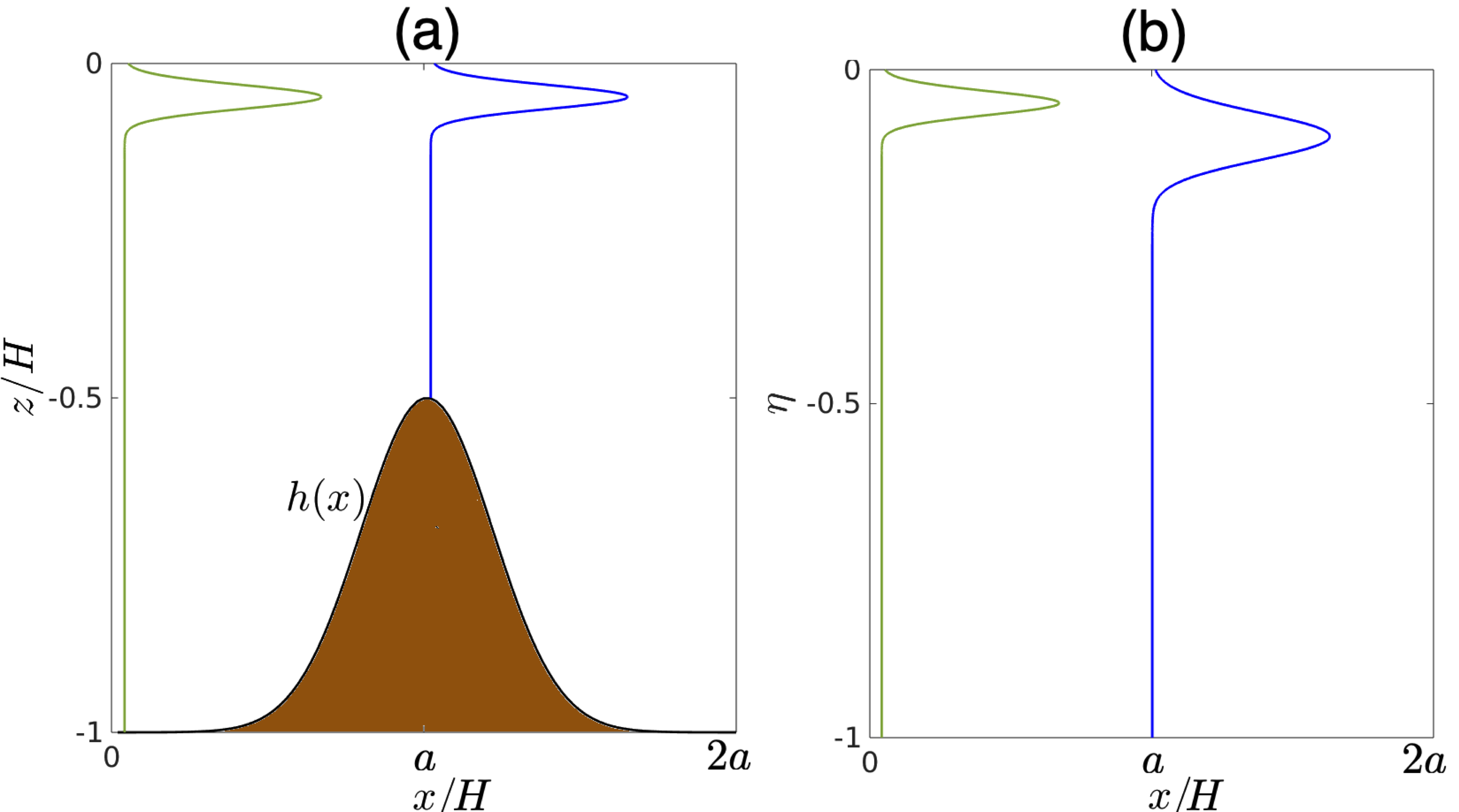}}
  \caption{The effective change in the stratification profile $N(z)$  when the coordinates are changed from (a) $x$--$z$ to (b) $x$--$\eta$ in the presence of bathymetry. For the latter case, if $N$ is a function of $z$ in $x$--$z$, it becomes a function of both $\eta$ and $x$ in $x$--$\eta$. $N$ profiles corresponding to the top of a seamount and an abyssal plain region have been respectively denoted by blue and green colors.}
  \label{fig:Strat_change_sc}
\end{figure}

Next we substitute \eqref{eqn:stream_ans} in \eqref{eqn:combined_eta}. %\textcolor{red}{Due to slow variation of the topography, differential operators containing derivatives of topography are at least an order of magnitude smaller in comparison to the operator ${\partial^2}/{\partial x^2}$ in \eqref{eqn:operator_def_1}--\eqref{eqn:operator_def_3}. As a result, at leading order, $L_{xx} = {\partial^2}/{\partial x^2}$}. 
At the leading order $(\mathcal{O}(\epsilon_{a}))$, the governing equation \eqref{eqn:combined_eta} reduces to:
\begin{equation}
     \left(\frac{\partial^2}{\partial x^2} +  \frac{1}{ h^{2}}\frac{\partial^{2}}{\partial \eta^{2}}\right) \frac{\partial^2 \Xi_j}{\partial t^{2}} +  N^2(- h(x)\eta) \frac{\partial^2 \Xi_j}{\partial x^2}  +  \frac{f^2}{ h^{2}}\frac{\partial^2 \Xi_j}{\partial \eta^{2}} =  0.
\label{eqn:leading_order}
\end{equation}
Hereafter we drop the argument of $N^2$, assuming it is implied. Furthermore assuming  $\Xi_j = \widehat{\Xi}_j(x,\eta)\ee^{-\ii\omega_j t}$, where $\omega_j\in \mathbb{R}^+$ is the angular frequency of the $j$-th internal wave,  (\ref{eqn:leading_order}) simplifies to
\begin{equation}
     \left[(N^2-\omega_j^{2})\frac{\partial^2}{\partial x^2} -  \frac{\omega_j^{2}-f^2}{h^{2}}\frac{\partial^2}{\partial \eta^2}\right]\widehat{\Xi}_j  =  0.
\label{eqn:leadingorder_xdirection}
\end{equation}
For a mild slope bathymetry (see appendix \ref{app:A} for details), we can use variable separation to solve (\ref{eqn:leadingorder_xdirection}) at the leading order. To this end we assume  $\widehat{\Xi}_j = {\phi}_j(\eta;x) {P}_j(x)$, which leads to
 \begin{equation}
     \frac{h^{2}}{{P}_j}\frac{\partial^2 {P}_j}{\partial x^2} = \frac{\omega_j^{2}-f^2}{N^2-\omega^{2}_j} \frac{1}{{\phi}_j}\frac{\partial^2 {\phi}_j}{\partial \eta^2} = -\mathcal{K}_{j}^{2},
\label{eqn:eigen_step1}
\end{equation}
where ${\phi}_j$ parametrically depends on $x$ via $h$. We emphasize that in the $x$--$\eta$ coordinates, the presence of bathymetry makes $N$ to also be a function of $x$; see figure \ref{fig:Strat_change_sc} for clarity.

Two separate equations, one for ${P}_j$ and the other for ${\phi}_j$, can be formed from  (\ref{eqn:eigen_step1}):
\begin{subequations}
\begin{align}
 \hspace{0.5cm} \left[\frac{\partial^2}{\partial x^2} + \frac{\mathcal{K}_{j}^{2}}{h^{2}}\right]{P}_j &= 0,\label{eqn:eigen1} \\
\mathcal{L}_j {\phi}_j \equiv \left[\frac{\partial^2}{\partial \eta^2} + \mathcal{K}_{j}^{2} \chi_j^2 \right] {\phi}_j &= 0,
\label{eqn:eigen2}
\end{align}
\end{subequations}

\noindent where $\chi_j \equiv \sqrt{\left({N^2-\omega^{2}_j)}/{(\omega_j^{2}-f^2}\right)}$ is defined for convenience. The boundary conditions for \eqref{eqn:eigen2} are ${\phi}_j = 0$ at $\eta = 0,-1$.  The nondimensional horizontal wavenumber of the $j-$th wave, i.e. $\mathcal{K}_j$, is the set of eigenvalues obtained from \eqref{eqn:eigen2}, which can vary in $x$ when $N$ is a function of $x$ in $x$--$\eta$ coordinates.
%Note that $\mathcal{K}_{j}\equiv k_j h$ 
%is the nondimensional horizontal wavenumber of the $j-$th wave. }
% Furthermore, the eigenvalue $\mathcal{K}_{j}\equiv k_j h$ 
% is a nondimensional horizontal wavenumber that can vary in $x$ because of the bathymetry (through $h$), and is found by solving (\ref{eqn:eigen2}).  
{ An important point to note is the convention used in our study. While positive (negative) $k_j$ implies waves propagating along $+x$ ($-x$), owing to the fact that $h$ is negative, $\mathcal{K}_{j}$ follows the exactly opposite convention.  This means that a negative (positive) $\mathcal{K}_{j}$ implies that the wave is traveling along $+x$ ($-x$) direction}. 
 {Moreover we  notice that \eqref{eqn:eigen2} does not explicitly depend on $h$. The only way \eqref{eqn:eigen2} can be influenced by $h$ is through $N$ when the latter varies in the $z-$direction (in $x$--$z$ coordinates).
%leading to $N(-h(x)\eta)$ in the $x$--$\eta$ coordinates. 
%This effect in \eqref{eqn:eigen2} is encapsulated in the $\chi_j$ term.  
 %This is because $N$ will become a function of $h$ if $N$ varies in $z-$direction in $x-z$ coordinates. This will result in $\mathcal{K}_j, \phi_j$ also changing as $h$ varies. 
However for a uniform stratification, i.e. $N=\textrm{constant}$, eigenvalues of \eqref{eqn:eigen2} are independent of $h$.} In this case the eigenvalues are given by
\begin{equation}
    \mathcal{M}_j \equiv \mathcal{K}_j \chi_j = n \pi,
 \label{eqn:lambda_unif}    
\end{equation}
where $n\in \mathbb{Z}^+$. We also observe that the quantity $\mathcal{M}_j$ behaves like the vertical wavenumber of the wave that is {nondimensionalised} by the local {bathymetry} $h$. 
 
%Meanwhile, at the leading order, the solution of ${P}_j$ can be found using
Meanwhile, ${P}_j$ at the leading order of the WKB approximation is given by
\begin{equation}
    {P}_j = \exp{\ii \int_0^{x} \frac{\mathcal{K}_j(x')}{h(x')} \hspace{0.1cm} dx'}.  %\textnormal{where} \hspace{1cm} k_j = \frac{\mathcal{K}_j}{h}
\label{eqn:firstorder_wkb}
\end{equation}
% We introduce a function $\beta_j (\epsilon_k x)$ { such that  $\hat{P}_j$ is corrected to} $P_j \equiv \hat{P}/\beta_j$.
{We introduce a function $\beta_j (\epsilon_k x)$  such that ${P}_j$ is corrected to $P_j/\beta_j$.} This slow varying function $\beta_j (\epsilon_k x)$, which acts as a correction to the first order WKB solution (\ref{eqn:firstorder_wkb}), is given in \eqref{eqn:WKB_cr_final}. {We note in passing that $P_j/\beta_j$ {is still a solution} of (\ref{eqn:eigen1}) in the leading order even after the above-mentioned {correction}. 
% where $k_j \equiv \mathcal{K}_j/h$ is the horizontal wavenumber of the internal wave. We introduce a function $\beta_j (\epsilon_k x)$  \Omit{ such that  $\hat{P}_j$ is corrected to}{for normalizing $\hat{P}_j$, i.e.} $P_j \equiv \hat{P}/\beta_j$. This slow varying function $\beta_j (\epsilon_k x)$, which acts as a correction to the first order WKB solution (\ref{eqn:firstorder_wkb}), is derived later in \eqref{eqn:WKB_cr}--\eqref{eqn:WKB_cr_final}. We note in passing that $P_j$ will still solve  (\ref{eqn:eigen1}) in the leading order even after the above-mentioned \Omit{correction}{normalization}. 
To normalize the {eigenfunction} of the waves obtained from \eqref{eqn:eigen2}, every wave's ${\phi}_j$ is constrained to satisfy:
\begin{equation}
    \frac{1}{2} \int_{-1}^{0} \frac{1}{h^2} \left[\mathcal{K}_j^2 {\phi}_j^2 + \left( \frac{\partial {\phi}_j}{\partial \eta} \right)^2 \right] \partial \eta = 1.
    \label{eqn:nondim_energy}
\end{equation}}
% To normalize the vertical structure of the waves obtained from \eqref{eqn:eigen2}, every wave's vertical function ${\phi}_j$ is divided by the square-root of the energy density $(\mathcal{E}_j)$ at a given $h$, which is as follows:
% \begin{equation}
%   {\phi}_j \equiv \frac{\hat{\phi}_j}{\sqrt{\mathcal{E}_j}}, \hspace{1cm} \mathcal{E}_j(h) \equiv \frac{1}{2} \int_{-1}^{0} \frac{1}{h^2} \left[\mathcal{K}_j^2 \hat{\phi}_j^2 + \left( \frac{\partial \hat{\phi}_j}{\partial \eta} \right)^2 \right] \partial \eta.
%     \label{eqn:nondim_energy}
% \end{equation}
\noindent After this  normalisation, waves having the same amplitude ($a_j$) will also have the same energy density at a given $h$, provided $\beta_j=1$. 
%\textcolor{red}{Hence the final $\Psi_j$ field after the normalisation is:
% \begin{equation}
%     \Psi_j = a_{j} \phi_j P_{j}\ee^{-\ii\omega_{j}t} + \mathrm{c.c.},
%     \label{eqn:final_Stream_ans}
% \end{equation}}

The meridional velocity and the buoyancy perturbation at the leading order can be obtained by respectively converting (\ref{eqn:corilios}) and (\ref{eqn:material_cons}) into the $x$--$\eta$ coordinates and then 
substituting the streamfunction ansatz \eqref{eqn:stream_ans}:
%in (\ref{eqn:material_cons}) after converting (\ref{eqn:material_cons}) into the $x$--$\eta$ coordinates, which is as follows:
\begin{align}
\mathcal{V}_{j} &= \ii\frac{ f}{h\omega_{j}}\frac{a_{j}}{\beta_j}\frac{\partial \phi_j}{\partial \eta} P_{j}\ee^{-\ii\omega_{j}t} +  \mathrm{c.c.}, 
\label{eqn:v_ans}\\
B_{j} &= \ii\frac{N^{2}}{\omega_{j}} \frac{a_{j}}{\beta_j}\frac{\partial P_{j}}{\partial x}\phi_j \ee^{-\ii\omega_{j}t} +  \mathrm{c.c}. \label{eqn:buo_ans} 
\end{align}
% In a similar way, the meridional velocity at the leading order is obtained by converting (\ref{eqn:corilios}) into the $x$--$\eta$ coordinates and then substituting the streamfunction ansatz \eqref{eqn:stream_ans}:
%in (\ref{eqn:corilios}) after
%which is given below:
% \begin{equation}
% \mathcal{V}_{j} = \ii\frac{ f}{h\omega_{j}}\frac{a_{j}}{\beta_j}\frac{\partial \phi_j}{\partial \eta} P_{j}\ee^{-\ii\omega_{j}t} +  \mathrm{c.c}. 
% \label{eqn:v_ans}
% \end{equation} 
\subsection{Second order analysis}\label{subsec:hos}
\subsubsection{ Amplitude evolution equations for a resonant triad in non-uniform stratification} \label{sec:2.2.1}

%Triad interaction between three internal waves occurs at $\mathcal{O}(\epsilon^2)$, whose amplitude evolution equations are given in \eqref{eqn:wave1}--\eqref{eqn:wave3}.

%hence we isolate the $\mathcal{O}(\epsilon^2)$ terms after substituting the streamfunction ansatz (\ref{eqn:stream_ans}) and the buoyancy perturbation \eqref{eqn:buo_ans} in \eqref{eqn:combined_eta}. 
Triad interaction between three internal waves occurs at $\mathcal{O}(\epsilon^2)$. Below we describe the detailed derivation that finally leads to the   amplitude evolution equations \eqref{eqn:wave1}--\eqref{eqn:wave3} of the waves constituting a triad. 
%, whose detailed derivation procedure is given below.

After substituting the streamfunction (\ref{eqn:stream_ans}),  meridional velocity \eqref{eqn:v_ans}, and buoyancy perturbation \eqref{eqn:buo_ans} in \eqref{eqn:combined_eta}, the equation for the $j$-th wave can be written as:
\begin{equation}
 a_{j} \frac{\mathfrak{P}_{j}}{\beta_j} \mathcal{L}_j \phi_j  = -\mathcal{F}_j,
 \label{eqn:main}
\end{equation} 
where  $\mathfrak{P}_{j} \equiv P_j\ee^{-\ii \omega_j t}$, $\mathcal{L}_j$ has been defined in \eqref{eqn:eigen2}, and
\begin{align}
% \mathcal{L}_j &\equiv   \frac{\partial^2 }{\partial \eta^2} + \mathcal{K}_{j}^{2} \chi_j^2, \label{eqn:linear_op} \\
\mathcal{F}_j &\equiv \hspace{0.1cm} \underbrace{ \ii \frac{\partial a_j}{\partial t}\left(\phi_j \mathcal{K}_{j}^{2} - \frac{\partial^{2} \phi_j}{\partial \eta^{2}} \right)\frac{2 \omega_j}{h^{2}}\left(\frac{\mathfrak{P}_{j}}{\beta_j}\right) + 2\ii(N^{2}-\omega_j^2)\left(   \frac{\mathcal{K}_{j}}{h} \phi_j \frac{\partial a_j}{\partial x}\right)\left(\frac{\mathfrak{P}_{j}}{\beta_j}\right)}_\text{Linear term-1} \nonumber\\
&+ \underbrace{ \ii(N^{2}-\omega_j^2)\frac{\mathcal{K}_{j}}{h}\left[ 2\frac{\partial \phi_j}{\partial x} +   \frac{\phi_jh}{\mathcal{K}_{j}}\frac{\partial }{\partial x}\left(\frac{\mathcal{K}_{j}}{h} \right) -\frac{2\eta}{h}\frac{\partial h}{\partial x}\frac{\partial \phi_j}{\partial \eta}-2\frac{\phi_j}{\beta_j}\frac{d (\beta_j)}{dx}\right]\left( a_{j} \frac{\mathfrak{P}_{j}}{\beta_j}\right)}_\text{Linear term-2} - \textnormal{NL}_j. \label{eqn:complete_f}
 \end{align}

\noindent {$\mathcal{F}_j$ is the collection of all the linear and nonlinear ($\textnormal{NL}_j$) terms at $\mathcal{O}(\epsilon^2)$ which have the phase of the $j-$th wave.} 
%\textcolor{red}{Linear terms at $\mathcal{O}(\epsilon^2)$ contain $\mathcal{O}(\epsilon_a\epsilon_t)$, $\mathcal{O}(\epsilon_a\epsilon_x)$, and $\mathcal{O}(\epsilon_a\epsilon_h\epsilon_k)$ terms. Note that for large amplitude topographies, $\epsilon_h \sim 1$. Hence  $\mathcal{O}(\epsilon_a\epsilon_h\epsilon_k)$ would still be an $\mathcal{O}(\epsilon^2)$ term. `Linear term-2' purely arises because of the variation of $h$, and these terms ensure the conservation of energy of a packet as it moves through varying fluid depth. Note that in the absence of nonlinear terms, the wave packets' energy have to be individually conserved since we assume scattering cannot occur. `Linear term-2' would be identically zero if $h$ is a constant.}
Equation \eqref{eqn:main} can have a non-trivial solution when $\mathcal{F}_j$ is orthogonal to the adjoint solutions of the linear operator $\mathcal{L}_j$, and this procedure is outlined in \cite{craik_1971}. The complete mathematical proof for using such condition is given in detail in \citet[\S 9.34]{Ince}. Following \cite{craik_1971}, $\mathcal{F}_j$ is multiplied by $\phi_j$ (since $\mathcal{L}_j$ is a self-adjoint operator, $\phi_j$ is also the solution of the adjoint  of $\mathcal{L}_j$) and then  integrated in the $\eta$ direction inside the boundary limits. This would result in:
 \begin{align}
&\hspace{-0.5cm} 2\left[\ii \omega_j\frac{\partial a_j}{\partial t}\left(\gamma^{(1)}_j \mathcal{K}_{j}^{2} - \gamma^{(2)}_j \right)\frac{1}{h^{2}} + \ii\gamma^{(3)}_j\left(   \frac{\mathcal{K}_{j}}{h} \frac{\partial a_j}{\partial x}\right)\right]\frac{\mathfrak{P}_{j}}{\beta_j}\nonumber\\
&\hspace{-0.5cm}+ \ii\frac{\mathcal{K}_{j}}{h}\left[ 2\gamma^{(4)}_j + \frac{h\gamma^{(3)}_j}{\mathcal{K}_{j}}  \frac{\partial }{\partial x}\left(\frac{\mathcal{K}_{j}}{h} \right) -\gamma^{(5)}_j\frac{2}{h}\frac{\partial h}{\partial x} - \frac{2\gamma^{(3)}_j}{\beta_j}  \frac{d \beta_j}{dx} \right]a_j\frac{\mathfrak{P}_{j}}{\beta_j} = \int_{-1}^{0} \textnormal{NL}_j \phi_j d\eta, 
 \label{eqn:LHS_triad_after_ortho}
  \end{align}
%{ \noindent  where $\gamma^{(n)}_j$, defined below, are functions which vary in the $x$-direction and are obtained after integration in the $\eta$ direction:}
\noindent  where $\gamma^{(n)}_j$ are functions that vary in the $x$-direction, and are obtained after integration in the $\eta$ direction. $\gamma^{(n)}_j$ are provided in appendix \ref{app:C}.
% \begin{equation*}
%  \gamma^{(1)}_j = {\int^{0}_{-1}\phi_j^2 d \eta}  \hspace{1cm} \gamma^{(2)}_j = {\int^{0}_{-1}\phi_j \frac{\partial^{2} \phi_j}{\partial \eta^{2}} d \eta} \hspace{1cm} \gamma^{(3)}_j = {\int^{0}_{-1} (N^{2}-\omega_j^2) \phi_j^2 d \eta}
% \end{equation*}
% \begin{equation*}
%  \gamma^{(4)}_j = {\int^{0}_{-1} (N^{2}-\omega_j^2) \phi_j \frac{\partial \phi_j}{\partial x} d \eta} \hspace{1cm} \gamma^{(5)}_j = {\int^{0}_{-1}\eta (N^{2}-\omega_j^2) \phi_j \frac{\partial \phi_j}{\partial \eta} d \eta} 
% \end{equation*}
% \begin{equation*}
% \gamma^{(6)}_j = {\int^{0}_{-1}\eta^2 (N^{2}-\omega_j^2) \phi_j \frac{\partial^2 \phi_j}{\partial \eta^2} d \eta} 
% \end{equation*} 
%\sout{The quantities $\gamma^{(6)}_j$, $\gamma^{(7)}_j$ and $\gamma^{(8)}_j$ will be used later in the paper in \S \ref{Section_6} and in appendices \ref{app:A} and \ref{app:B}.} 
Up to this point, $\beta_j$ is an arbitrary function, and for convenience, we define $\beta_j$ such that the second square-bracketed term in the LHS of \eqref{eqn:LHS_triad_after_ortho} vanishes identically. It  also implies that `Linear term-2' in \eqref{eqn:complete_f} also vanishes identically.  In mathematical terms this means, 
% \begin{equation}
%      \frac{\gamma^{(3)}_j}{\beta_j} \frac{2\mathcal{K}_{j}}{h} \frac{d \beta_j}{dx} = \left[\frac{2\mathcal{K}_{j}}{h} \gamma^{(4)}_j +  \gamma^{(3)}_j  \frac{\partial }{\partial x}\left(\frac{\mathcal{K}_{j}}{h} \right) -\gamma^{(5)}_j\frac{2}{h}\frac{\partial h}{\partial x}\left(\frac{\mathcal{K}_{j}}{h} \right) \right],
% \label{eqn:WKB_cr}
% \end{equation}
% \noindent solving which we obtain
%the solution for $\beta_j$ is obtained which is given below:
\begin{equation}
   \beta_j  = \exp \left\{ \int^{x}_0 \frac{h}{2\mathcal{K}_{j}} \frac{1}{\gamma^{(3)}_j} \left[ \left(2\gamma^{(4)}_j - \gamma^{(5)}_j\frac{2}{h}\frac{\partial h}{\partial x} \right) \left(\frac{\mathcal{K}_{j}}{h} \right)  +  \gamma^{(3)}_j  \frac{\partial }{\partial x}\left(\frac{\mathcal{K}_{j}}{h} \right)  \right]dx \right\}. 
\label{eqn:WKB_cr_final}
\end{equation}
{For constant $N$, $\beta_j$ can be analytically simplified to $\beta_j = h(x)/h(0) = -h(x)/H$, where it is assumed that $h(0)=-H$.} {We note in passing that the equivalent of $\beta_j$ functions was derived in \cite{leg_lahaye} using a different approach.} {For this particular choice of $\beta_j$, if the amplitudes $a_j$ are $x-$invariant, the energy flux will be  $x-$invariant as well, regardless of the modal shape, or depth.} 
%however significant attention on the functions' behaviour was not explored. This was because the primary focus of their work was on the topographic scattering.
%After using \eqref{eqn:WKB_cr_final  
%Substituting  the definition of $\beta_j$, i.e. \eqref{eqn:WKB_cr_final} in \eqref{eqn:LHS_triad_after_ortho}, the $\mathcal{O}(\epsilon^2)$ terms in LHS of \eqref{eqn:LHS_triad_after_ortho} is reduced to:
% \begin{align}
% \textnormal{LHS} &= 2\ii \left[\frac{\partial a_j}{\partial t}\left(\gamma^{(1)}_j \mathcal{K}_{j}^{2} - \gamma^{(2)}_j \right)\frac{\omega_j}{h^{2}} + \gamma^{(3)}_j\left(   \frac{\mathcal{K}_{j}}{h} \frac{\partial a_j}{\partial x}\right)\right]  \frac{\mathfrak{P}_{j}}{\beta_j}.
% % &+ \left\{\left[\frac{1}{h^{2}}\left(\frac{\partial h}{\partial x}\right)^{2}\gamma^{(6)}_j - \gamma^{(5)}_j\frac{1}{h}\frac{\partial^{2} h}{\partial x^{2}} + \gamma^{(5)}_j\frac{2}{h^2} \left(\frac{\partial h}{\partial x}\right)^{2}  \right] a_j \right\} \frac{\mathfrak{P}_{j}}{\beta_j}
%  \label{eqn:LHS_triad_final}
%  \end{align}
% If we ignore the nonlinear physics and solve LHS$=0$ (LHS given by \eqref{eqn:LHS_triad_final}), the maximum value of $a_j$ will not change. 
{More importantly, a wave packet's maximum amplitude does not change when $\beta_j$, given by \eqref{eqn:WKB_cr_final}, is used in \eqref{eqn:LHS_triad_after_ortho}. This invariance of the maximum value of the $a_j$ with varying $h$ is very useful in estimating wave growth rates in our study, in which a major focus is on  wave interactions in a region of varying $h$.}

Next we outline the procedure to obtain $\int_{-1}^{0} {\textnormal{NL}}_{j} \phi_j d \eta$ in \eqref{eqn:LHS_triad_after_ortho} to complete the amplitude evolution equations. The streamfunction, meridional velocity and buoyancy frequency ansatz are substituted in the nonlinear terms of \eqref{eqn:combined_eta}.
%\textcolor{red}{The nonlinear terms in terrain following coordinates (given in Equation (2.5)) contain significant number of terms. However, while evaluating nonlinear terms, differential operators having coefficients $dh/dx, d^2h/dx^2$ are neglected. Moreover, the term ${\partial^{2}}/{\partial x^{2}}$ is only assumed to operate on the phase of the wave. This is because amplitude ($a_j$), eigenfunction ($\phi_j$) are slow functions of $x$, and the length scale of their variation would be an order of magnitude higher than wavelength of the waves. Hence, the following relations are used while evaluating nonlinear terms: 
% \begin{align}
% {L}_{x}\left(\frac{a_j\phi_j}{\beta_j}P_j\right) &= \frac{a_j\phi_j}{\beta_j}\left(\frac{\partial P_j}{\partial x}\right) = \frac{a_j\phi_j}{\beta_j}\left( {\ii k_jP_j}\right), \label{eqn:LX}  \\ {L}_{xx}\left(\frac{a_j\phi_j}{\beta_j}P_j\right) &= \frac{a_j\phi_j}{\beta_j}\left(\frac{\partial^2 P_j}{\partial x^2}\right) = \frac{a_j\phi_j}{\beta_j}\left( -{k_j^2P_j}\right) \label{eqn:Lxx}
% \end{align}
% Note that using \eqref{eqn:LX}, the Poisson bracket ($\mathcal{J}\{\}$) can also be simplified easily.}
{As a result, the resultant resonant nonlinear terms, after omitting non-resonant terms, can be written in a compact form as given below:}
\begin{subequations}
\begin{align}
       \int_{-1}^{0}  \textnormal{NL}_1 \phi_1 d \eta &=   \left[\int_{-1}^{0}\left(\widehat{\textnormal{NL}}_{(\Psi,1)}  + \widehat{\textnormal{NL}}_{(B,1)} + \widehat{\textnormal{NL}}_{(\mathcal{V},1)}\right)\phi_1 d \eta\right]  \frac{a_3\bar{a}_2}{\beta_2\beta_3}\mathfrak{P}_3\bar{\mathfrak{P}}_2\\
       \int_{-1}^{0} \textnormal{NL}_2 \phi_2  d \eta &=   \left[\int_{-1}^{0}\left(\widehat{\textnormal{NL}}_{(\Psi,2)}  + \widehat{\textnormal{NL}}_{(B,2)} + \widehat{\textnormal{NL}}_{(\mathcal{V},2)}\right)\phi_2 d \eta\right] \frac{a_3\bar{a}_1}{\beta_1\beta_3}\mathfrak{P}_3\bar{\mathfrak{P}}_1\\
      \int_{-1}^{0} \textnormal{NL}_3 \phi_3 d \eta &=   \left[\int_{-1}^{0}\left(\widehat{\textnormal{NL}}_{(\Psi,3)}  + \widehat{\textnormal{NL}}_{(B,3)} + \widehat{\textnormal{NL}}_{(\mathcal{V},3)}\right)\phi_3 d \eta\right] \frac{a_1a_2}{\beta_1\beta_2}\mathfrak{P}_1\mathfrak{P}_2 %\mathfrak{a}_1\mathfrak{{a}}_2
\label{eqn:NL_def}
\end{align} 
\end{subequations}
{We define ${\textnormal{NL}}_{(*,j)} \equiv\int_{-1}^{0}\widehat{\textnormal{NL}}_{(*,j)} \phi_j d \eta$ for convenience, and their expressions are provided in appendix \ref{app:C}. Note that ${\textnormal{NL}}_{(*,j)}$ is directly used in amplitude evolution equations given just below in equation \eqref{eqn:coupling_coefficient_j}.}

The amplitude evolution equations for the three internal gravity waves are finally obtained after equating the LHS of  \eqref{eqn:LHS_triad_after_ortho} with its RHS; where the latter has been expressed in terms of \eqref{eqn:NL_big_s2_1}--\eqref{eqn:NL_big_s2_3}:
{\begin{subequations}
 \begin{align}
  \frac{\partial a_{1}}{\partial t} +   c^{(g)}_{(x,1)}\frac{\partial a_{1}}{\partial x}  =  \mathfrak{N}_1{a}_{3}\bar{a}_{2} \exp{\int_{0}^x\ii(\mathcal{K}_3-\mathcal{K}_1-\mathcal{K}_2)/h \hspace{0.1cm} dx' + \ii \Delta \omega t }, \label{eqn:wave1}\\
 \frac{\partial a_{2}}{\partial t}  +   c^{(g)}_{(x,2)}\frac{\partial a_{2}}{\partial x}  =  \mathfrak{N}_2{a}_{3}\bar{a}_{1} \exp{\int_{0}^x\ii(\mathcal{K}_3-\mathcal{K}_1-\mathcal{K}_2)/h \hspace{0.1cm} dx' + \ii \Delta \omega t}, \label{eqn:wave2}\\
  \frac{\partial a_{3}}{\partial t} +   c^{(g)}_{(x,3)}\frac{\partial a_{3}}{\partial x}  =  \mathfrak{N}_3{a}_{1}{a}_{2}  \exp{\int_{0}^x\ii(\mathcal{K}_1+\mathcal{K}_2-\mathcal{K}_3)/h \hspace{0.1cm} dx' - \ii \Delta \omega t }, 
\label{eqn:wave3}
\end{align}
\end{subequations}}
%\frac{{{\mathfrak{P}}_{3}}{{\mathfrak{P}}_{1}}}{{{\mathfrak{P}}_{2}}}
where  
%The functions $ c^{(g)}_{(x,j)}$ and $\mathfrak{N}_{j}$ are given by:
\begin{subequations}
 \begin{align}
& c^{(g)}_{(x,j)} =   \left[\frac{2\ii\mathcal{K}_j\gamma^{(3)}_j }{h\mathfrak{D}_j}\right], \hspace{0.5cm}\mathrm{in\,which}\hspace{0.5cm}  \mathfrak{D}_j = 2{\ii\omega_j} \left(\gamma^{(1)}_j \mathcal{K}_{j}^{2} - \gamma^{(2)}_j \right)/h^2,
 %\hspace{1cm} \mathcal{P}_j = \left(\frac{dh}{dx}\right)^2\frac{\gamma_6}{\mathfrak{D}_j}
 \label{eqn:group_speed}\\
 & \mathfrak{N}_{j} =  \frac{1}{\mathcal{D}_j}\left[\textnormal{NL}_{(\mathcal{V},j)} + \textnormal{NL}_{(B,j)} + \textnormal{NL}_{(\Psi,j)} \right]. \label{eqn:coupling_coefficient_j} 
 \end{align}
\end{subequations}

\noindent In the above equation
\begin{equation}
\mathcal{D}_1 =     \mathfrak{D}_1  \frac{\beta_2\beta_3}{\beta_1}, \hspace{0.5cm} \mathcal{D}_2 =     \mathfrak{D}_2  \frac{\beta_1\beta_3}{\beta_2}, \hspace{0.5cm} \mathcal{D}_3 =     \mathfrak{D}_3   \frac{\beta_1\beta_2}{\beta_3}.
%\mathcal{D}_3 =     \mathfrak{D}_3 h^2 \frac{\beta_1\beta_2}{\beta_3}, \hspace{0.5cm} \mathcal{D}_2 =     \mathfrak{D}_2 h^2 \frac{\beta_1\beta_3}{\beta_2}, \hspace{0.5cm} \mathcal{D}_1 =     \mathfrak{D}_1 h^2 \frac{\beta_2\beta_3}{\beta_1}.
\label{eqn:reduced_d}
\end{equation}
 
The coefficient $c^{(g)}_{(x,j)}$ denotes the (weakly varying) horizontal group speed and $\mathfrak{N}_j$ denotes the nonlinear coupling coefficient  of the $j$-th wave; $\mathfrak{N}_j$  determines the rate of energy transfer between the waves.  $\Delta \omega \equiv \omega_1+\omega_2 - \omega_3$ denotes the detuning in the frequency. The argument of the exponential terms in (\ref{eqn:wave1})--(\ref{eqn:wave3}) denote both the detuning in the horizontal wavenumber condition and the frequency condition. For a pure resonant triad, $\mathcal{K}_3-\mathcal{K}_1-\mathcal{K}_2 = 0$ and $\Delta \omega = 0$. When $\mathcal{K}_3-\mathcal{K}_1-\mathcal{K}_2 \neq 0$ or $\Delta \omega \neq 0$, the triad is said to be \emph{detuned}. The equations are only valid when both $\Delta \omega/\omega_j \ll 1$ and $\Delta \mathcal{K}/\mathcal{K}_j \ll 1$ are satisfied, that is, the equations are valid only near the vicinity of resonance. {Analytical methods have also been developed for studying wave-wave interactions in non-resonant regimes in the presence of a slowly varying background shear flow (for example: \cite{aky,grimshaw_1988,grimshaw_1994}), where the wave train can pass through non-resonant regimes and resonant regimes. However, this is not in the scope of this paper.} To summarize, amplitude ($a_j$) in the wave amplitude equations can vary because of the group speed term, or the nonlinear term.
The group speed term is responsible the advection of a wave packet, while the nonlinear term is responsible for energy transfer among the waves. Moreover, waves' energy density changes because of its motion through a region of varying $h$. $\phi_j$ and $\beta_j$ are heavily involved in the change in energy density that occur in a wave due to its motion through a region of varying $h$. Note that the evolution of $a_j$ does not provide complete information of the changes in a wave's quantities. This is because $\Psi_j = a_j\phi_j/\beta_j P_j\ee^{-\ii\omega_j t}$, where $\phi_j$ and $\beta_j$ themselves are functions of $x$.

For a triad, the parent wave is always the wave-3, while the daughter waves (subharmonic waves) are wave-1 and wave-2. For self-interactions, we use a different convention; see \S \ref{sec:2.2.2}.
{To determine how fast the daughter waves grow, a growth rate parameter ($\sigma$) is defined as follows:
\begin{equation}
    \sigma \equiv \sqrt{\mathfrak{N}_1\mathfrak{N}_2 A_3^2},
    \label{eqn:GR_definition}
\end{equation}
where $A_3$ is the parent wave's amplitude, which is held constant. To obtain this expression, the pump wave approximation of \cite{Craik_pump} is used. {Pump wave approximation is a strong assumption which is only valid at initial times where the parent wave has much more energy than the daughter waves.} Equation \eqref{eqn:GR_definition} reveals that growth rate is directly dependent on the nonlinear coupling coefficients.}
If we ignore the nonlinear terms,  equations (\ref{eqn:wave1})--(\ref{eqn:wave3}) model the movements of internal wavepackets over a mild-slope bathymetry. We emphasize here that wave scattering is not included in these equations. {The amplitude variation of internal waves was recently analyzed by \cite{leg_lahaye} (the authors focused on internal wave scattering, which is essentially a linear mechanism).} While we have restricted our study to mild-slope conditions,  we have extended the previous works by including {the physics of} (i) finite width wave packets, (ii) nonlinearity, and (iii) detuning in the horizontal wavenumber condition and hence investigation of both resonant (zero detuning) and near-resonant conditions. In this paper, we mainly focus on the variation of detuning, and growth rates (using pump wave approximation) with $h/H$ for wave-wave interactions. Even though equations \eqref{eqn:wave1}--\eqref{eqn:wave3} allows finite width wave packets, we do not discuss it significantly since these have been studied in \cite{gururaj_guha_2020}. The combined effect of nonlinear coupling coefficients, group speed and detuning have been discussed in \cite{gururaj_guha_2020}.

% \textcolor{purple}{The primary results from scaling analysis given in appendix \ref{app:A} is provided here in a brief manner. The relation between the small parameters are given by:
% \begin{equation}
%  \epsilon_{t} \sim \frac{{\mathfrak{N}}}{\omega } \epsilon_a  -    {\widehat{c}_g}  \epsilon_x  \label{eqn:summary_epsilon}
% \end{equation}
% where ${\widehat{c}_g}$ is a non-dimensional term which gives a scale of gropu speed. Its definition is given in Appendix \ref{app:A}. 
% Equation \eqref{eqn:summary_epsilon} provides the scaling for `Linear term-1' and $\textnormal{NL}_j$ given in \eqref{eqn:complete_f}. These are also the final terms which are given in wave amplitude equations \eqref{eqn:wave1}--\eqref{eqn:wave3}.  The wave amplitude $a_j$ can evolve due to the group speed term or the nonlinear term. Note that if $\mathfrak{N}$ or $\epsilon_a$ is reduced (implying that nonlinear coupling coefficients or amplitude is reduced), then we can expect nonlinear effects to decrease. However, if $\epsilon_x$ is reduced (which means packet width is increased), then the effect of packet's movement is reduced. The scaling of `Linear term-2', which are terms arising due to a non-constant depth, are provided in detail in appendix \ref{app:A}. These terms, which depend on small parameters $(\epsilon_h, \epsilon_k)$, influences $\mathfrak{N}_j$ and $\beta_j$. }

{The {main} results of scaling analysis, detailed in appendix \ref{app:A}, is summarized here. The relation between the small parameters are given by:
\begin{equation}
 \epsilon_{t} \sim \frac{{\mathfrak{N}}}{\omega } \epsilon_a  -    {\widehat{c}_g}  \epsilon_x  
 \label{eqn:summary_epsilon}
 \end{equation}
where ${\widehat{c}_g}$ is a non-dimensional term that gives a scale of the group speed. 
Equation \eqref{eqn:summary_epsilon} provides the scaling for `Linear term-1' and $\textnormal{NL}_j$ given in \eqref{eqn:complete_f}. These are also the final terms which are present in wave amplitude equations \eqref{eqn:wave1}--\eqref{eqn:wave3}. The wave amplitude $(a_j)$ can evolve due to the group speed term or the nonlinear term. Note that if $\mathfrak{N}$ or $\epsilon_a$ is reduced (implying that nonlinear coupling coefficients or amplitude is reduced), then we can expect nonlinear effects to decrease. However, if $\epsilon_x$ is reduced (which means packet width is increased), then the effect of group speed, which advects the packets, is reduced.}

\subsubsection{ Amplitude evolution equations for self-interaction in non-uniform stratification} \label{sec:2.2.2}
 
%The amplitude evolution equations \eqref{eqn:wave1}--\eqref{eqn:wave3} is for interaction between three waves. 
{Self-interactions can be considered as a special case of triad interactions}. During resonant self-interactions, an internal wave spontaneously {gives its energy} to  another internal wave which has twice its frequency and horizontal wavenumber \citep{wunsch}. In non-uniform stratification, a resonant self-interaction occurs when both $(\omega, k)$ and $(2\omega, 2k)$ satisfy the dispersion relation.
{The evolution equations for the self-interaction of a mode  can be obtained from the set of equations \eqref{eqn:wave1}--\eqref{eqn:wave3} after some straightforward modifications.} The complete set of governing equations for the self-interaction of a mode in the presence of a mild-slope bathymetry $h$ is given below: 
{\begin{subequations}
 \begin{align}
 \frac{\partial a_{(3,s)}}{\partial t}  +   c^{(g)}_{(x,3)}\frac{\partial a_{(3,s)}}{\partial x}     &=  \mathcal{N}_3{a}_{(1,s)}^2 \exp{{\int_{0}^x\ii(2\mathcal{K}_1-\mathcal{K}_3)/h \hspace{0.1cm} dx' + \ii \Delta \omega_s t}},  \label{eqn:wave_self_prim}\\
  \frac{\partial a_{(1,s)}}{\partial t} +   c^{(g)}_{(x,1)}\frac{\partial a_{(1,s)}}{\partial x}    &=  \mathcal{N}_1{a}_{(3,s)}\bar{a}_{(1,s)} \exp{{\int_{0}^x\ii(\mathcal{K}_3-2\mathcal{K}_1)/h \hspace{0.1cm} dx' - \ii \Delta \omega_s t }},
\label{eqn:wave_self_super}
\end{align}
\end{subequations}}
where the subscript `$s$' denotes self-interaction. {Moreover, $\Delta \omega_s = 2\omega_1 - \omega_3$}. Unlike the triad case, the parent wave for self-interaction is wave-1 while the daughter (superharmonic) wave is wave-3. The notation throughout this paper follows the convention that wave-3 always has the highest frequency (hence for triads, wave-3 becomes the parent wave).
The functions $c^{(g)}_{(x,j)}$ are the same as the expressions given in \eqref{eqn:group_speed}.
The functions $\mathcal{N}_j$, which are the nonlinear coupling coefficients for the self-interaction process, are given by:
\begin{subequations}
 \begin{align}
  \mathcal{N}_{1} &= \mathfrak{N}_2,\label{eqn:coupling_coefficient_s2}\\
   \mathcal{N}_{3} =  \frac{2\mathcal{K}_{1}^3}{h^4\mathcal{D}_3}\left( \frac{\Gamma^{(4)}}{\omega_1}  \right) - \frac{2f^2}{h^4\mathcal{D}_3}\left(\frac{\Gamma^{(3)}_1\mathcal{K}_1}{\omega_1} \right) &+ \frac{\mathcal{K}_{1}\omega_{3}}{h^4\mathcal{D}_3}\left(\zeta_1\omega_1^2\Gamma^{(1)}_1 - \zeta_1\Gamma^{(2)}_1-\Gamma^{(3)}_1\right).  \label{eqn:coupling_coefficient_s3}
  \end{align}
\end{subequations} 
{where $\zeta_j \equiv \mathcal{K}_j^2/{(\omega_j^2-f^2)}$ is defined for convenience.} Here all $\Gamma, \mathcal{D}_j$ terms in equations \eqref{eqn:coupling_coefficient_s2} and \eqref{eqn:coupling_coefficient_s3} are evaluated using \eqref{eqn:gammas_nonlinear} and \eqref{eqn:reduced_d} by simply considering all `2' subscripts as `1'. For example, substituting $\beta_1$ for $\beta_2$ in $\mathcal{D}_j$ and similarly substituting $\phi_1$ for  $\phi_2$ in $\Gamma$ expressions. This is because in self-interaction, wave-2 is the same as wave-1.  

The equations  \eqref{eqn:wave_self_prim}--\eqref{eqn:wave_self_super} can predict the growth of the daughter wave and the consequent decay of the parent wave. For obtaining the growth rate of the daughter waves, we use the pump wave approximation and hence treat the parent wave's amplitude $a_{(1,s)}$ as constant. This yields (assuming plane waves in the $x-$direction):
%However we simplify the problem by using the pump wave approximation, i.e. the parent wave's amplitude $a_{(1,s)}$ is assumed to be constant. Furthermore,  assuming both waves as plane waves in the $x-$direction (thus making $a_j$ independent of $x$), the superharmonic wave's amplitude is given by:
%we mainly focus on the growth of the daughter wave, that is we use the pump wave approximation where the parent wave's amplitude ($a_{(1,s)}$) is assumed to be constant. After the pump wave approximation, and assuming both waves are plane waves in the $x-$direction (thus making $a_j$ independent of $x$), the superharmonic wave amplitude is given by:
\begin{equation}
    a_{(3,s)} = \left[\mathcal{N}_3{a}_{(1,s)}^2 \right] t,
    \label{eqn:GR_self_s5}
\end{equation}
where the term in square brackets denote the growth rate. From the above equation it is evident that $\mathcal{N}_3$ acts as a proxy to the growth rate.
%Since ${a}_{(1,s)}$ is constant here, $\mathcal{N}_3$ becomes a proxy for the growth rate. 

%Here we extend the previous linear theory studies in the following ways: 1) In this study, finite width wave packets can be modelled 2) Nonlinearity is included. 

%\textcolor{blue}{The length scale of variation of the amplitude function, bottom bathymetry is assumed to  
%be at least an order of magnitude more than the length scale (horizontal wavenumber) of the waves. } 
%\textcolor{red}{I think we can omit this statement since we are already saying this in derivation. THAT'S WHAT I ALSO THINK}  

%The functions $c^{(g)}_{(x,j)}, S_{j}, \mathfrak{N}_j$ of (\ref{eqn:wave1})--(\ref{eqn:wave3}) influence the energy transfer, and also create amplitude variations in the $x$-direction, even if the waves' amplitudes are initialized with no $x$-dependence. This is precisely due to the non-uniformity of the bottom bathymetry in the bounded domain.

\subsection{Energy evaluation}

% The evolution of energy for these three waves is calculated by considering the total energy, where total density at an instant is given by:
% \begin{equation}
% \textnormal{TE}_{j} = \frac{\rho_{0}}{2}\left[ \left( \frac{\partial \psi_{j}}{\partial z}\right)^{2} + \left(\frac{\partial \psi_{j}}{\partial x}\right)^{2} + v_j^2 + \frac{b_j^2}{N^2} \right]
% \end{equation}
\noindent The time average energy density for an internal gravity wave over its time period is given by:  
\begin{equation}
\langle \textnormal{TE}_{j}\rangle = \frac{\omega_{j}}{2\pi}\int_{0}^{{2\pi}/{\omega_{j}}} \frac{\rho_{0}}{2}\left[\left( \frac{\partial \psi_{j}}{\partial z}\right)^{2} + \left(\frac{\partial \psi_{j}}{\partial x}\right)^{2} + v_j^2 + \frac{b_j^2}{N^2} \right] dt.
\end{equation}
The domain integrated total energy is given by:
\begin{equation}
\widehat{\textnormal{TE}}_{j} =\intop_{0}^{D} \intop_{h}^{0}\langle \textnormal{TE}_{j}\rangle dz dx = \intop_{0}^{D} \intop_{-1}^{0} \langle \textnormal{TE}_{j}\rangle(-h(x)) d\eta dx.
\end{equation}
After some simplification, we arrive at:
\begin{equation}
\widehat{\textnormal{TE}}_{j} = \intop_{0}^{D} \intop_{-1}^{0} -\frac{2}{h} \left[ \mathcal{K}_{j}^2 \phi_j^{2} + \left(\frac{\partial \phi_j}{\partial \eta}\right)^2  \right]  |a_j|^2 \frac{\rho_0}{\beta_j^{2}} d\eta dx,
\end{equation}
where $D$ is the length of the domain in the $x$-direction.
We non-dimensionalize $\widehat{\textnormal{TE}}_{j}$ with the initial energy of parent wave (abbreviated as `$Pw$') :  ${E}_{j} = {\widehat{\textnormal{TE}}_{j}}/{\widehat{\textnormal{TE}}_{Pw}|_{t = 0}}$. Note that $Pw=3$ (i.e. wave-3) for triads and $Pw=1$ (i.e. wave-1) for self-interactions.

%The non-dimensionalised growth rate of a wave based on TKE is defined as: 
% \begin{equation}
% \sigma_{j} \equiv \frac{1}{\omega_{3}}\frac{1}{(\textnormal{TKE}_{j})}\frac{d (\textnormal{TKE}_{j})}{dt}.
% \label{eqn:growth_rate}
% \end{equation} 
\section{Triad interactions in a uniform stratification in the presence of a mild-slope bathymetry }
 \label{Section_3}
In this section, we consider resonant and near-resonant triads in a  uniform background stratification in the presence of a mild-slope bathymetry. Here, we will briefly consider the horizontal wavenumber triad condition in uniform stratification as $h$ is varied. Without any loss of generality, the triad condition for the horizontal wavenumber is:
\begin{equation}
 \mathcal{K}_{3} = \mathcal{K}_{1} + \mathcal{K}_{2}, 
\end{equation}
However, using \eqref{eqn:eigen2}, it can be seen that $\mathcal{K}_j$ are constants for uniform stratification. As a result, triad conditions are satisfied everywhere in the domain, provided the conditions are perfectly satisfied for any given domain height.

\subsection{Effect of bathymetry on the nonlinear coupling coefficients of resonant triads} 
Here we focus on the    {nonlinear coupling coefficients in}   resonant triad interactions (i.e. no detuning) in the presence of a uniform stratification and a weakly varying bathymetry. Equation \eqref{eqn:GR_definition} revealed that the growth rate of the daughter waves is dependent on the nonlinear coupling coefficients.   {For constant $N$}, the nonlinear coupling coefficients ($\mathfrak{N}_j$) in \eqref{eqn:coupling_coefficient_j} can be further simplified:
 
 \begin{subequations}
 \begin{align}
 \mathfrak{N}_{1} &= \frac{\ii H}{2h^2\omega_1\kappa_1\kappa_2\kappa_3}\bigg[ {N^{2}(\mathcal{K}_{3}-\mathcal{K}_{2})}\left\{\left(\frac{\mathcal{K}_{3}}{\omega_{3}}-\frac{\mathcal{K}_{2}}{\omega_{2}}\right)\left(\mathcal{K}_2\mathcal{M}_3-\mathcal{K}_3\mathcal{M}_2\right) \right\} \nonumber \\ 
          &+\omega_1\left\{(\mathcal{K}_{2}\mathcal{M}_{3}-\mathcal{K}_{3}\mathcal{M}_{2})\left({\mathcal{M}^{2}_{2}}+\mathcal{K}^{2}_{2}-\mathcal{K}^{2}_{3}-{\mathcal{M}^{2}_{3}}\right) \right\}, \nonumber \\
           &+{f^2}\left(\mathcal{M}_3-\mathcal{M}_2\right)
          \left\{\left( \frac{\mathcal{K}_3}{\omega_3}+\frac{\mathcal{K}_2}{\omega_2} \right) \left(\mathcal{M}_3\mathcal{M}_2\right)-\left( \frac{\mathcal{K}_2}{\omega_3}+\frac{\mathcal{K}_3}{\omega_2} \right) \left(\mathcal{M}^2_2+\mathcal{M}^2_3\right)   \right\}\bigg],
          \label{eqn:uni_coupling_coefficient_1}
 \end{align}
 \begin{align}
 \mathfrak{N}_{2} &= \frac{\ii H}{2h^2\omega_2\kappa_1\kappa_2\kappa_3}\bigg[ {N^{2}(\mathcal{K}_{3}-\mathcal{K}_{1})}\left\{\left(\frac{\mathcal{K}_{3}}{\omega_{3}}-\frac{\mathcal{K}_{1}}{\omega_{1}}\right)\left(\mathcal{K}_1\mathcal{M}_3-\mathcal{K}_3\mathcal{M}_1\right) \right\} \nonumber \\ 
        &+\omega_2\left\{(\mathcal{K}_{1}\mathcal{M}_{3}-\mathcal{K}_{3}\mathcal{M}_{1})\left({\mathcal{M}^{2}_{1}}+\mathcal{K}^{2}_{1}-\mathcal{K}^{2}_{3}-{\mathcal{M}^{2}_{3}}\right) \right\}, \nonumber\\
        &+{f^2}\left(\mathcal{M}_3-\mathcal{M}_1\right)
          \left\{\left( \frac{\mathcal{K}_1}{\omega_1}+\frac{\mathcal{K}_3}{\omega_3} \right) \left(\mathcal{M}_1\mathcal{M}_3\right)-\left( \frac{\mathcal{K}_1}{\omega_3}+\frac{\mathcal{K}_3}{\omega_1} \right) \left(\mathcal{M}^2_1+\mathcal{M}^2_3\right)   \right\}\bigg],
          \label{eqn:uni_coupling_coefficient_2}
 \end{align}
 \begin{align} 
  \mathfrak{N}_{3} &=\frac{\ii H}{2h^2\omega_3\kappa_1\kappa_2\kappa_3}\bigg[ {N^{2}(\mathcal{K}_{1}+\mathcal{K}_{2})}\left\{\left(\frac{\mathcal{K}_{1}}{\omega_{1}}-\frac{\mathcal{K}_{2}}{\omega_{2}}\right)\left(\mathcal{K}_2\mathcal{M}_1-\mathcal{K}_1\mathcal{M}_2\right) \right\} \nonumber \\
          &+\omega_3
          \left\{(\mathcal{K}_{2}\mathcal{M}_{1}-\mathcal{K}_{1}\mathcal{M}_{2})\left({\mathcal{M}^{2}_{1}}+\mathcal{K}^{2}_{1}-\mathcal{K}^{2}_{2}-{\mathcal{M}^{2}_{2}}\right)\right\} \nonumber \\
          &+{f^2}\left(\mathcal{M}_1+\mathcal{M}_2\right)
          \left\{\left( \frac{\mathcal{K}_1}{\omega_2}+\frac{\mathcal{K}_2}{\omega_1} \right) \left(\mathcal{M}^2_1+\mathcal{M}^2_2\right) - \left( \frac{\mathcal{K}_1}{\omega_1}+\frac{\mathcal{K}_2}{\omega_2} \right) \left(\mathcal{M}_1\mathcal{M}_2\right) \right\}\bigg], \label{eqn:uni_coupling_coefficient_3}
 \end{align}
 \end{subequations} 
\noindent where $\kappa_j = \sqrt{ \mathcal{M}^2_j+\mathcal{K}^2_j}$. {Note that the above expressions are obtained only when the vertical wavenumber condition is satisfied.} The terms inside the square brackets are constant and hence do not vary with the {bathymetry} $h$ (the fact that $\mathcal{K}_j$ and $\mathcal{M}_j$ are constants for a constant $N$ is given in \eqref{eqn:lambda_unif}).  
{For constant $N$, $\beta_j = -h(x)/H$, which has been used in \eqref{eqn:uni_coupling_coefficient_1}--\eqref{eqn:uni_coupling_coefficient_3}, and this finally results in $\mathfrak{N}_j \propto 1/h^2$. Hence for waves traveling from a given fluid depth to a lesser depth (i.e. as the waves climb up a seamount), the nonlinear coupling coefficients, and hence the growth rates, increase following the inverse square rule.} 

%\textcolor{blue}{Note that evaluating growth rates by using constant $h$ theory at different depths (where the parent wave's energy density at different depths is fixed) would lead to $\sigma \propto 1/h$.}

In summary, for a uniform stratification, if three modes satisfy the resonant triad condition at a particular domain height, then they would satisfy the resonant triad condition for any domain height. Moreover, we also showed that the nonlinear coupling coefficients increase (decrease) as the fluid depth decreases (increases) following an inverse square law. 

\section{Triad and self interactions in a non-uniform stratification in the presence of a  mild-slope bathymetry: detuning effects} 
\label{section:4} 
In \S \ref{Section_3}, it was shown that in the presence of a uniform stratification, if the triad condition is satisfied between three modes at a particular $h$, then they are satisfied for all $h$. 
%This is because, in uniform stratification, all the horizontal wavenumbers are inversely proportional to the bathymetry function. Hence this leads to energy transfer without any mismatch in the triad wavenumber condition even if the waves (or the same modes) interact in a region of varying $h$. 
However in non-uniform stratification, such a simple outcome is not possible. In certain types of triads, there can be a heavy mismatch in the horizontal wavenumber condition as the waves involved in the triad interact in a region of varying domain height. This may affect the energy transfer between the waves.
 
In this section we study the factors that decide the detuning (or mismatch)  between the horizontal wavenumber of the waves  as $h$ is varied in the presence of non-uniform stratification. Here as well as in the rest of this paper, we will consider a Gaussian function to represent the buoyancy frequency:
\begin{equation}
N(z) = N_{b} + N_{\textnormal{max}}\exp[-\left\{\left(z-z_{c}\right)/W_{p}\right\}^{2}],
\label{eqn:strat_profile}
\end{equation} 
where the parameters $N_b, N_{\textnormal{max}}, W_{p}$, and $z_{c}$ are varied. This kind of profile (see figure \ref{fig:sc}(b)) is a simplified representation of oceanic stratification and is widely used in the literature; see \cite{grisouard_staquet_gerkema_2011}, \cite{Mathur_14}, and \cite{varma}. {We choose stratification profiles such that the pycnocline is above the topography. If the topography cuts the pycnocline, internal wave scattering may be significant as shown in \cite{Hall_2013}}.

\subsection{Effect of varying $h$ on the horizontal wavenumber condition for waves satisfying $f \ll \omega_j \ll N_b$ \label{section:4.1}}  
First we study the class of triads for which the angular frequencies of the constituent waves obey the condition $f \ll \omega_j \ll N_b$. It is assumed that the parent wave (angular frequency $\omega_3$) gives its energy to two subharmonic daughter waves of angular frequencies $\omega_1$ and $\omega_2$ respectively, that is, the condition $\omega_1\!<\!\omega_3$ and ${\omega_2}\!<\!{\omega_3}$ is always assumed. 
A parameter $\alpha \in (0,1)$ is defined such that $\omega_1 = \alpha\omega_3$ and $\omega_2 = (1-\alpha)\omega_3$. 
Two different types of interactions, Class-1 and Class-2, are defined for which a parent wave can form a triad with the subharmonic daughter waves.

\subsubsection{Class-1 interactions}\label{sec:4.1.1}
We consider three waves with angular frequencies ($\omega_3, \omega_1, \omega_2 $) such that $\omega_3 = \omega_1  + \omega_2 $. Furthermore we assume that at a particular $h$, the horizontal wavenumber condition is satisfied between mode $i$ of wave-1, mode $j$ of wave-2, mode $k$ of wave-3, i.e.
%This means mathematically: 
% \begin{equation}
% \mathcal{K}_{3(k)} = \mathcal{K}_{1(i)} + \mathcal{K}_{2(j)} \hspace{1cm} \Rightarrow \hspace{1cm} \Tilde{\mathcal{K}}_{(k)}\omega_3 = \Tilde{\mathcal{K}}_{(i)}\omega_1  + \Tilde{\mathcal{K}}_{(j)}\omega_2     
% \end{equation}
\begin{equation}
\mathcal{K}_{3(k)} = \mathcal{K}_{1(i)} + \mathcal{K}_{2(j)}, 
\label{eqn:hor_cond_4}
\end{equation} 
where $i$, $j$ and $k$ are not all equal. 
%For example, if mode-1 with $\omega_1$, mode 4 with $\omega_2$ and mode-3 with $\omega_3$ form a triad, it is classified as a Class-2 interaction. 
This constitutes a Class-1 interaction. 
%Unlike the Class-1 case,  Class-2 interactions are not restricted to the parameter regime $f \ll \omega_j \ll N_b$; however this subsection focuses  on this regime only so as to compare with Class-1 interactions.
 Now if the stratification profile changes (the stratification profile will change in $x$--$\eta$ coordinates provided $h$ is varying), then the wavenumbers ${\mathcal{K}}_{1(i)}, {\mathcal{K}}_{2(j)}, {\mathcal{K}}_{3(k)}$ will also change. However, for a given change in $h$, all the wavenumbers need not change in a way such that the condition \eqref{eqn:hor_cond_4} is satisfied.
For example, if $\mathcal{K}_{1(i)} = \mathrm{func}(h)$, then it is possible that $\mathcal{K}_{2(j)} \neq c\,\mathrm{func}(h)$, where $c$ and $\mathrm{func}(\,)$  denote an arbitrary constant and function respectively.
 %For example, ${\mathcal{K}}_{(i)}$, ${\mathcal{K}}_{(j)}$ may be different functions of $h$. 
 Therefore, even though the triad condition  may be satisfied at a particular $h$, it may not be satisfied for all $h$. Hence Class-1 triads might get detuned as {they interact in a region of varying $h$}. 
 %This behavior strongly differentiates Class-2 interactions with Class-1.}
 
 %for class-1 interaction where $i=j=k$, all three eigenvalues are the same functions of $h$. Thus triad condition will be still be satisfied even if $h$ is varied significantly.
 
To measure the detuning (or mismatch) in the horizontal wavenumber, we define a new variable $\Delta \mathcal{K}$:
\begin{equation}
\Delta \mathcal{K} \equiv \frac{\mathcal{K}_{3(k)}-\mathcal{K}_{1(i)}-\mathcal{K}_{2(j)}}{\mathcal{K}_{\textnormal{min}}},
\label{eqn:detun_def}
\end{equation}
where 
%$i,j,k$ are some integers and
$\mathcal{K}_{\textnormal{min}}$ is the minimum wavenumber of the three wavenumbers at a particular $x$-coordinate. $\Delta \mathcal{K}$ basically acts as a non-dimensional measure of the detuning between the waves, and for a resonant triad, $\Delta \mathcal{K} = 0$. 

 \begin{figure}
  \centering{\includegraphics[width=\linewidth]{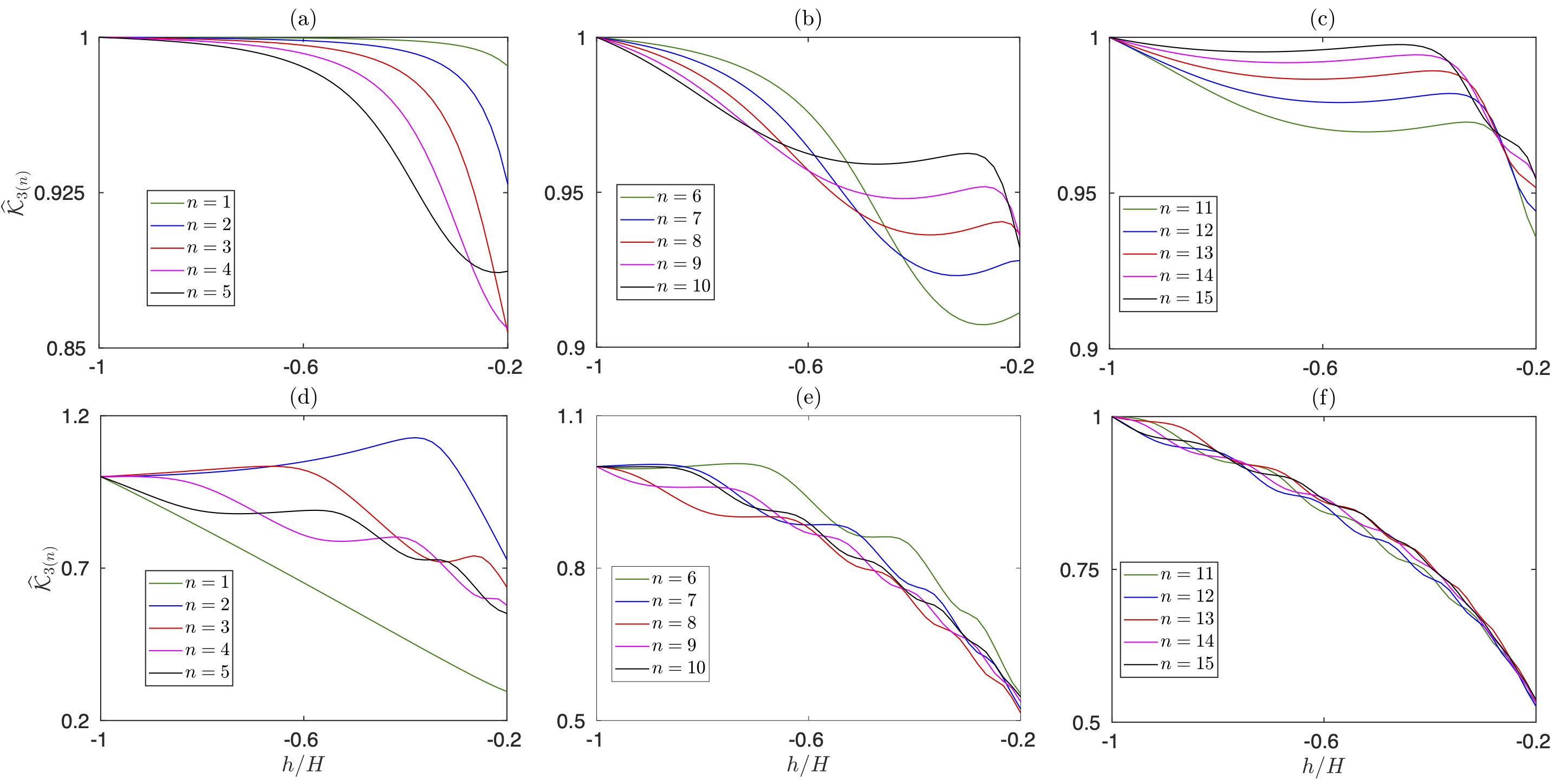}}
   \caption{ Variation of $\widehat{{\mathcal{K}}}_{3(n)}$ with $h/H$ for two different stratification profiles.
   For stratification profile $N^{(1)}$: (a) modes 1-5, (b) modes 6-10 and (c) modes 11-15. For stratification profile $N^{(2)}$: (d) modes 1-5, (e) modes 6-10 and (f) modes 11-15.
    }
   \label{fig:eig_dif_evolution_c}
 \end{figure}

% \begin{figure}
%  \centering{\includegraphics[width=\linewidth]{new_detuning_modes_2.png}}
%   \caption{ Variation of detuning ($\Delta \mathcal{K}$) with $h/H$ for various modal interactions pertaining to three stratification profiles -- (a)  $N^{(3)}$, (b) $ N^{(4)},$ and (c) $N^{(5)}$. 
%   %(a), (b) and (c) uses the parameter set $N^{(3)}, N^{(4)}$ and $N^{(5)}$ respectively.
%   The legends indicate what secondary/daughter waves were involved in the triad interactions. 
%   $\alpha \equiv \omega_1/\omega_3$.}
%   \label{fig:mode1_detun}
% \end{figure}

% \begin{figure}
%  \centering{\includegraphics[width=130mm]{detuning_higher_vs_lower_comparison.png}}
%   \caption{ Variation of detuning ($\Delta \mathcal{K}$) with domain height for various modal interactions. (a) Interaction involving only lower modes  (b) Interaction which involves higher modes.}
%   \label{fig:highervslower}
% \end{figure}
 
We now study how different (nondimensional) wavenumbers ${\mathcal{K}}_{3(n)}$ of frequency $\omega_3$ change as $h$ is varied in the presence of a non-uniform stratification. {To obtain ${\mathcal{K}}_{3(n)}$ for a given stratification profile, we solve \eqref{eqn:eigen2} for $h/H \in [-1,-0.2]$.} The functional form of $h$, as long it is mildly varying, does not influence the wavenumbers or detuning at a particular $h$.
The non-uniform stratification profile, given by (\ref{eqn:strat_profile}) is used throughout this paper.
The stratification profiles are chosen such that $N_{\textnormal{max}}=(2 N_b, 4 N_b, ... 12N_b)$, $W_p=(H/200, 2H/200... 5H/200)$, and $z_c=(H/80, H/40, H/20, H/10)$; and we consider all possible ($120$) combinations. Moreover, $\omega_3 = 0.1N_b$ and $f=0$ is used consistently for all combinations.
% Figures \ref{fig:eig_dif_evolution_c}(a)--(c) show the variation of $\widehat{{\mathcal{K}}}_{3(n)} \equiv {\mathcal{K}}_{3(n)}(h)/{\mathcal{K}}_{3(n)}(H)$ with $h/H$ for different modes $n$.
Figure \ref{fig:eig_dif_evolution_c} show the variation of $\widehat{{\mathcal{K}}}_{3(n)} \equiv {\mathcal{K}}_{3(n)}(h)/{\mathcal{K}}_{3(n)}(H)$ with $h/H$ for different modes $n$. Figures \ref{fig:eig_dif_evolution_c}(a)--\ref{fig:eig_dif_evolution_c}(c) uses the stratification profile $N^{(1)}$ with the following parameters:  $N_{\textnormal{max}} = 2N_{b}$, $W_{p} = H/200$,  and $z_c = H/80$. Moreover,  figures \ref{fig:eig_dif_evolution_c}(d)--\ref{fig:eig_dif_evolution_c}(f) uses the profile $N^{(2)}$ given by:  $N_{\textnormal{max}} = 10N_{b}$, $W_{p} = H/50$,  and $z_c = H/10$. Note that $N^{(1)}$ has a sharp pycnocline, while $N^{(2)}$ has a larger $W_p$ resulting in a wider pycnocline. 
For profiles where all three parameters are low (e.g.\, $N^{(1)}$), $\widehat{{\mathcal{K}}}_{3(1)}$ is nearly constant for some range of $h/H$ and then starts decreasing.
This can be seen in figure \ref{fig:eig_dif_evolution_c}(a), where the first five modes exhibit this behaviour.  Moreover, for profiles where all $z_c, W_p, N_{\textnormal{max}}$ are high (e.g.\, $N^{(2)}$), $\widehat{{\mathcal{K}}}_{3(1)}$ decreases almost linearly with $h/H$, as  can be clearly seen in figure \ref{fig:eig_dif_evolution_c}(d). For any profile, $\widehat{{\mathcal{K}}}_{3(1)}$ always decreases as the fluid depth is reduced for $h/H \in [-1, -0.2]$.
However this behaviour does not hold for any mode other than mode $1$. For example, for $z_c = H/10$ (regardless of $W_p, N_{\textnormal{max}}$),  $\widehat{{\mathcal{K}}}_{3(2)}$ increases for some $h/H$ as fluid depth is reduced, see  figure \ref{fig:eig_dif_evolution_c}(d) (blue curve). Similar behaviour is also observed for modes $3$,$4$ and $5$ when $W_p$ is low. In summary, the variation of $\widehat{{\mathcal{K}}}_{3(1)}$ with $h/H$ can be different from that of the higher modes' wavenumber, which can result in detuning.

For profiles with high $W_p$, $\widehat{{\mathcal{K}}}_{3(n)}$ for $n>10$ starts to collapse on each other, see figure \ref{fig:eig_dif_evolution_c}(f). In such kind of scenarios, since $\widehat{{\mathcal{K}}}_{3(n)}$ remains nearly the same, $\Delta \mathcal{K}$ will not be induced by the difference in higher modes' $\widehat{{\mathcal{K}}}_{3(n)}$.  In general it was observed that as $W_p$ is reduced, {$n$ has to be higher for the modes to collapse on each other}.

\begin{figure}
 \centering{\includegraphics[width=0.85\linewidth]{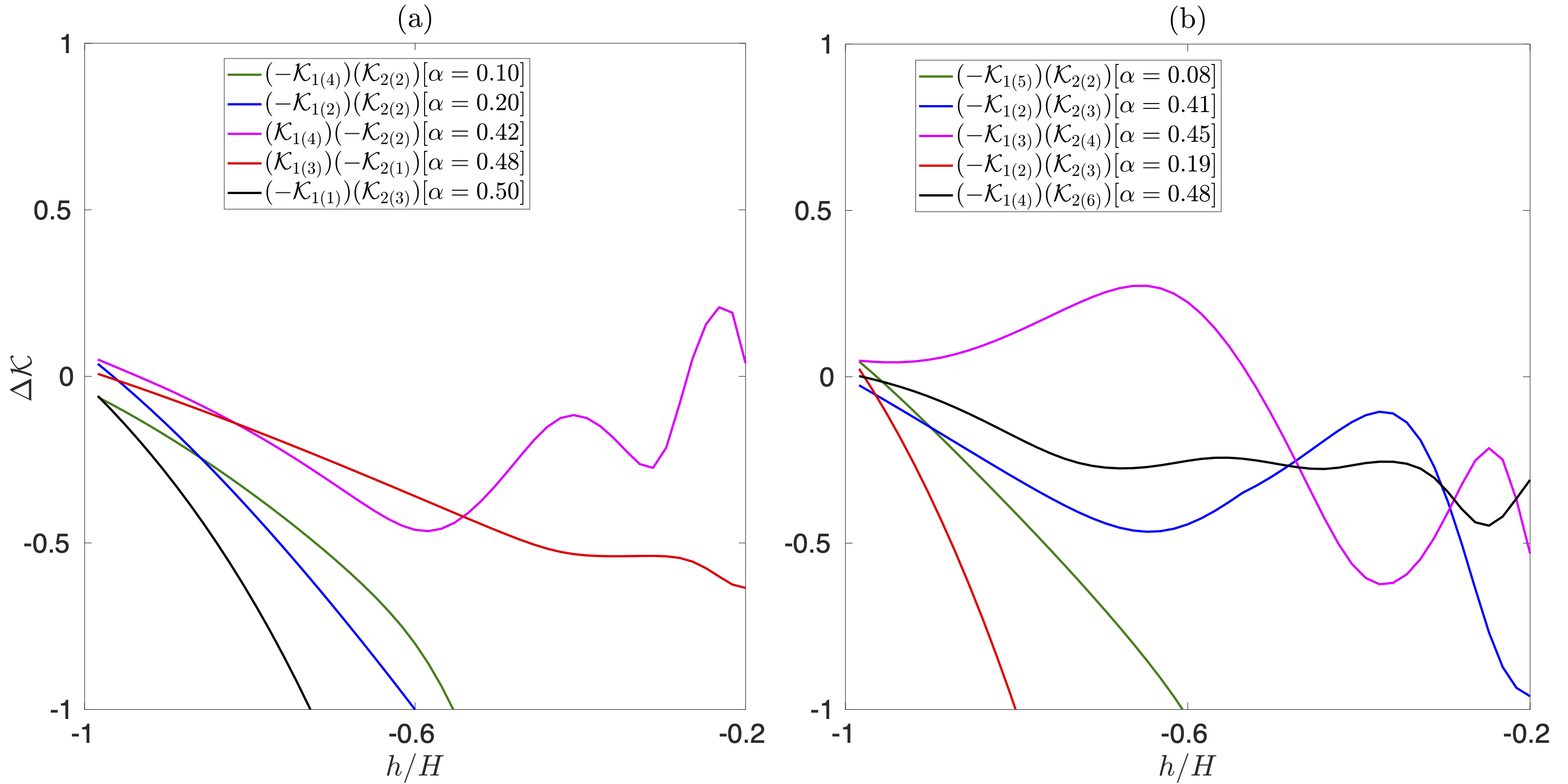}}
  \caption{ Variation of detuning ($\Delta \mathcal{K}$) with $h/H$ for various modal interactions pertaining to the stratification profiles (a)  $N^{(3)}$, and (b) $ N^{(4)}$.
  The legends indicate what daughter waves were involved in the triad interactions. 
  $\alpha \equiv \omega_1/\omega_3$.}
  \label{fig:mode1_detun}
\end{figure}

The interaction of mode-1 internal  wave (wave-3) with different modes in the presence of two different non-uniform stratification profiles is  considered next. These profiles are a part of the $120$ profiles that we already mentioned. Sample results are shown in figure \ref{fig:mode1_detun} in which the
 frequencies and stratification profile parameters are as follows: 
\begin{itemize}
    \item $N^{(3)}$: $\omega_3 = 0.1N_b$, $f=0$,$N_{\textnormal{max}} = 10N_b$, $W_p = H/100$ and $z_c = H/10$.
    \item $N^{(4)}$: $\omega_3 = 0.1N_b$, $f=0$,$N_{\textnormal{max}} = 10N_b$, $W_p = H/50$ and $z_c = H/20$.
    % \item $N^{(5)}$: $\omega_3 = 0.1N_b$, $f=0$, $N_{\textnormal{max}} = 15N_b$, $W_p = H/100$ and  $z_c = H/20$.
\end{itemize}
For each profile, we have shown $5$ different modal interactions. Figure \ref{fig:mode1_detun}  clearly reveals that the detuning can be quite sensitive to the changes in the domain height.

% \begin{figure}
%  \centering{\includegraphics[width=0.85\linewidth]{Fig_4.png}}
%   \caption{ Variation of detuning ($\Delta \mathcal{K}$) with $h/H$ for various modal interactions pertaining to the stratification profiles (a)  $N^{(3)}$, and (b) $ N^{(4)}$.
%   The legends indicate what daughter waves were involved in the triad interactions. 
%   $\alpha \equiv \omega_1/\omega_3$.}
%   \label{fig:mode1_detun}
% \end{figure}

\subsubsection{Class-2 interactions: A special case of triad interactions} \label{section:4.1.2}
%Class-2 interaction constitutes all kinds of triadic interactions excluding the situation mentioned in Class-1. 
The interaction in Class-2 is between the  $n$-th modes (where $n\in \mathbb{Z}^+$) of different waves constituting a triad.  For example, if mode $1$ with frequency $\omega_1$, mode $1$ with frequency $\omega_2$ and mode $1$ with frequency $\omega_3$ form a triad,  it is classified as a Class-2 interaction. This kind of triad is  possible when $f \ll \omega_j \ll N_b$.
\noindent To show how this interaction is possible, we consider the eigen-problem concerning the $n-$th mode of the $j-$th wave: %for the  for the $n$-th eigenvalue ($\mathcal{K}_{j(n)}$) and the $n$-th eigenfunction $\phi_{j(n)}$ for the $j-$th wave is given below: 
\begin{equation}
\frac{\partial^2 \phi_{j(n)} }{\partial \eta^2} + \mathcal{K}_{j(n)}^{2} \chi_j^2\phi_{j(n)} \approx \frac{\partial^2 \phi_{j(n)} }{\partial \eta^2} +  \bigg(\frac{\mathcal{K}_{j(n)}}{\omega_j}\bigg)^2N^2\phi_{j(n)}= 0,
\label{eqn:eigen_cr}
\end{equation}
where we used $\chi_j \approx N/\omega_j$ (under the approximation $f \ll \omega_j \ll N_b$), and the system is solved using the boundary conditions: $\phi_{j(n)}  = 0$ at $\eta = 0$ and $\eta = -1$.  { However, by Sturm Liouville theory, for a given operator (here $ \partial^2/\partial \eta^2$) and weight function (here $N(z)^2$), the $n-$th eigenvalue (here $\mathcal{K}_{j(n)} /\omega_j$) is unique, i.e.,  $\mathcal{K}_{j(n)} /\omega_j=\mathrm{constant}$ $\forall j$. Therefore, if the triad condition for frequency: $\omega_3 = \omega_2 + \omega_1$ is valid, this automatically implies validity of the wavenumber condition
$ \mathcal{K}_{3(n)} =  \mathcal{K}_{1(n)} +  \mathcal{K}_{2(n)}$.}

The situation mentioned above is true for all  stratification profiles satisfying $f \ll \omega_j \ll N(z)$ (at all $z$ locations). This is especially important because in the presence of a bathymetry, the stratification profile changes in the $x$ direction in $x-\eta$ coordinates. However, for Class-2 interaction, all three non-dimensional wavenumbers (eigenvalues) are the same functions of $h$ since they are the same eigenvalues divided by their frequency. Thus the resonant triad condition will be still be satisfied even if $h$ is varied significantly (i.e. variation of $h$ will not cause detuning). {Note that, in the parameter regime of $f \ll \omega_j \ll N(z)$, only class-2 self interactions were observed in numerical experiments of \cite{sutherland}, hence class-2 triads may always be dominated by self interactions, resulting in parent wave's energy transfer to the superharmonics instead of subharmonics. As a result, class-2 triads may not be practically as relevant as class-2 self interactions.}

\subsection{Effect of bathymetry on horizontal wavenumber condition for Class-1 self-interaction} \label{section:4.2}
Detuning can also be introduced during a self-interaction process as $h$ is varied.  Following the same terminology as before, we  classify self-interactions as Class-1 and Class-2. 
%Similar to triad classification, we define Class-1 and Class-2 for self-interactions. Class-1 self-interaction is between the  $n$-th modes (where $n\in \mathbb{Z}^+$) of frequencies $\omega$ and $2\omega$. 
%Class-2 self-interactions is between different modes of frequencies $\omega$ and $2\omega$. 
As shown by \cite{wunsch}, Class-2 self-interactions will always be slightly detuned, where the detuning increases as $f$ increases. This is due to the fact that in a non-uniform stratification, if the $n-$th mode of $(\omega)$ satisfies the dispersion relation, then the $n$-th mode of $(2\omega)$ will be able to satisfy it  only approximately.

Following \eqref{eqn:detun_def}, the detuning for a self-interaction process is defined as
%To measure the detuning in the self-interaction process, we define a non-dimensional variable $\Delta \mathcal{K}_s$ is defined which is given by:  
\begin{equation}
\Delta \mathcal{K}_s = \frac{\mathcal{K}_{3(k)}-2\mathcal{K}_{1(i)}}{\mathcal{K}_{1(i)}},
\label{eqn:detun_def_s}
\end{equation}
where  wave-3 is the superharmonic (daughter) wave, while wave-1 is the parent wave, i.e.\,$\omega_1=\omega_3/2$ (following the convention used throughout this paper that wave-3 has the highest frequency). The amplitude evolution equations for a self-interaction process is discussed in \S \ref{sec:2.2.2}.

\begin{figure}
 \centering{\includegraphics[width=0.9\textwidth]{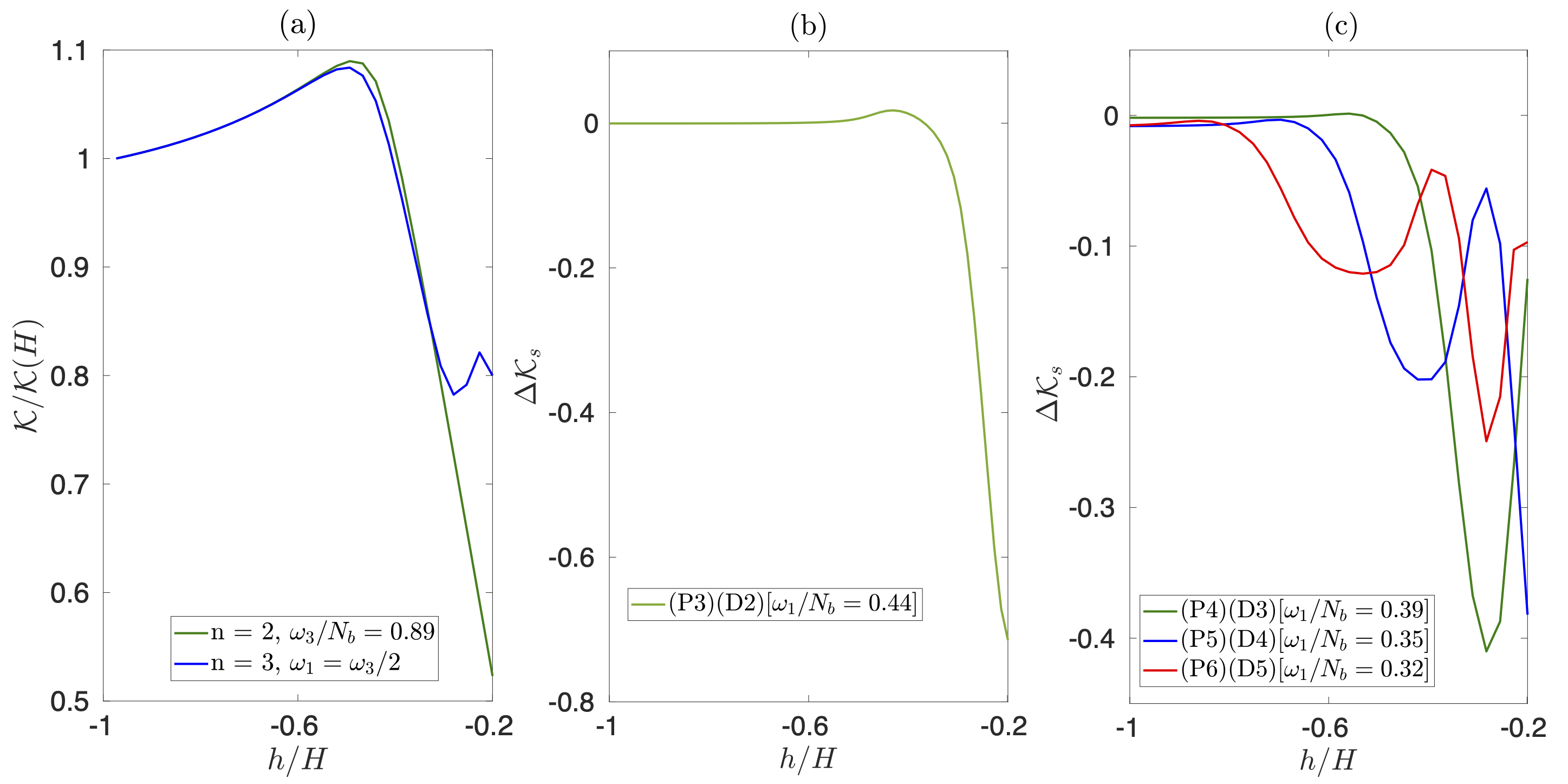}}
  \caption{ (a) Variations of $\mathcal{K}_{3(2)}$ and $\mathcal{K}_{1(3)}$ with $h/H$ for the parameter set $N^{(6)}$. (b) Variation of detuning with $h/H$ for the same case. (c) The detuning for three different self-interaction combinations for the parameter set $N^{(7)}$. Here the notation (Pa)(Db)  implies  `Parent wave' with `mode-a' and `Daughter wave' with `mode-b'. } 
  \label{fig:self_s4_2}
\end{figure}

{For the range $f \ll \omega_j \ll N(z)$, Class-2 self-interaction process will follow similar principles outlined in \S \ref{section:4.1.2}. 
%Self-interactions for $f \ll \omega \ll N(z)$ (for flat bathymetry) range has been previously studied by \cite{wunsch}.
As mentioned in \S \ref{section:4.1.2}, significant variations in $h$ for this frequency range will not introduce detuning in Class-2 self-interactions.}   
%\sout{Since internal waves in oceans usually have very small slope, the approximation $\omega_j \ll N(z)$ is largely valid.} 
{We now study the other end of the parameter space where $(N^{2}-\omega_j^{2})\approx N^{2}$, which was the basic approximation used in much of our analysis is \S \ref{section:4.1}, is no longer valid.}
% However for completeness, we study the other end of the parameter space, i.e., $\omega_j \approx N(z)$ $\forall z$. In this limit,  $(N^{2}-\omega_j^{2})\approx N^{2}$, which was the basic approximation used in much of our analysis is \S \ref{section:4.1}, is no longer valid.
Hence, out of Class-1 and Class-2 self-interactions, only the latter is possible. This would mean that as the domain height changes, the detuning introduced could be significant. Interestingly though, if the wavenumbers involved in the self-interaction change with $h/H$ in a similar way, the detuning is insignificant, see figure \ref{fig:self_s4_2}. 
The frequencies and the stratification profile parameters used here are:  
\begin{itemize}
    \item $N^{(6)}$: $N_{\textnormal{max}} = 10N_b$, $W_p = H/100$ and $z_c = H/10$.
    \item $N^{(7)}$: $N_{\textnormal{max}} = 10N_b$, $W_p = H/100$ and $z_c = H/20$,
\end{itemize} 
and $f=0$ always.
Figure \ref{fig:self_s4_2}(a) uses the set $N^{(6)}$, and   shows the variation of the horizontal wavenumber of mode-2 (of $\omega_3 = 0.89N_b$) and mode-3 (of $\omega_1 = \omega_3/2$). These modes satisfy the condition for resonant self-interaction. We observe that these two wavenumbers behave quite similarly  for a wide range of $h/H$, and hence the detuning, shown in figure \ref{fig:self_s4_2}(b), is small (and constant for an appreciable range), in spite of the fact that it is a Class-1 interaction. The same phenomenon is also shown for several other self-interaction combinations in figure \ref{fig:self_s4_2}(c), where the parameter set $N^{(7)}$ is used. 
%The following frequencies are used in figure \ref{fig:self_s4_2}(c): $\omega_3/N_b = 0.39, 0.35,$ and $0.32$.

The detuning for all the combinations shown stays constant for a certain range of $h/H$. We note in passing that Class-1 \emph{triad} interactions may also give rise to a small detuning for a range of $h/H$, provided all the modes involved behave in a similar way. However, this is a more stringent condition than a self-interaction process, where only two waves are involved. {Even though equations derived in \S \ref{Section:2} are only valid when $\Delta \mathcal{K} \ll 1$ (or $\Delta \mathcal{K}_s \ll 1$), there are significant number of interactions where $\Delta \mathcal{K}$, or $\Delta \mathcal{K}_s$, is a small quantity even for $\mathcal{O}(1)$ changes in depth and the wavenumber. For example, for interactions shown in this section, and for class-2 interactions, detuning can stay as a small quantity even for $\mathcal{O}(1)$ changes in depth. However, we do note that in several triad interactions detuning can be sensitive to $h$ and in those cases $\mathcal{O}(1)$ changes in depth cannot be accurately modeled by the wave-amplitude equations.}
 
To summarize, in the presence of a non-uniform stratification, we divide triad and self-interactions into two classes: Class-1 and Class-2. { Class-1 interactions contain waves whose mode numbers are all not the same, while Class-2 interactions contain waves which are the $n-$th modes of their respective frequencies.
Class-1 interactions, may undergo detuning with the variation in $h$, irrespective of the frequency. However, interestingly, certain Class-1 self-interactions do not undergo detuning as $h$ is varied inside a certain range.
For both triads and self-interactions, Class-2 interactions can only exist for $f\ll \omega_j \ll N(z)$, and does not get detuned as $h$ is varied.}

%due to similarity of the wavenumbers' evolution with $h$.
%This is because each wavenumber can be a unique function of $h$, thus resulting in horizontal wavenumber condition being potentially violated as $h$ is varied.
% triad interactions that do not consist of the same $n-$th mode of their angular frequencies may get detuned significantly as the fluid depth changes.

\section{Variation of growth rates and nonlinear coupling coefficients with depth for non-uniform stratification \label{section_5}}  
% In this section, we focus on the effects of domain height variation on the nonlinear coefficients ($\mathfrak{N}_j$),  and hence its effect on the rate of energy transfer between the waves for both triad and self interactions. The non-uniform stratification profile (\ref{eqn:strat_profile}) will be used in this section.
In this section, we focus on the effects of domain height variation on the growth rate ($\sigma$) of triads, and the nonlinear coupling coefficient $\mathcal{N}_3$, which provides a measure of the growth of the daughter wave in a self-interaction. The non-uniform stratification profile (\ref{eqn:strat_profile}) will be used in this section.

\subsection{Variation of growth rates with domain height for triads}
\label{section:5.1} 

%Here we study the variation of growth rate in triads with mode-1 as the parent wave in the presence of varying $N(z)$ and $h$. 
Triad interactions are important for the decay of internal waves  near the $28.9^{\circ}$ latitude \citep{winter,mac_2013}, specifically the mode-1 wave, which is the most energy containing mode \citep{vic2019deep}. Here we study this phenomena in the high latitude region ($f/\omega_3\geq0.3$) for varying $N(z)$ and $h$. The mode-1 wave (which, being the parent wave, is wave-3) can decay forming various triad combinations; we restrict the subharmonic daughter waves (wave-1 and wave-2) up to mode-50.
 Moreover, for studying growth rates in the presence of varying $h$, the triads are identified separately at different $h/H$. This is because a triad combination at a particular $h/H$ value may not satisfy the horizontal wavenumber condition at a different $h/H$ (as explained in \S \ref{section:4}). 
 %At every $h/H$ value, the waves' vertical function $\phi_j$ is divided with $\mathcal{E}_j$ at that particular $h/H$ value. 
Three main branches of triads are considered here for the  mode-1 internal wave: 
\begin{equation}
 \underbrace{\abs{\mathcal{K}_3} = \abs{\mathcal{K}_2} - \abs{\mathcal{K}_1}}_{
 \textnormal{Branch-1}} 
    \hspace{0.5cm} \textnormal{or}\hspace{0.5cm}  \underbrace{\abs{\mathcal{K}_3} = \abs{\mathcal{K}_1} - \abs{\mathcal{K}_2}}_{
 \textnormal{Branch-2}}  \hspace{0.5cm}  \textnormal{or} \hspace{0.5cm}  \underbrace{\abs{\mathcal{K}_3} \approx \abs{\mathcal{K}_1} + \abs{\mathcal{K}_2}}_{
 \textnormal{Branch-3}}.
    \label{eqn:triad_possiblity_s5}
 \end{equation}
{For Branch-1(2) triads, wavenumber of wave-2(1) is larger in magnitude than that of wave-1(2)}. The only possible  Branch-3 interaction is a Class-2 interaction, where both the daughter waves are also mode-1 of their respective frequencies.  However this interaction, like the Class-2 self-interaction, also undergoes heavy detuning for high $f$ values.  Therefore  Branch-3 being an inefficient energy transfer pathway, we restrict our focus to Branch-1 and Branch-2. Triads are studied for $f/\omega_3 = (0.3, 0.4, 0.45)$ in the presence of various stratification profiles. The triads are computed for $\alpha \in [0.31,0.5]$, $\alpha \in [0.41,0.5]$ and $\alpha \in [0.455,0.5]$ for $f/\omega_3 = 0.3, 0.4$ {and} $0.45$ respectively (see \S \ref{section:4.1} for the definition of $\alpha$).  

%Within the specified $\alpha$ limits, all possible triad combinations up to mode-50 are identified and studied.  \\
%An important point to notice is that the wavenumbers of the daughter waves is lower than parent wave hence this will behave as an inverse cascade.

% \begin{figure}
%  \centering{\includegraphics[width=130mm]{15_025_8_04_02.png}}
%   \caption{ Plots of non-dimensionalised growth rates of Mode-1 triads for profile $N^{(11)}$ with $f/w_d = 0.4$ and $\omega_3/N_b = 0.2$. (a) Line $(n,n)$ of Branch-1. (b) Line $(n+1,n)$ of Branch-1. (c) Line $(n,n+1)$ of Branch-1. (d) Line $(n+6,n)$ of Branch-2. }
%   \label{fig:GR_3}
% \end{figure}

\begin{figure}
 \centering{\includegraphics[width=130mm]{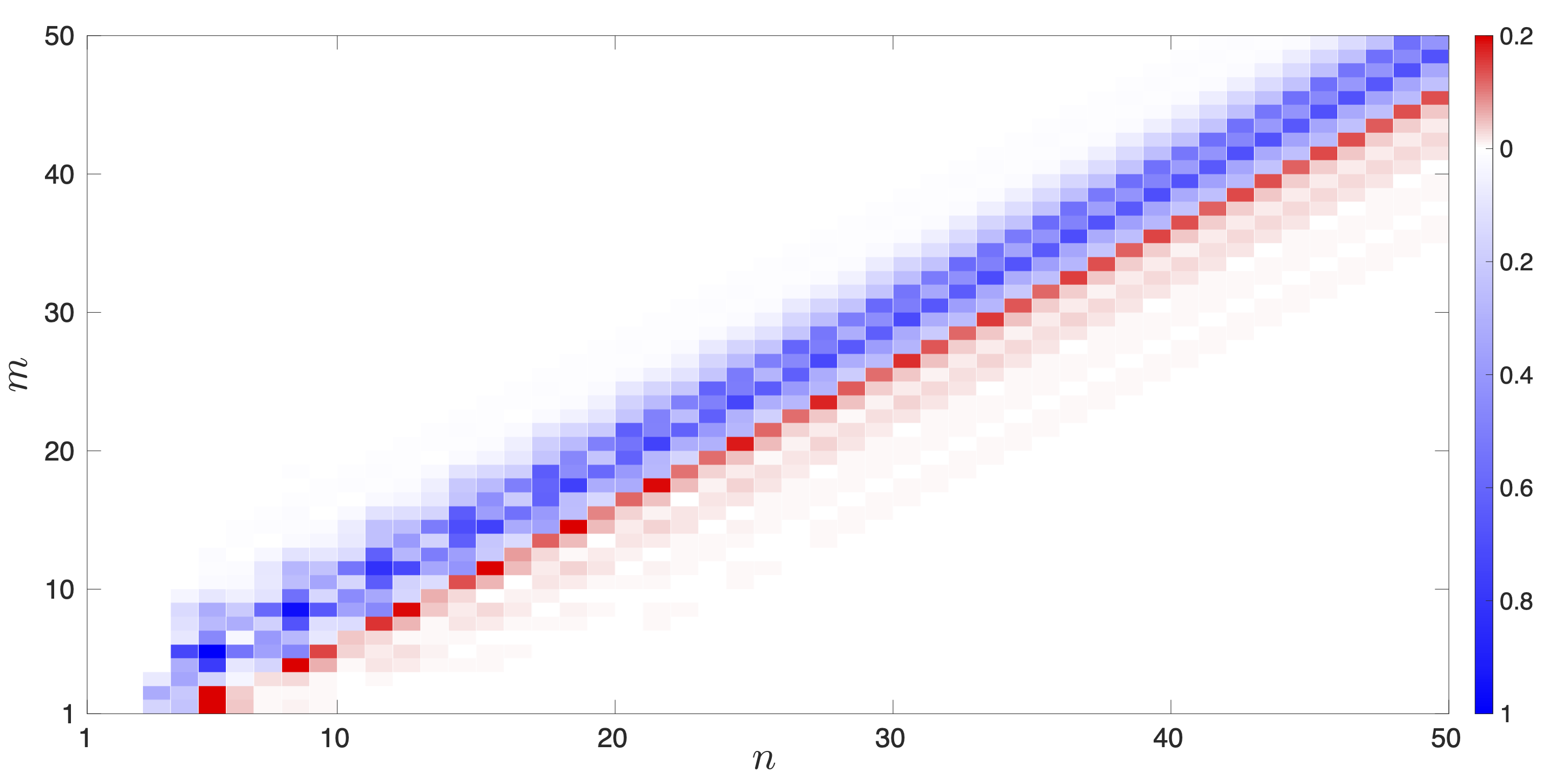}}
  \caption{ Contours of non-dimensional growth rate ($\sigma/\sigma_{\textnormal{ref}}$) of triads formed between  mode-$n$, mode-$m$, and mode-1 (i.e., wave-3). Blue and red colors respectively represent Branch-1 and Branch-2 triads, and both colors represent positive values. For Branch-1 triads,  mode-$m(n)$ is wave-1(2), while for Branch-2 triads, mode-$n(m)$ is wave-1(2).
 % The convention used in this figure is as follows: $n (m)$ represents the modenumber of wave-2 (wave-1) for Branch-1 triads. However, it is the exact opposite for Branch-2 triads where $n (m)$ represents wave-1's (wave-2's) modenumber.
  }
  \label{fig:GR_model_plot}
\end{figure}

% \textcolor{blue}{\sout{When an internal mode travels in a region of varying $h$, the maximum amplitude of the mode ($a_j/\beta_j$) will get altered through $\beta_j$. However, we do not take this factor into account  since the focus here is on which daughter wave combination is most unstable, and this is not affected by the $\beta_j$ function.}}
Figure \ref{fig:GR_model_plot} shows the non-dimensionalised growth rate contour for a mode-1 wave. All growth rates $\sigma$ are non-dimensionalised with a reference growth rate value $\sigma_{\textnormal{ref}}$, where the latter denotes  the maximum growth rate for all Branch-1 triads at $h=-H$ (hence the value of $A_3$ does not impact the results shown). The frequency of the mode-1 wave is $\omega_3/N_b = 0.2$, while $f/\omega_3 = 0.4$ is taken. The stratification profile is given by
\begin{itemize}
    \item $N^{(8)}$: $N_{\textnormal{max}} = 10N_b$, $W_p = H/50$, $z_c = H/20$.
\end{itemize} 
Branch-1(2) triads have the higher(lower) frequency daughter wave propagating in the same direction as the parent wave.
%Branch-2 triads have the lower frequency daughter wave propagating the same direction as the parent wave, whereas for Branch-1 the higher frequency daughter wave propagates the same direction as the parent wave.  
Figure \ref{fig:GR_model_plot} reveals that from both branches, the highest growth rates are centered around $n \approx m$. However, majority of the white region contains resonant triads, but their growth rates are significantly lower in comparison to that clustered around  $n \approx m$. 
%\textcolor{blue}{Note that the central region is asymmetric between Branch-1 and Branch-2 triads, and this is purely a consequence of internal wave's dispersion relation. When the lower frequency daughter wave (wave-1) travels in the same direction as the parent wave (i.e. Branch-2), 
%wave-1's modenumber should be higher than that of wave-2 for the triad condition to be satisfied.
%However in the reverse scenario, where wave-2 travels in the same direction as the parent wave (i.e. Branch-1), the difference between the modenumber of wave-2 and wave-1 need not be that high compared to Branch-2 triads. }
{Note that the central region is asymmetric between Branch-1 and Branch-2 triads, and this is purely a consequence of internal wave's dispersion relation. When the lower frequency daughter wave (wave-1) travels in the same direction as the parent wave (i.e. Branch-2), 
wave-1's modenumber ($n$) should always be higher than wave-2's modenumber ($m$) for the triad condition to be satisfied. However for Branch-1, where wave-2 travels in the same direction as the parent wave, the modenumber of wave-2 ($n$) need not be higher than wave-1's modenumber ($m$). }

\begin{figure}
 \centering{\includegraphics[width=120mm]{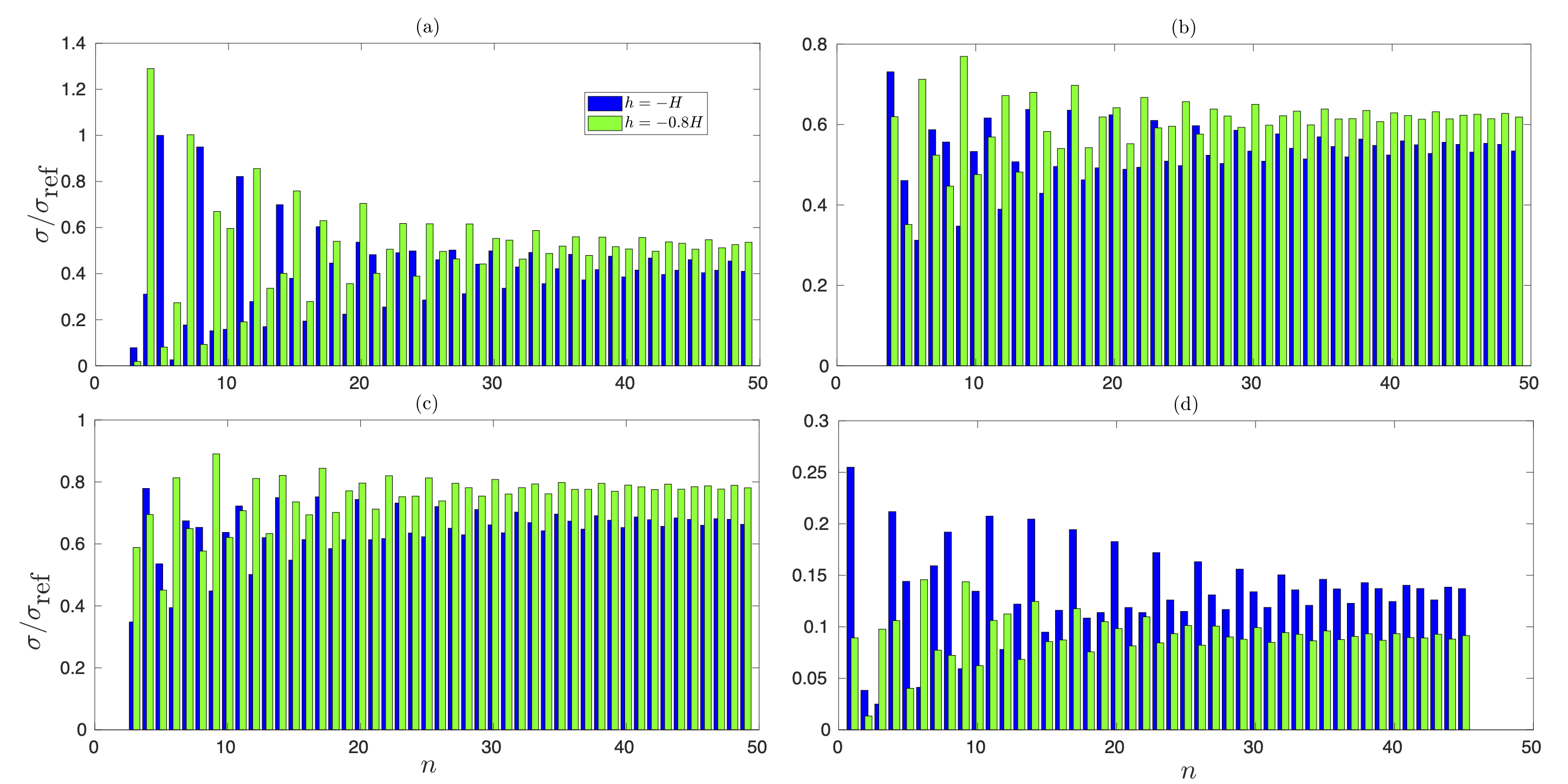}}
  \caption{ Plots of non-dimensional  growth rates of mode-1 triads for profile $N^{(8)}$  with $f/\omega_3 = 0.4$ and $\omega_3/N_b = 0.2$. (a) Line $(n,n)$ of Branch-1, (b) line $(n+1,n)$ of Branch-1, (c) line $(n,n+1)$ of Branch-1, and (d) line $(n+4,n)$ of Branch-2. }
  \label{fig:GR_1}
\end{figure}

\begin{figure}
 \centering{\includegraphics[width=120mm]{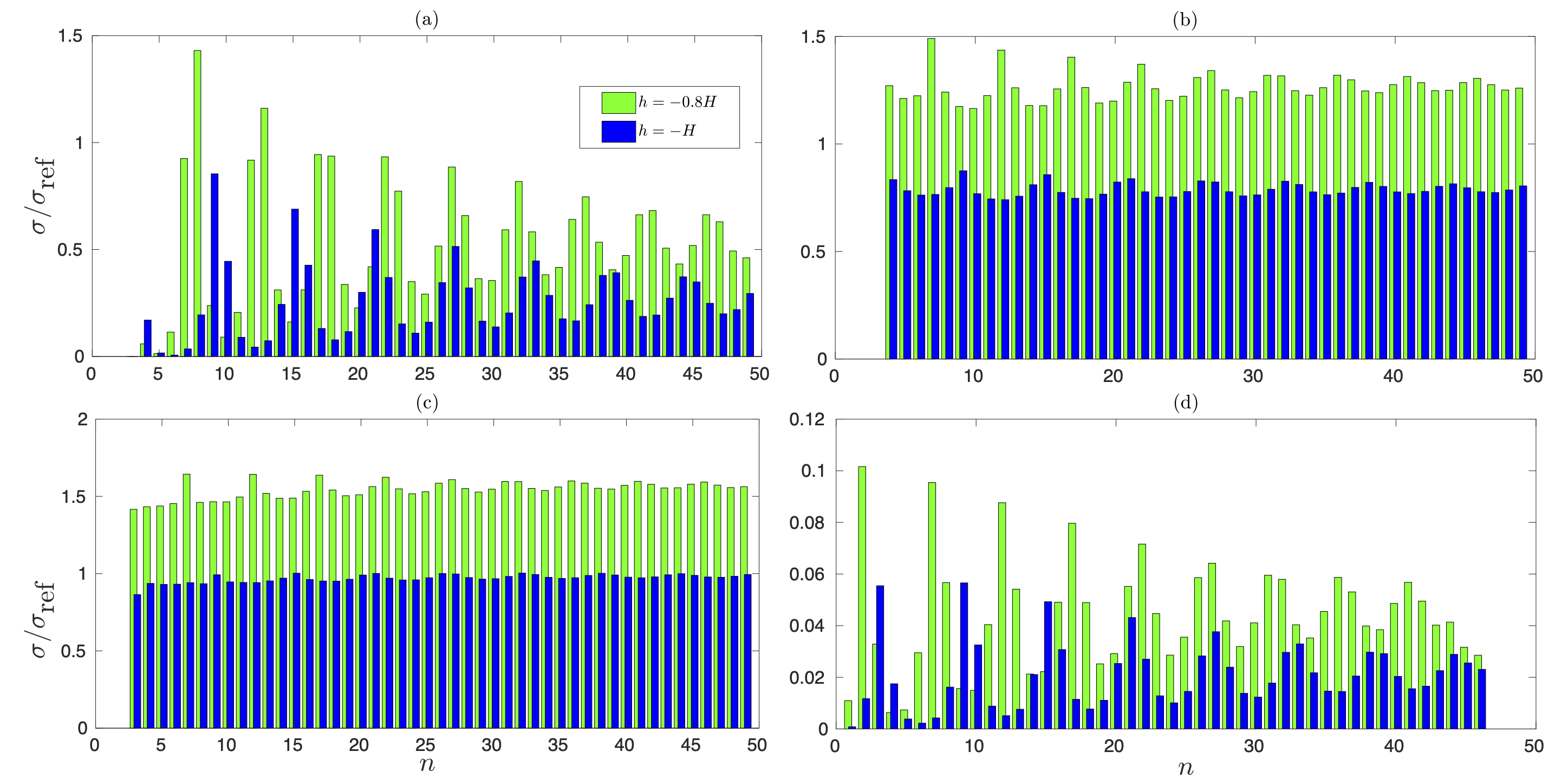}}
  \caption{ Plots of non-dimensional  growth rates of mode-1 triads for profile $N^{(9)}$ with $f/\omega_3 = 0.4$ and $\omega_3/N_b = 0.2$. (a) Line $(n,n)$ of Branch-1, (b) line $(n+1,n)$ of Branch-1, (c) line $(n,n+1)$ of Branch-1, and (d) line $(n+6,n)$ of Branch-2. }
  \label{fig:GR_2}
\end{figure}
The clustering around $n \approx m$ is consistently observed for any setting or stratification profile considered in our study. As a result instead of focusing on all possible triads, we choose specific lines of interaction near the central region and plot the growth rate along that line of interaction. For example, the interaction lines ($n$,$n$), ($n+1$,$n$) and ($n$,$n+1$)  are plotted for $n\in(1,50)$ in figures \ref{fig:GR_1}(a)--\ref{fig:GR_1}(c) for Branch-1 triads, and   ($n+4$,$n$) in figure \ref{fig:GR_1}(d) for Branch-2 triads.  {The notation (a,b) means wave-1(2) is mode-a(b). The notation is same for both branches.} The dominant nature of the interaction lines ($n$,$n$), ($n$,$n+1$), ($n+1$,$n$) has also been observed in \cite{young_g} while studying the stability of mode-1 internal wave in the presence of near inertial daughter waves (with frequency $f$). Furthermore, figure \ref{fig:GR_1} also reveals that the different lines are sensitive to $h$.
For completeness, we explore another stratification profile given by:
\begin{itemize}
    \item  $N^{(9)}$: $N_{\textnormal{max}} = 10N_b$, $W_p = H/50$, $z_c = H/80$,
\end{itemize}
and the corresponding plots are in figure \ref{fig:GR_2}.
Both figures \ref{fig:GR_1} and \ref{fig:GR_2} show that the growth rates along different lines of interaction have a significant oscillatory nature with $n$. In general, line $(n,n)$ has the largest amplitude of oscillations. More importantly, the growth rate of a modal combination can significantly change as $h$ changes. For example,  figure \ref{fig:GR_1}(a) shows that the most unstable modal combination at $h=-H$ is ($5$,$5$). However, for $h=-0.8H$, the most unstable triad is the modal combination ($4$,$4$). Moreover, the combination ($5$,$5$) has approximately $0.25$ times of ($4$,$4$) growth rate at $h=-0.8H$. This behavior can be seen for the line $(n,n)$ in both figures  \ref{fig:GR_1} and \ref{fig:GR_2}. This effectively means that the growth rate of certain daughter wave combinations can be sensitive to changes in $h$ (especially the combinations which involve lower modes). Such combinations may not be effective in a region of varying $h$ because of the significant drop in the growth rates. However, sensitivity to $h$ is slowly reduced as the modenumber is increased for both the branches. Even though Branch-2 triads have considerably less growth rates for the profiles $N^{(8)}$ and $ N^{(9)}$, for different profiles (not displayed here) Branch-2 can have $\sigma$ comparable to that of the Branch-1 triads.

%(yes yes \Psi. the Psi in terrain follwoing coordinates.).

\subsubsection{Effect of variation of $f/\omega_3$ and $\omega_3/N_b$ on different Branches.}
 
For both stratification profiles used in \S \ref{section:5.1}, $f/\omega_3 = (0.3, 0.45)$ for $\omega_3/N_b = (0.2,0.7)$ is explored (hence total of $4$ different combinations). 
For $N^{(8)}$, in all $4$ cases, the qualitative behaviour of all Branch-1 lines are similar to figure \ref{fig:GR_1}. However, the maximum growth rate has a significant increase from $h=-H$ to $h=-0.8H$ for $\omega_3/N_b = 0.7$. Moreover, Branch-2 triads' maximum growth rate significantly increases in $2$ cases of $f/\omega_3 = 0.3$ in comparison to $f/\omega_3 = 0.4$. 
%\textcolor{blue}{This means that triads that are not on the central lines $(n,n+1), (n,n), (n+1,n)$ can also have significant growth rates (similar growth rates to the central lines).}
For $N^{(9)}$, the qualitative behaviour of line $(n,n)$ is similar to what was observed in $f/\omega_3 = 0.4$. In general, the maximum growth rate is the $(n,n)$ modal combination. Interestingly, it is found that the maximum growth rate among all triads increased nearly twice from $h=-H$ to $h=-0.8H$ for $(f/\omega_3 = 0.45, \omega_3/N_b = 0.7)$. Significant increase in maximum growth rate is also found for $\omega_3/N_b = 0.2$ for the same $f$. For $(f/\omega_3 = 0.3, \omega_3/N_b = 0.7)$ the behaviour of line $(n+1,n)$ has a significant oscillation with $n$, similar to line $(n,n)$. Note that this is different from the line $(n+1,n)$ shown in figure \ref{fig:GR_2}.  In general, it is also observed that reducing the fluid depth increases the maximum growth rate of all possible triads even without considering the $\beta$ term of the parent wave amplitude. 

\subsection{Variation of nonlinear coupling coefficient with domain height for self-interaction process}
Here we restrict to self-interaction of internal gravity waves that do not experience significant detuning $\Delta \mathcal{K}$ with changes in $h$. In this subsection, we mainly focus on the superharmonic wave's nonlinear coupling coefficient $\mathcal{N}_3$ given in \eqref{eqn:coupling_coefficient_s3}.

\subsubsection{Class-1 interactions} \label{section:5.1.1} 
As previously mentioned in \S \ref{section:4.2}, some Class-1 self-interactions can have negligible detuning even for a finite range of $h/H$. We study the variation of $\Tilde{\mathcal{N}}_3$ under such circumstances; 
%for interactions whose detuning ($\Delta \mathcal{K}_s$) is such that $|\Delta \mathcal{K}_s| < 0.01$ for a given range of $h/H$. 
the different interactions considered (denoted by $\mathcal{I}_p$) are given below: 
\begin{center}
\begin{tabular}{ |c|c|c|c|c| } 
 \hline
 $\mathcal{I}_1$ - $[P3,D2]$ & $\mathcal{I}_2$ - $[P4,D2]$ & $\mathcal{I}_3$ - $[P4,D3]$ & $\mathcal{I}_4$ - $[P5,D3]$ & $\mathcal{I}_5$ - $[P5,D4]$ \\ 
 \hline
 $\mathcal{I}_6$ - $[P6,D3]$ & $\mathcal{I}_7$ - $[P6,D4]$ & $\mathcal{I}_8$ - $[P6,D5]$ & $\mathcal{I}_9$ - $[P7,D4]$ & $\mathcal{I}_{10}$ - $[P7,D5]$  \label{tab:1}\\ \hline
\end{tabular}
\end{center}
Here the notation $[Pm,Dn]$ denotes that the parent wave is the $m-$th mode and daughter (superharmonic) wave is the $n-$th mode. 

\begin{figure}
 \centering{\includegraphics[width=0.95\textwidth]{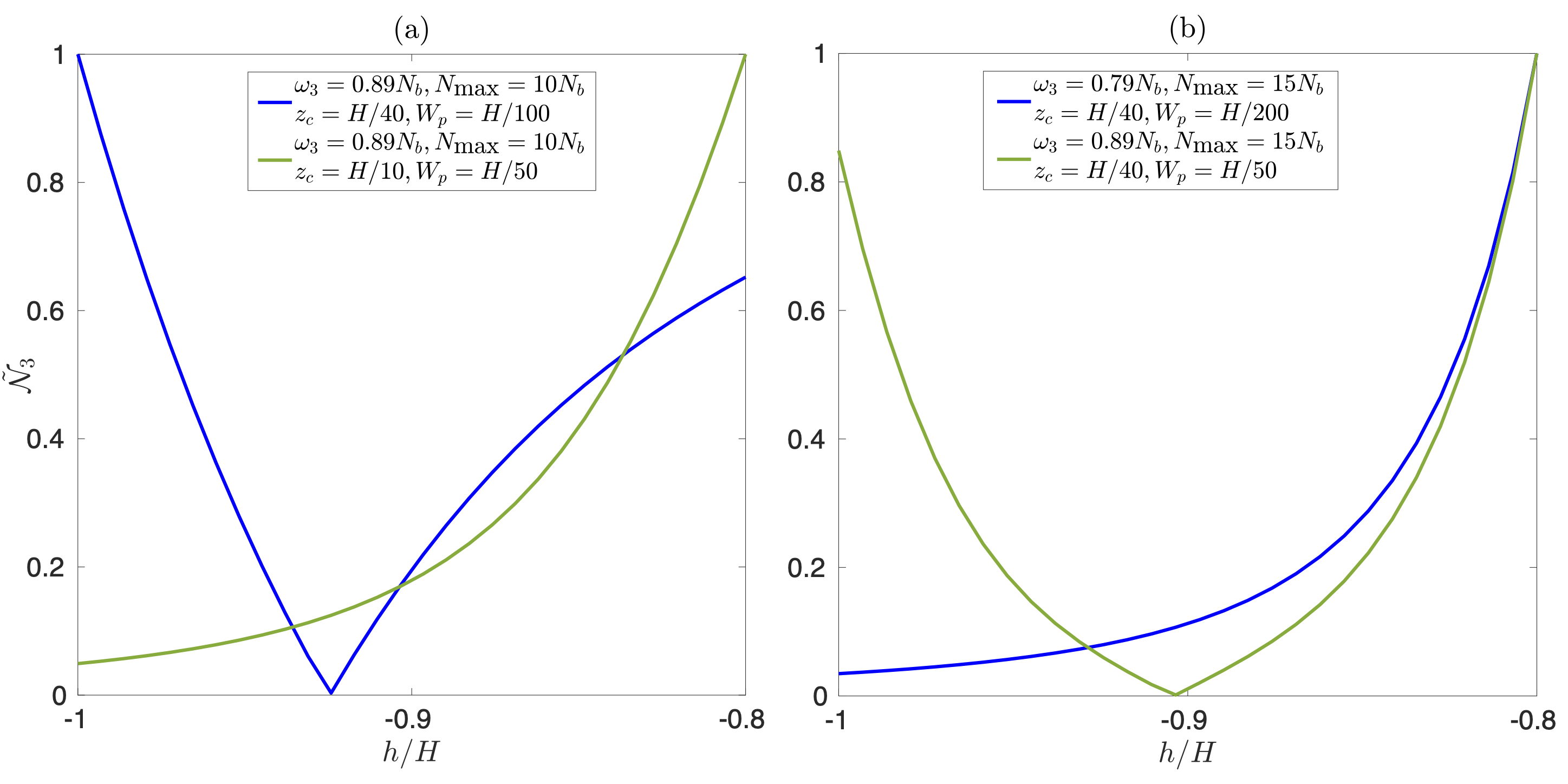}}
  \caption{ Variation of $\Tilde{\mathcal{N}}_3$ with $h$ for Class-1 self-interactions (a) $\mathcal{I}_1$ and (b) $\mathcal{I}_{10}$. Each sub-figure is plotted for two different stratification profiles (for details see  legend).   %Interaction $\mathcal{I}_1$ is shown in (a) for two stratification profiles, while (b) shows $\mathcal{I}_{10}$ for two different profiles. 
  }
  \label{fig:self_NL_2}
\end{figure}

{The stratification profiles are chosen such that $N_{\textnormal{max}}=(2 N_b, 5 N_b, 10N_b)$, $W_p=(H/200, H/100, H/50)$, and $z_c=(H/40, H/20, H/10)$.} For the profiles considered, we study variations of $\Tilde{\mathcal{N}}_3$ for interactions that strictly satisfy  $|\Delta \mathcal{K}_s| < 0.01$ for $h/H \in [-1,-0.8]$. %$\Tilde{\mathcal{N}}_3$ has the same definition as given in \S \ref{section:5.1.1}.
{Figure \ref{fig:self_NL_2} shows variations of $\Tilde{\mathcal{N}}_3\equiv|\mathcal{N}_3|/\textnormal{max}(|\mathcal{N}_3|)$ for two Class-1 self-interactions. Figure \ref{fig:self_NL_2} reveals that interactions can have a non-monotonic variation of $\Tilde{\mathcal{N}}_3$ with $h/H$.} Moreover, the figure reveals that even relatively small changes in $h/H$ can lead to significant variations in $\Tilde{\mathcal{N}}_3$. For all interactions considered in table \ref{tab:1}, small changes in $h/H$ can cause significant change in $\Tilde{\mathcal{N}}_3$.  Some generic features are summarised below. Sensitivity of $\Tilde{\mathcal{N}}_3$ to $h/H$ increases as $z_c$ is increased for a given $W_p, N_{\textnormal{max}}$ \footnote{For every interaction, there are  specific combinations of $W_p, N_{\textnormal{max}}$ where this behaviour is not exhibited.}.  Moreover, increasing $W_p$ also increases the sensitivity of $\Tilde{\mathcal{N}}_3$ to changes in $h/H$. Increasing $N_{\textnormal{max}}$ for a given $(z_c, W_p)$ also increases the sensitivity of $\Tilde{\mathcal{N}}_3$ to $h/H$.  

\begin{figure}
 \centering{\includegraphics[width=\textwidth]{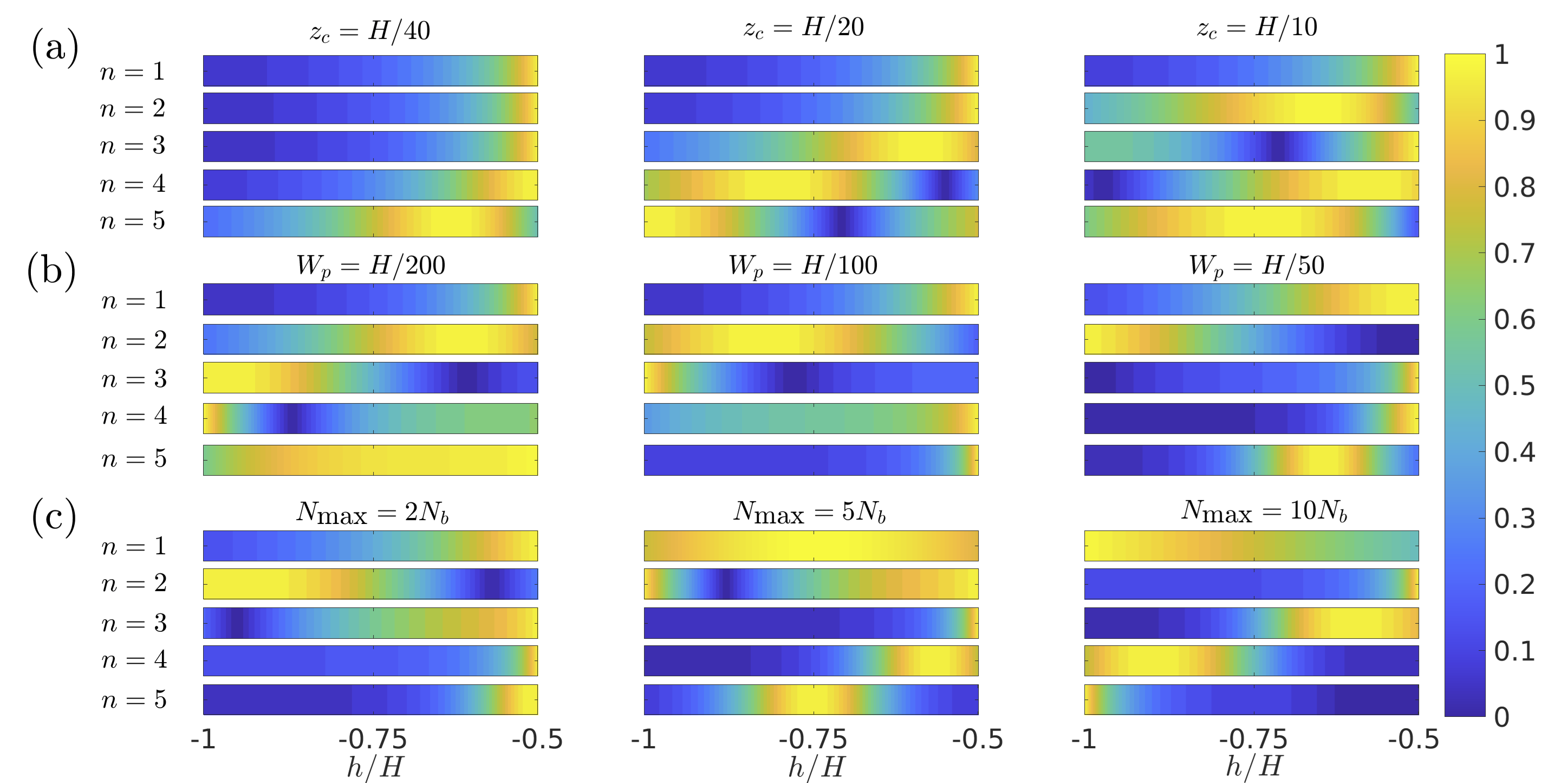}}
  \caption{ The variation of non-dimensionalised nonlinear coupling coefficient ($\Tilde{\mathcal{N}}_3$) of wave-3 (the superharmonic wave) as $h$ is varied. Altogether there are $9$ blocks, each consisting of $5$ units, and each unit represents the modenumber (given by $n$). For each block, $(N_{\textnormal{max}},W_p,z_c)$ is   fixed. The figure is further subdivided into three horizontal panels (each panel consisting of three blocks): (a) $N_{\textnormal{max}}=2N_b$, $W_p = H/200$, and $z_c$ is varied, (b) $N_{\textnormal{max}}=5N_b$, $z_c = H/20$ and $W_p$ is varied, and (c) $z_c = H/10$, $W_p=H/50$ and $N_{\textnormal{max}}$ is varied. 
  }
  \label{fig:self_NL}
\end{figure}

\subsubsection{Class-2 interactions}
\label{section:5.1.2} 
We initially study the variation of $\mathcal{N}_3$ with $h$ for Class-2 self interactions. To this end, we consider $\mathcal{N}_3$  of the first $5$ modes for $27$ different stratification profiles. {Similar to the case of Class-1 self-interactions in \S \ref{section:5.1.1}, the stratification profiles are chosen such that $N_{\textnormal{max}}=(2 N_b, 5 N_b, 10N_b)$, $W_p=(H/200, H/100, H/50)$, and $z_c=(H/40, H/20, H/10)$, where we consider all possible ($3^3$) combinations.}
Out of the $3^3=27$ combinations, $9$ profiles are chosen for plotting  figure  \ref{fig:self_NL} and thereby elucidating the effect of each individual parameter in the stratification profile. For all cases, $\omega_3 = 0.1N_b$ and $f=0$. 
%In figure \ref{fig:self_NL}, variation of first 5 modes' $\Tilde{\mathcal{N}}_3$ is shown for 9 stratification profiles. $\Tilde{\mathcal{N}}_3$ is a non-dimensionalised quantity that is given by $\Tilde{\mathcal{N}}_3 = \mathcal{N}_3/\textnormal{max}(\mathcal{N}_3)$, where $\textnormal{max}(\mathcal{N}_3)$ is the maximum growth rate of the superharmonic mode in the range $h/H \in [-1, -0.5]$. Figure \ref{fig:self_NL} is divided into 3 sub-categories (a,b,c) in which $z_c, W_p, N_{\textnormal{max}}$ is varied respectively while keeping the other two parameters constant. 
For some higher modes, the nonlinear coupling coefficient has a band like structure; there exists some range of $h/H$ where  $\Tilde{\mathcal{N}}_3$ is significantly higher in magnitude than that corresponding to other values of $h/H$. For example, the mode $n=5$ corresponding to $N_{\textnormal{max}}=5N_b$ in figure \ref{fig:self_NL}(c) reveals a large increase in $\Tilde{\mathcal{N}}_3$ near $h/H \approx -0.75$, while it is much lower at either ends.  \cite{wunsch} also observed such banded structure in the self-interaction of different modes as the stratification profile was changed. The reason behind the direct analogy between our observations and that of \cite{wunsch} is as straightforward --  when an internal wave travels to a different domain height, it essentially travels to a different stratification profile.

%  \textcolor{blue}{[The two paragraphs below have too much unnecessary information.. if you can combine the whole thing into something like for this set of constraints, we have a monotonic behavior that goes like $1/h^4$, and refer to some example situations in fig 6, that's all that is needed. Then say that such monotonic nature is not observe for other parameter regimes, and also allude to the fact that you have examined 18 more cases.=>] }
  
% For $N_{\textnormal{max}} = 2N_b$, for all combinations of $z_c$ and $W_p$ except  $z_c=H/10,W_p=H/50$, $\Tilde{\mathcal{N}}_3$ is proportional to $1/h^4$. In fact for mode 5 the profile with lowest values of $z_c,W_p,N_{\textnormal{max}}$ does not have the $\Tilde{\mathcal{N}}_3 \propto 1/h^4$ property. For $N_{\textnormal{max}} = 2N_b$, for any value of $z_c$ and $W_p$  $\Tilde{\mathcal{N}}_3$ monotonically increases with as $h/H$ is increased. For higher modes, behaviour of monotonically increasing $\Tilde{\mathcal{N}}_3$ with $h/H$ does not have a clear pattern as $z_c, W_p, N_{\textnormal{max}}$ is increased similar to mode-1 internal wave.
 
{For mode-1, when $z_c, W_p, N_{\textnormal{max}}$ are all on the lower side, we observe that $\Tilde{\mathcal{N}}_3 \propto 1/h^4$. For higher $N_{\textnormal{max}}$, even lower values of $z_c$ and $W_p$ do not have the property of $\Tilde{\mathcal{N}}_3 \propto 1/h^4$.  In general for modes $>1$, proportionality to $1/h^4$ is lost faster as $z_c, W_p, N_{\textnormal{max}}$ is increased. In several profiles, $\Tilde{\mathcal{N}}_3$ of higher modes is also more sensitive to changes in $h$ than $\Tilde{\mathcal{N}}_3 \propto 1/h^4$.}

\section{Higher order self-interactions in the presence of a small amplitude monochromatic {topography} }
\label{Section_6}
The focus of this section is on higher order self-interactions between a parent wave of frequency $\omega_1$ and a superharmonic  daughter wave of frequency $\omega_3=2\omega_1$ in the presence of a small amplitude monochromatic {topography}. 
In such kind of scenarios, the {topography} can act as a `zero frequency wave', which can lead to resonant higher order interactions, and is similar to the Class-2 studied in \cite{alam_2009} for surface gravity waves.  Such kind of higher order self-interactions might be important for mode-1 internal waves propagating in regions where $f>\omega_d/2$ ($\omega_d$ is the semidiurnal frequency), since triad interactions involving two
(subharmonic) daughter waves is not possible. 
%with infinitesimal energy cannot break down a mode-1 internal wave of semidiurnal frequency \citep{olbers_2020}.   
Moreover, resonant self-interaction for mode-1 internal wave of frequency $\omega_d$ is also not possible when $f>\omega_d/2$ for any stratification profile as a consequence of its dispersion relation \citep{wunsch}. This arises from the fact that a  mode-1 parent wave with frequency $\omega_d$ and a superharmonic daughter wave with frequency $2\omega_d$ fail to satisfy the horizontal wavenumber condition for a self-interaction process. {Note that higher order interactions are different from Bragg resonance focused in \cite{buhler_2011}, which is also a mechanism via which a parent mode--$1$ wave can decay by transferring its energy to the higher modes. In a standard Bragg resonance, resonant wave-topography interaction occurs if the bottom topography has a wavenumber $k_b$ such that $(\omega_1,k_b \pm k_1)$  satisfies the dispersion relation.}
%\cite{buhler_2011} showed that via Bragg resonance, a parent mode--$1$ wave can decay by transferring its energy to the higher modes.

To study higher order self-interactions, 
 we follow the streamfunction ansatz used in \cite{couston_2017} for studying internal wave Bragg resonance, and in \cite{leg_lahaye} for studying internal wave scattering due to interaction with a large amplitude {topography}.
 %use a different form of ansatz for the streamfunction $\psi$ (and hence also for buoyancy frequency $b$ and meridional velocity $v$). The type of streamfunction, used in this section, has been used in \cite{couston_2017} for studying internal wave Bragg resonance, and in \cite{leg_lahaye} for studying internal wave scattering due to interaction with a large amplitude bathymetry.
% Moreover this type of ansatz can also be used to study systems where $\Delta \mathcal{K}_s \sim \mathcal{O}(1)$ in the presence of a flat or slowly varying bathymetry. 
This ansatz for the  streamfunction of the $j-$th wave is as follows:
%given according to the following ansatz: 
\begin{equation}
\Psi_{j} =  \mathcal{A}_{j}(x)\phi_j(\eta;x)\ee^{-\ii\omega_j t}  + \mathrm{c.c}.. 
\label{eqn:s6_stream_ans}
\end{equation}
The corresponding buoyancy frequency and meridional velocity is given by
\begin{equation}
B_{j} =   \frac{\ii N^{2}}{\omega_{j}} \frac{\partial \mathcal{A}_j}{\partial x}\phi_j \ee^{-\ii\omega_{j}t} +  \mathrm{c.c}..,
\label{eqn:s6_buo_ans}
\end{equation}
%while the corresponding meridional velocity is:
\begin{equation}
\mathcal{V}_{j} = \frac{\ii f}{\omega_{j}} \frac{\mathcal{A}_j}{h}\frac{\partial \phi_j}{\partial \eta} \ee^{-\ii\omega_{j}t} +  \mathrm{c.c}..
\label{eqn:s6_v_ans}
\end{equation}
The above-mentioned ansatz can also be used to study systems where the detuning $\Delta \mathcal{K}_s \sim \mathcal{O}(1)$ in the presence of a flat or slowly varying bathymetry. The functions $\phi_j$ are same as the functions used in \S 2 and are given by solving \eqref{eqn:eigen2}. Here we only consider small amplitude {topography} whose wavenumber is comparable to the parent wave, i.e., $\epsilon_h \ll \mathcal{O}(1)$ and $\epsilon_k \sim \mathcal{O}(1)$. 

% The streamfunction ansatz is substituted in \eqref{eqn:combined_eta}, and similar to \S \ref{Section:2}, the linear terms of \eqref{eqn:combined_eta} is multiplied by $\phi_j$ and integrated in the $\eta$ direction, which leads to:

{To study higher order interactions of a parent wave propagating in the presence of a small amplitude topography, we also consider the linear scattering of the parent wave. Note that the linear scattering of the parent wave on its own is not resonant (we do not consider topography wavenumbers which allow resonant Bragg scattering) and hence over a long distance has a negligible effect on the parent wave's amplitude. However, even the non-resonant linear interaction of the parent wave with the topography, which leads to higher modes with $\omega_1$ frequency, can significantly impact the growth of the superharmonic wave. To derive the linear scattering of the parent wave as it moves through a topography, we assume the streamfunction of the waves to be:}
{
\begin{equation}
    \Psi_{1} =  \sum_{n=1}^{n=M_n} \mathcal{A}_{(1,n)}(x)\phi_{(1,n)}(\eta;x)\ee^{-\ii\omega_1 t}  + \mathrm{c.c}..
    \label{eqn:STR_P_wave}
\end{equation}}
\noindent {where $M_n$ is the maximum mode number after which the series is truncated, and $\phi_{(1,n)}$ is the $n-$th eigenfunction of $\omega_1$ frequency. The streamfunction ansatz \eqref{eqn:STR_P_wave} is substituted in \eqref{eqn:combined_eta}, and similar to \S \ref{Section:2}, the linear terms of \eqref{eqn:combined_eta} is multiplied by $\phi_{(1,n)}$ and integrated in the $\eta$ direction. This leads to $M_n$ ordinary differential equations, where the $n-$th differential equation is given by:}
{\begin{align}
\gamma_{n}^{(3)}\left[  \frac{\partial^2 \mathcal{A}_{(1,n)}}{\partial x^2}  + \mathcal{K}^2_{(1,n)}\frac{\mathcal{A}_{(1,n)}}{h^2}  \right] = -&\sum_{m=1}^{m=M_n} \left[\frac{2\gamma_{(m,n)}^{(5)}+\gamma_{(m,n)}^{(6)}}{h^2}\left(\frac{\partial h}{\partial x}\right)^2 - \frac{2\gamma_{(m,n)}^{(7)}}{h}\frac{\partial h}{\partial x}\right]   \mathcal{A}_{(1,m)}  \nonumber\\
-&\sum_{m=1}^{m=M_n} \left[ \gamma_{(m,n)}^{(8)} - \frac{\gamma_{(m,n)}^{(5)} }{h}\left(\frac{\partial^2 h}{\partial x^2} \right) \right]   \mathcal{A}_{(1,m)}  \nonumber\\
-&\sum_{m=1}^{m=M_n} 2\left[\gamma_{(m,n)}^{(4)} - \frac{\gamma_{(m,n)}^{(5)}}{h}\frac{\partial h}{\partial x}\right] \frac{\partial \mathcal{A}_{(1,m)}}{\partial x},
\label{eqn:S6_LHS_main_P_Wave}
 \end{align} 
% where $\mathcal{T}_j \equiv \ee^{-\ii\omega_{j}t}$, and
% $\textnormal{LIN}_j$ denotes the linear terms corresponding to the $j-$th wave. 
%The bathymetry is considered to be of small amplitude ($\epsilon_h \ll 1$), and hence the vertical eigenfunctions can evaluated in a similar way as \S \eqref{Section:2}.
where $\mathcal{K}_{(1,n)}$ is the corresponding eigenvalue of $\phi_{(1,n)}$. Moreover $\gamma^{(*)}_{(m,n)}$ are evaluated using the expressions given in appendix \ref{app:C}.
The above set of equations are similar to the equations derived in \cite{leg_lahaye}, except that we do not consider waves that travel in the  direction opposite to the parent wave since they are assumed to be negligible. Now that we have the full wave spectrum with $\omega_1$ frequency by solving \eqref{eqn:S6_LHS_main_P_Wave}, we model the evolution of the superharmonic wave. 
For simplicity, the feedback to the parent wave is neglected, which is analogous to the pump-wave approximation used in \S \ref{Section:2}.}

{The streamfunction of the superharmonic wave $\Psi_3$ is substituted in \eqref{eqn:combined_eta}, and the linear terms are multiplied by $\phi_{3}$ and integrated in the $\eta$ direction. This leads to:}
{
 \begin{align}
\noindent\textnormal{LIN}_3 &\equiv \hspace{0.1cm}  \left[  \left(\gamma_3^{(3)} \frac{\partial^2 \mathcal{A}_3}{\partial x^2} \right) + \mathcal{K}^2_3\left(\frac{\mathcal{A}_3}{h^2} \gamma_3^{(3)} \right) \right] \ee^{-\ii\omega_3 t} + 2\left[\frac{\gamma_3^{(5)}}{h^2}\left(\frac{\partial h}{\partial x}\right)^2 - \frac{\gamma_3^{(7)}}{h}\frac{\partial h}{\partial x}\right]   \mathcal{A}_3    \ee^{-\ii\omega_3 t}   \nonumber\\
&+ \left[ 
\frac{\gamma_3^{(6)}}{h^2}\left(\frac{\partial h}{\partial x}\right)^2 -  \frac{\gamma_3^{(5)} }{h}\left(\frac{\partial^2 h}{\partial x^2} \right)+ \gamma_3^{(8)} \right]\mathcal{A}_3\ee^{-\ii\omega_3 t} + 2\left(\gamma_3^{(4)} - \frac{\gamma_3^{(5)}}{h}\frac{\partial h}{\partial x}\right) \frac{\partial \mathcal{A}_3}{\partial x}  \ee^{-\ii\omega_3 t},
\label{eqn:S6_LHS_main}
 \end{align} 
}
\noindent {$\textnormal{LIN}_3$ only contains linear terms, and models the propagation of superharmonic wave in the presence of a topography. Note that superharmonic wave cannot exchange energy with higher modes of $\omega_3$. Now we move on to deriving the nonlinear terms which force the superharmonic wave.} Since we are focusing on higher order interactions, all nonlinear terms (including terms containing $x-$direction derivatives of $\phi$ and $h$), which have the same angular frequency as the superharmonic wave, are retained. 
%This is because higher order interactions in the presence of small amplitude topographies is caused because of the small, periodic changes in the $\mathcal{A}$, $k$ and $\phi$ of the parent wave due to the wave's propagation over the bathymetry. Hence in nonlinear terms, terms containing $x-$direction derivatives of $\phi$ and $h$ also have to considered for accurate modelling. 
In the terrain following coordinates, this would however lead to  a large number of terms that need to be evaluated. This issue can be circumvented following the procedure outlined below.
%Therefore to circumvent this issue the nonlinear terms are evaluated in the following way which is explained below.
%\begin{equation}
% \textnormal{NL} = -\frac{\partial}{\partial t} \left[\mathcal{J}\{({L}_{xx}+{L}_{\eta\eta})\Psi,\Psi\}\right] + {L}_{x}\left(\mathcal{J}\{B,\Psi\}\right) - f{L}_{z}\left(\mathcal{J}\{\mathcal{V},\Psi\}\right),
% \label{eqn:s6_rhs_NS}
% \end{equation}
The nonlinear terms in the terrain following coordinates are given by right hand side of \eqref{eqn:combined_eta}.

{We assume that the superharmonic wave is forced nonlinearly by the $\omega_1$ spectrum. To model this, we substitute $\Psi_{1}, B_{1}, \mathcal{V}_{1}$ into the nonlinear terms of \eqref{eqn:combined_eta}. Note that we can obtain $B_{1}$ and $\mathcal{V}_{1}$ from \eqref{eqn:s6_buo_ans} and \eqref{eqn:s6_v_ans} respectively.} 
% The superharmonic wave is assumed to be forced nonlinearly by parent waves with $\omega_1$ frequency. 
% (subscript `$1$' denote the complete set of waves with frequency ${\omega1}$)
% However, we need to evaluate $\mathcal{A}_1$, which is found by solving
% \begin{equation}
%      \textnormal{LIN}_{1} = 0.
%      \label{eqn:s6_p_wave_eq1}
% \end{equation}
%However, we need to this first entails the evaluation $\mathcal{A}_1$, which is found by solving
%\eqref{eqn:s6_p_wave} given below. 
% The nonlinear terms in the terrain following coordinates are given by right hand side of \eqref{eqn:combined_eta}, which is:
% \begin{equation}
% \textnormal{NL} = -\frac{\partial}{\partial t} \left[\mathcal{J}\{({L}_{xx}+{L}_{\eta\eta})\Psi,\Psi\}\right] + {L}_{x}\left(\mathcal{J}\{B,\Psi\}\right) - f{L}_{z}\left(\mathcal{J}\{\mathcal{V},\Psi\}\right),
% \label{eqn:s6_rhs_NS}
% \end{equation}
% Next we substitute $\Psi_1, B_1, \mathcal{V}_1$ (subscript $1$ denotes parent wave), given by \eqref{eqn:s6_stream_ans}--\eqref{eqn:s6_v_ans}, into \eqref{eqn:s6_rhs_NS} but this first entails the evaluation $\mathcal{A}_1$, which is found solving \eqref{eqn:s 
After the substitution, similar to the linear terms of wave-3, the nonlinear terms are multiplied by $\phi_3$ and integrated in $\eta$ direction within the domain limits. The resultant expression obtained is as follows:
\begin{equation}
\hspace*{-0.2cm}\langle \textnormal{NL}_3 \rangle = \int_{-1}^{0}\phi_3 \left[\ii \omega_3 \mathcal{J}\{({L}_{xx}+{L}_{\eta\eta})\Psi_1,\Psi_1\} + {L}_{x}\left(\mathcal{J}\{B_1,\Psi_1\}\right) - f{L}_{\eta}\left(\mathcal{J}\{\mathcal{V}_1,\Psi_1\}\right) \right] d \eta.
\label{eqn:s6_rhs_main}
\end{equation}
% In the $\langle \textnormal{NL}_3 \rangle$, the terms are evaluated numerically for the evaluated fields $\Psi_1, B_1, \mathcal{V}_1$, instead of analytically evaluating derivatives by considering $\Psi_1, B_1, \mathcal{V}_1$
{Therefore the final superharmonic wave equation can be written in a compact form:
\begin{align}
\textnormal{LIN}_{3} &= \langle \textnormal{NL}_3 \rangle.
\label{eqn:s6_d_wave}
\end{align}}
%where $\textnormal{LIN}_{j}$ is defined in \eqref{eqn:S6_LHS_main} and $\langle \textnormal{NL}_3 \rangle$ is defined in \eqref{eqn:s6_rhs_main}.
In equations \eqref{eqn:S6_LHS_main} and \eqref{eqn:s6_d_wave}, instead of splitting $\mathcal{A}_j$ into a product of slowly varying amplitude and rapidly varying phase part, we simply solve the equations numerically by retaining $\mathcal{A}_j$ as it is. This is mainly because, as mentioned above, the number of nonlinear terms would be significantly high in terrain following coordinates.
For high ratios of $f/\omega_1$ (for example, north of critical latitude), the parent wave cannot resonantly self interact with the superharmonic wave in the presence of a flat bottom. However, a resonant higher order self-interaction can occur provided the {topography} has a wavenumber $k_b$ such that:
\begin{equation}
    k_b = k_3 - 2k_1,
    \label{eqn:HR_cond_1}
\end{equation}
where $k_3$ is the wavenumber of the superharmonic wave and the $k_1$ is the wavenumber of the parent wave. In such scenarios, the daughter wave's amplitude will consistently grow. However this being a higher order interaction, the growth rate of daughter wave (consequently, the decay of the parent wave) can be expected to be slower than a resonant self interaction.  

{To elucidate and validate the higher order self interaction process, we perform numerical simulations by solving the complete 2D Boussinesq equations and comparing the output with the results of the reduced order model derived in this section. We run three simulations where the parent and daughter waves' frequencies are held fixed. They are denoted by Case-1, Case-2, and Case-3.}
For all the simulations, the parent wave frequency is $\omega_1/N_b = 0.2$, where $\omega_1$ is the semi-diurnal frequency, i.e. $\omega_1 = 1.4 \times 10^{-4} \textnormal{s}^{-1}$. Both the parent and daughter waves are mode-1 of their respective frequencies ($\omega_1$ and $2\omega_1$). These parameters would result in a significant detuning between the two waves at high $f$ values.
%is chosen such that $\omega_3/N_b = 0.2$, where $\omega_1 = 1.4 \times 10^{-4} \textnormal{s}^{-1}$ (semidiurnal frequency)
%The primary wave frequency $\omega_1$ is chosen such that $\omega_1/N_b = 0.2$ for all simulations, where $\omega_1 = 1.4 \times 10^{-4} \textnormal{s}^{-1}$ (semidiurnal frequency) is used for all simulations. Moreover $f/\omega_1 = 0.6$ is chosen for all the simulations.
 \begin{figure}
 \centering{\includegraphics[width=1.0\textwidth]{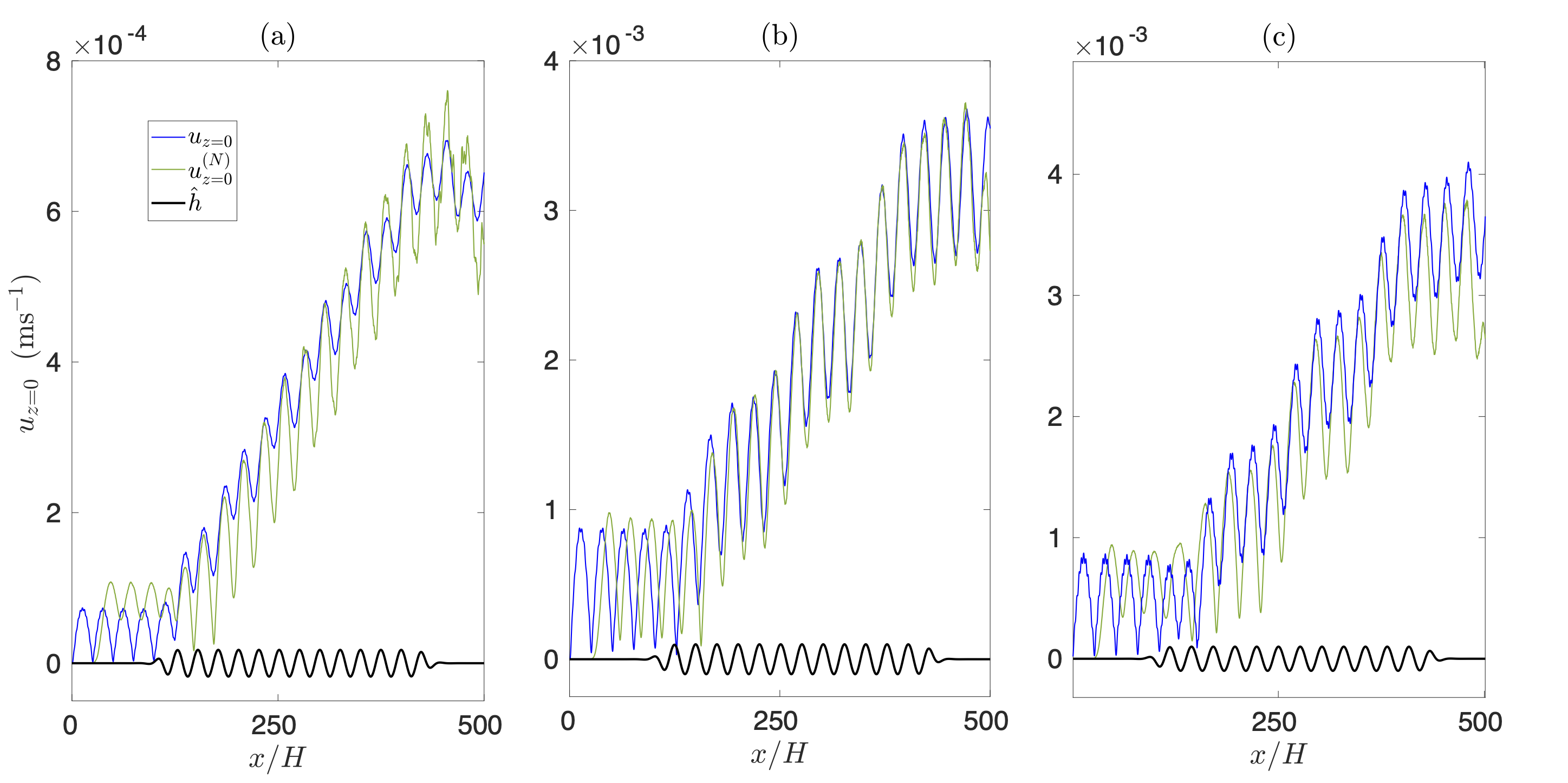}}
  \caption{ {Higher order self-interaction of mode-1 internal wave in the presence of a monochromatic bathymetry for three different cases: (a) Case-1, (b)  Case-2, and (c) Case-3. The topography profile (not to scale) is shown for all three cases.}
  %$\lambda_b = 2\pi/k_b$ is the wavelength of bottom bathymetry.
  }
  \label{fig:S6_higher_order}
\end{figure} 
{The bathymetry profile is given by:
\begin{equation}
    h = -H + \left[\epsilon_h H \sin{(k_b( x-x_c))}\right]\times\frac{1}{(1+(x-x_c)^{32}/W_T^{32})}, 
\end{equation} 
where $\epsilon_h = 0.01$,  $H=3000$m, domain length $L=540H$, and $x_c=L/2$ are held fixed across all three simulations. The stratification profile parameters, $f$ value, incoming maximum velocity of the parent mode ($u_{\textnormal{max}}$), and $W_T$ for the three simulations are given in table \ref{tab:2}.}
\begin{table}
    \centering
{\begin{tabular}{ c c c c c c c } 
  & $W_p/H$ & $z_c/H$ & $N_{\textnormal{max}}/N_b$ & $f/\omega_d$ & $u_{\textnormal{max}} (\textnormal{ms}^{-1})$ & $W_T/L$  \\ 
 \hline
 Case-1 \hspace{0.1cm} & 1/16 \hspace{0.1cm}& 1.5/1000   &\hspace{0.1cm} 4.5   &\hspace{0.1cm} 0.58   & \hspace{0.1cm} 0.0120   &\hspace{0.1cm} 0.30 \\ 
 \hline
Case-2 \hspace{0.1cm}& 29.5/400 \hspace{0.1cm}& 1.5/1000   &\hspace{0.1cm}  7  \hspace{0.1cm}& \hspace{0.1cm} 0.64  &\hspace{0.1cm}  0.0227  &\hspace{0.1cm} 0.30 \\ 
 \hline
Case-3 \hspace{0.1cm}& 1/20 \hspace{0.1cm}& 1.5/1000   &\hspace{0.1cm}  10  \hspace{0.1cm}& \hspace{0.1cm} 0.60  &\hspace{0.1cm}  0.0232  &\hspace{0.1cm} 0.31 \\  
\end{tabular}}
\caption{{ The stratification profile parameters, Coriolis frequency, $W_T$, and the velocity amplitude of the three waves for Case-1, Case-2, and Case-3.}}
\label{tab:2}
\end{table}
% \begin{itemize}
%     \item  {$N^{(10)}$:} $W_p = $, $z_c = 1.5H/1000$, $N_{\textnormal{max}} = 4.5N_b$, and $f=0.58\omega_d$
%     \item  {$N^{(11)}$:} $W_p = 29.5H/400$, $z_c = 1.5H/1000$, $N_{\textnormal{max}} = 7N_b$, and $f=0.64\omega_d$.
%     \item  {$N^{(12)}$:}  $W_p = H/20$, $z_c = 1.5H/1000$, $N_{\textnormal{max}} = 10N_b$, and $f=0.60\omega_d$.
% \end{itemize} 
% $W_T=0.3L$ is used for $N^{(10)}$ and $N^{(11)}$, while $W_T=0.31L$ is used for $N^{(12)}$. Note that for each simulation $k_b$ satisfies equation \eqref{eqn:HR_cond_1}. For $N^{(10)}$
% The initial parent wave amplitude at $x=0$ is chosen as $\mathcal{A}_1 = 0.005$ for all three cases. 
{The results after solving the reduced order model and 2D Boussinesq equations for the above mentioned parameters are shown in figure \ref{fig:S6_higher_order}. The 2D Boussinesq equations are solved using Dedalus.} {More details on the simulations are given in the end of \S \ref{Section:7}}. In all three sub-figures, the amplitude of the daughter wave is observed to be slowly increasing due to the higher order self-interaction.
%it can be seen that the amplitude of the daughter wave is slowly increasing.
Moreover, the daughter wave's amplitude also rapidly oscillates because of the non-resonant standard self-interaction process between the parent wave and the daughter wave. In the absence of a varying bathymetry, only the rapid non-resonant interaction would be present without any consistent growth in the daughter wave's amplitude. Therefore we have shown that for scenarios where Bragg resonances are not resonant, higher order interaction might be a possible mechanism that can scatter the energy of the mode--1 internal wave.

\section{Numerical Validation  \label{Section:7}} 
 
In this section, we provide numerical validations for the reduced--order equations \eqref{eqn:wave_self_prim}--\eqref{eqn:wave_self_super} derived through multiple-scale analysis for two different cases. This is done by solving the {2D Boussinesq equations} in terrain-following coordinates using an open--source, pseudo--spectral code Dedalus \citep{Dedalus}. The above mentioned equations in primitive variables along with viscous and hyperviscous terms {(the latter terms damping much smaller scales than the former)} are given below:
\begin{subequations}
\begin{align}
\frac{\partial u}{\partial t}  +  L_{x}(P)  + uL_{x}(u) + wL_{\eta}(u)   &= \nu L_{\eta\eta}(u) + \left(\frac{\nu_{6z}}{H^6}\frac{\partial^6 u}{\partial \eta^6}+{\nu_{6x}}\frac{\partial^6 u}{\partial x^6}\right), \label{eqn:u_vel_s7}\\
\frac{\partial w}{\partial t}  +  L_{\eta}(P)  + uL_{x}(w) + wL_{\eta}(w)   &=  \nu L_{\eta\eta}(w) + B, \label{eqn:w_vel_s7}\\
\frac{\partial B}{\partial t}  +  N^2w  + uL_{x}(B) + wL_{\eta}(B)   &= \nu L_{\eta\eta}(B) + \left(  \frac{\nu_{6z}}{H^6}\frac{\partial^6 B}{\partial \eta^6}+ {\nu_{6x}}\frac{\partial ^6 B}{\partial x^6}\right), \label{eqn:b_s7}\\
L_{x}(u) + L_{\eta}(w) &= 0.  \label{eqn:con_s7}
\end{align}
\end{subequations} 
{Here $(u,w)$ = $(L_{\eta}(\Psi),-L_{x}(\Psi))$, meaning that here the velocity field is defined in $x$--$\eta$ instead of $x$--$z$ coordinates}.
In all our simulations, $\nu = 10^{-5} \textnormal{m}^{2}\textnormal{s}^{-1}$, {$\nu_{6x} = 10^{8}\textnormal{m}^{6}\textnormal{s}^{-1}$ and  $\nu_{6z} = 81\textnormal{m}^{6}\textnormal{s}^{-1}$.} 
Equations (\ref{eqn:wave_self_prim})--(\ref{eqn:wave_self_super}) are solved  using RK4 method for time-stepping and second order accurate discretization scheme for the term ${\partial a_j}/{\partial x}$, where the scheme is forward or backward depending on the group speed direction of the particular wave. { We estimate the validity of two different cases.}
 
For Case 1, self interaction of a plane wave in the presence of a constant $h$ is simulated. The parameters of the simulation are as follows: $H=3000$m and $N_b=10^{-3}\textnormal{s}^{-1}$. The frequency of the parent wave is taken as $\omega_1/N_b = 0.447$, while $f=0$ is chosen. The stratification profile \eqref{eqn:strat_profile} is considered and the parameters used are:
\begin{itemize}
    \item $N^{(13)}$: $N_{\textnormal{max}} = 3N_b$, $W_p = 3H/100$, $z_c = H/10$. 
\end{itemize}
Mode-3 of parent wave frequency ($\omega_1$) in this scenario resonantly self interacts with mode-2 of $2\omega_1$ \citep{varma}. The parent wave streamfunction input (initial condition) to the full numerical simulation is given by:
\begin{equation}
        \Psi = A_1 {{\phi}_1} \sin(k_1 x) \exp(-(x-x_c)^2/W_1^2),
        \label{eqn:Psi_valid}
\end{equation} 
where  $|A_1| = 0.06$, $W_1 \rightarrow \infty$ is chosen (note that $x_c$ can be any value since $W_1 \rightarrow \infty$). $\Psi$ is used to obtain $(u,w)$. The numerical code is also initialised with the corresponding buoyancy frequency for the streamfunction given in \eqref{eqn:Psi_valid}.  
 
\begin{figure}
 \centering{\includegraphics[width=0.9\textwidth]{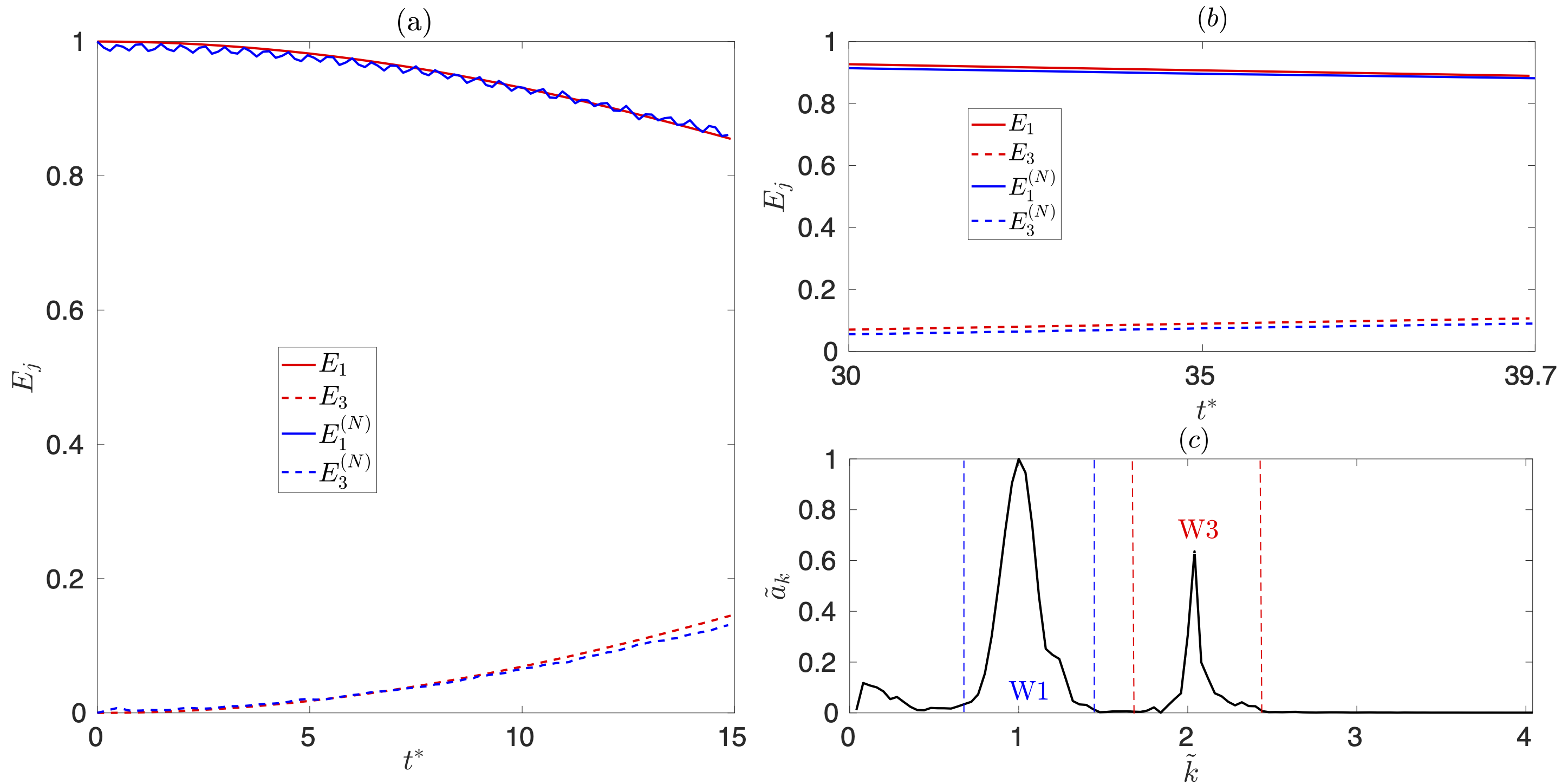}}
  \caption{(a) The energy evolution of waves for Case-1 from reduced--order equations and numerical simulations. The superscript ${(N)}$ denotes the results from numerical simulation of 2D Boussinesq equations. (b) The energy evolution of waves for Case-2 from reduced--order equations and numerical simulations in the time span $t^*=30$ to $t^*\approx40$, where $t^* \equiv \omega_1t/2\pi$. (c) Fourier transform of $B$ 
  at $\eta = -0.42$ and $t^{*}=40$.  }
  \label{fig:Valid_1}
\end{figure}

For estimating the energy of parent and daughter waves from the numerical simulations, only the potential energy of the waves is considered. This is valid because when $f=0$, energy is equally partitioned between potential energy and kinetic energy. For evaluating the potential energy of the two  waves, we take the Fourier transform of $B$ in the $x-$direction. Then by simply isolating the $k_3$ and $k_1$ wavenumbers, the respective fields due to wave-3 and wave-1 can be obtained for all time. The resulting energy evolution of the waves for Case 1 is shown in figure \ref{fig:Valid_1}(a). { At the end of the simulation, the parent wave energy was observed to be $86\%$ of the total energy in the Boussinesq equations simulation, while the reduced order model predicted that $85.5\%$ of the total energy will be contained in the parent wave at the specified time interval. Moreover, at the end of the simulation, the daughter wave's energy in Boussinesq equations simulation and the reduced order model are $13.0\%$ and $14.4\%$ respectively.}

\begin{figure}
 \centering{\includegraphics[width=\textwidth]{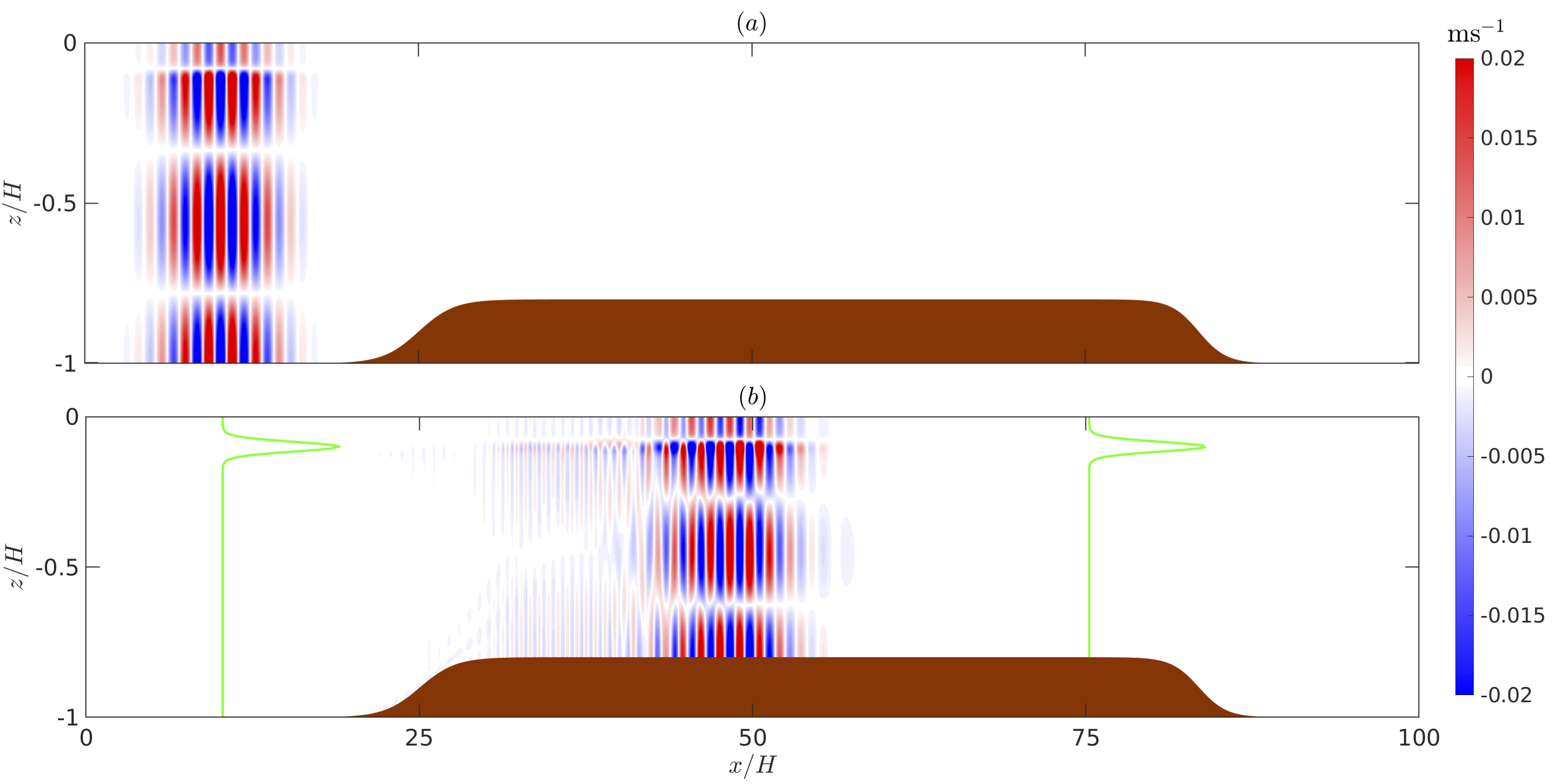}}
  \caption{ Horizontal velocity plot ($u$) from Case-2 simulation at (a) $t^{*}=0$ and (b) $t^{*}=30$. The stratification profile shape used is also shown for visual purposes. A faint but clear mode-2 trail left by the mode-3 parent wave can be seen at (b). The shape of the stratification profile used in Case-2 is  given by green curves in (b).  }
  \label{fig:Valid_contour}
\end{figure}

In Case 2, we consider the self interaction of a parent wave packet travelling in the presence of a slowly varying bathymetry. The parameters considered are as follows: $H=3000$m and $N_b=2.5\times10^{-3}\textnormal{s}^{-1}$. The following stratification profile is used:
\begin{itemize}
    \item $N^{(14)}$: $N_{\textnormal{max}} = 5N_b$, $W_p = 3H/100$, $z_c = H/10$. 
\end{itemize}
$\omega_1/N_b = 0.447$ with $f=0$ is chosen. Mode-3 of $\omega_1$ resonantly self interacts with mode-2 of $2\omega_1$ for $h/H \in [-1,-0.8]$. The bathymetry is given by:
\begin{equation}
    h = -H  + 0.1H \left[ \tanh((x-x_{t1})/W_{t1}) + \tanh((x_{t2}-x)/W_{t2}) \right],
    \label{eqn:top_valid}
\end{equation}
where $W_{t1} = 2.7H$, $W_{t2} = W_{t1}/1.3$, $x_{t1} = 25H$ and $x_{t2} = 83H$ was considered, where $100H$ is the domain length in $x-$direction. The bathymetry shape can be visualised in figure \ref{fig:Valid_contour}. 

The parent wave streamfunction of the form \eqref{eqn:Psi_valid} is used with $|A_1| = 0.022$, $W_1 = 3.57H$ and $x_c = 10.2H$ is chosen. Here $k_1$ is evaluated at $h=-H$ for the initial conditions. The $x_c$ value is chosen such that the wave packet is just at the bottom of the `plateau-like' {topography} at $t=0$ (as shown in figure \ref{fig:Valid_contour}(a)). The bathymetry is considered to be slowly varying so that the wave packet scattering by the bathymetry is negligible. Here the same procedure of energy evaluation as Case 1 is followed, except the energy is evaluated from $t^{*} = 30$, when the entire energy of both wave packets is almost confined to the top of the plateau region (where $h = -0.8H$). This makes the energy evaluation straightforward. We only consider $B$ in the range $x/H \in [33, 68]$ (where most of the energy is contained), and then again perform Fourier transform to separate the energy of the daughter and parent wave packets. 

%An important point to notice is that even though the energy evaluation graphs are plotted from $t^{*} = 30$ to $t^{*} = 40$, the parent wave starts to self interact from $t^{*} = 0$.\\
%For evaluating the energy in each wave, procedure used in Case 1 is also used in Case 2. The Fourier transform of $B$ at $t^*=30$, and $\eta = -0.375$ is shown in figure \ref{fig:Valid_1}(c).
In Case-2, since wave packets are considered, the Fourier transform of $B$ would not have a sharp peak at $k_1$ and $k_3$. Instead, a smoother peak in $k$--space would be produced as shown in figure \ref{fig:Valid_1}(c). We define a nondimensional wavenumber $\Tilde{k}$ as $\Tilde{k} \equiv k/k_1$. For evaluating the energy of both wave packets, amplitude ($|a_{\Tilde{k}}|$) in a finite range of $\Tilde{k}$ is considered. The energy contained in ($\Tilde{k}\in[0.6,1.4]$) is considered as the energy of the parent wave packet, while the energy in ($\Tilde{k}\in[1.6,2.4]$) is considered as the daughter wave's energy. For example, the $\Tilde{k}$ range considered for the parent and daughter waves are highlighted in figure \ref{fig:Valid_1}(c) using colored dotted lines for a specific $\eta$ and $t^{*}$.  For the parent wave, $|a_{\Tilde{k}}|$  between the blue dotted lines is considered. Similarly for evaluating the energy of the daughter wave packet, we consider the amplitude ($|a_{\Tilde{k}}|$)  between the red dotted lines. The energy evolution of the wave packets are shown in \ref{fig:Valid_1}(b). {The Parent wave packet energy in Boussinesq equations simulation and the reduced order model was observed to be $88.1\%$ and $88.9\%$ respectively at the end of the simulations. At the same time, the daughter wave's energy in Boussinesq equations simulation and the reduced order model are $8.95\%$ and $10.6\%$ respectively.} 

{To obtain the numerical results in figure \ref{fig:S6_higher_order}, equations \eqref{eqn:u_vel_s7}--\eqref{eqn:con_s7} along with equation for meridional velocity ($v$) were solved. Vertical hyperviscous term was not used, while a different horizontal hyperviscous term was used: $\nu_{12x}\partial^{12}/\partial x^{12}$, where $\nu_{12x} = 1.4\times10^{25}\textnormal{m}^{12}\textnormal{s}^{-1}$. The kinematic viscosity was chosen to be $\nu = 10^{-3} \textnormal{m}^{2}\textnormal{s}^{-1}$. The primary wave was forced by using a forcing function in the $u-$momentum equation, which sends a constant amplitude mode-1 wave train onto the small amplitude topography.}

%to obtain the results in Numerical simulations obtained by solving the Boussinesq given in section-6
 
\section{Summary and Conclusion  \label{Section:8}} 

\begin{figure}
 \centering{\includegraphics[width=1.0\textwidth]{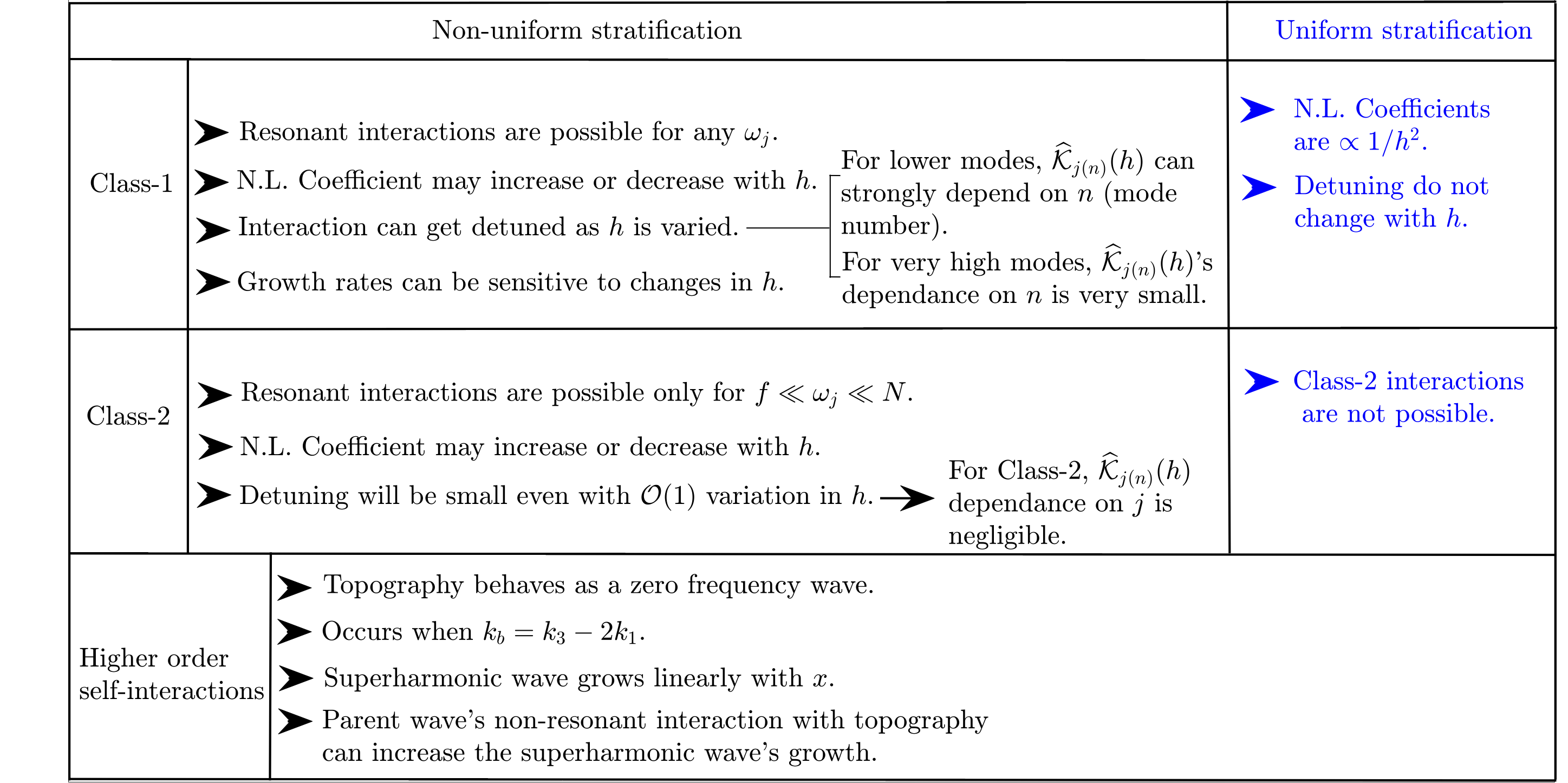}} 
 \caption{ A summary diagram shows how different factors such as detuning, nonlinear coefficients can vary for different classes of interaction in the presence of non-uniform stratification. It also provides a brief picture of the higher order self interaction studied in \S \ref{Section_6}. }
  \label{fig:table}
\end{figure}

Weakly nonlinear wave-wave interactions is one of the mechanisms through which internal gravity waves' energy cascade from large length scales (hundreds of kilometers) to small scales (centimeters to meters).
%($\mathcal{O}(1000\textnormal{m})$) to small length scales ($\mathcal{O}(\textnormal{cm})-\mathcal{O}(\textnormal{m})$).
{At small length scales, internal waves can give rise to convective or shear instabilities \citep{koudella_staquet_2006} and cause mixing, thus resulting in increased diffusion in oceans. }  
%Internal waves at small scales are prone to wave-breaking due to convective overturns or shear-insatbility \cite{koudella}.
%In this paper, we have derived the amplitude evolution equations for internal gravity waves undergoing weakly nonlinear wave-wave interactions in the presence of varying density stratification (resembling that of actual oceanic scenarios) as well as mild slope bathymetry in a bounded domain.
%{The incompressible, inviscid, 2D (in the $x$--$z$ plane), Boussinesq Navier-Stokes equations in the $f-$plane is written in terrain following coordinates ($x$--$\eta$).} 
{The 2D Boussinesq equations are written in terrain following coordinates ($x$--$\eta$).} 
Using multiple-scale analysis, we derive the amplitude evolution equations for internal gravity waves undergoing weakly nonlinear wave-wave interactions in the presence of varying density stratification (resembling that of actual oceanic scenarios) as well as mild slope bathymetry in a {vertically} bounded domain. {If the stratification varies with $z$ in the $x$--$z$ coordinates, then it becomes a function of both $x$ and $\eta$ in the $x$--$\eta$ coordinates when {bathymetry}, $h$, varies with $x$. In other words, the effective stratification profile varies with the ocean depth.}
%The effect of planetary rotation is also considered. 
Both triads and self-interactions are studied, and both pure resonant conditions as well as systems with wavenumber detuning are analyzed. The main results of this paper are given in a brief format in figure \ref{fig:table}.    

In the presence of uniform stratification, we show that the horizontal wavenumber triad condition, given by $k_{(1,a)}+k_{(2,b)}+k_{(3,c)}=0$, is not violated due to changes in $h$. Here $(a,b,c)$ are the modenumbers of waves $1$, $2$, and $3$ respectively. Moreover, in the presence of uniform stratification, the nonlinear coupling coefficients are inversely proportional to the square of the fluid depth ($\propto 1/h^2$).

For non-uniform stratifications, we define two classes of interaction for both triads and self-interactions. {
Class-1 involves weakly nonlinear interactions of waves that do not have the same mode number. Class-2 is a special situation that involves interactions of waves with the same modenumber, i.e. $a=b=c$.
% Interactions that do not fall into the definition of Class-1 are classified as Class-2. We observe the existence of 
Class-2 triad interactions can exist only in the parameter regime of $f \ll \omega_j \ll N$. Moreover, in the same parameter regime, near-resonant Class-2 self-interactions can exist with very low detuning even as $h$ is varied.}
%In such interactions, the horizontal wavenumber triad condition will be nearly satisfied for a finite range of $h$ without significant detuning.  
%This is because the wavenumbers involved in a triad, or a self-interaction, change in the same way as $h$ changes.
This is because the wavenumbers involved in a self-interaction, change in the same way as $h$ changes. For Class-1 interactions, detuning may be induced in triads and self-interactions if the waves interact in a region of varying $h$. This is because in a {vertically} bounded domain, the horizontal wavenumbers are not only a function of $h$ but is also a function of the modenumber. Moreover, the functional dependence of the wavenumber on $h$ may change as the modenumber changes. Therefore, in a weakly nonlinear interaction where different modenumbers are involved, there is no constraint for the wavenumbers to satisfy the triad condition in a given range of $h$.

The variation of the growth rate of the daughter waves in both triadic- and self-interactions is studied when $h$ is varied. For both Class-1 and Class-2 self-interactions, it is observed that small changes in $h$ may result in large changes in the growth rate of the daughter waves. This characteristic is especially observed for Class-1 self-interactions. Variation of growth rates with $h$ is studied for triads of a mode-$1$ parent wave in the presence of non-uniform stratification. Triads were identified such that the daughter waves can be up to mode-$50$. For relatively small changes in $h$, the growth rates can vary significantly for triads that involve only lower modes. Moreover, the most unstable daughter wave combination for the same parent wave can also change for relatively small changes in $h$. 
Unlike uniform stratification, in non-uniform stratification, the growth rates do not have a monotonic behavior with $h$. This was observed for both triadic- and self-interactions. 

Reduced-order equations for higher-order self-interactions of an internal wave in the presence of a small amplitude, monochromatic {topography} is also derived. In the higher-order self-interaction process, the small amplitude {topography} behaves as a zero frequency wave. It is shown that such higher-order interactions can cause resonant growth of the superharmonic wave. Such higher-order interactions can play a crucial role in the decay of the mode-$1$ internal wave at latitudes greater than $28.9^{\circ}$. This is because sum-type triad interactions are not possible \citep{olbers_2020} and a mode-$1$ internal wave cannot resonantly self-interact for high values of $f$ \citep{wunsch}.
 
\vspace{-0.2cm}
\section*{Declaration of interests}
The authors report no conflict of interest. 

\vspace{-0.4cm}
\appendix

\section{Expressions for $\gamma_j$, $\Gamma_j$ and $\textnormal{NL}_{(*,j)}$} \label{app:C}

%\subsection{Linear modal coefficients}

The expressions for $\gamma^{(n)}_{(j,i)}$, which are used in \S \ref{sec:2.2.1} and \S \ref{Section_6}, are provided below:
   \begin{alignat}{3}
 \gamma^{(1)}_{(j,i)} &= {\int^{0}_{-1}\left[\phi_j\right]\phi_i d \eta},  \hspace{3cm} &&\gamma^{(2)}_{(j,i)} = {\int^{0}_{-1} \left[\frac{\partial^{2} \phi_j}{\partial \eta^{2}} \right] \phi_i d \eta}, \nonumber\\
 \gamma^{(3)}_{(j,i)} &= {\int^{0}_{-1} \left[ (N^{2}-\omega_j^2) \phi_j \right] \phi_i d \eta},
   &&\gamma^{(4)}_{(j,i)} = {\int^{0}_{-1} \left[ (N^{2}-\omega_j^2) \frac{\partial \phi_j}{\partial x} \right] \phi_i d \eta},\nonumber\\ \gamma^{(5)}_{(j,i)} &= {\int^{0}_{-1} \left[\eta (N^{2}-\omega_j^2)  \frac{\partial \phi_j}{\partial \eta} \right] \phi_i d \eta},   
  &&\gamma^{(6)}_{(j,i)} = {\int^{0}_{-1} \left[ \eta^2 (N^{2}-\omega_j^2)  \frac{\partial^2 \phi_j}{\partial \eta^2} \right] \phi_i d \eta}, \nonumber \\   \gamma^{(7)}_{(j,i)} &= {\int^{0}_{-1} \left[ \eta (N^{2}-\omega_j^2)  \frac{\partial^2 \phi_j}{\partial x \partial \eta} \right] \phi_i d \eta}, \hspace{1cm}
&&\gamma^{(8)}_{(j,i)} = {\int^{0}_{-1} \left[ (N^{2}-\omega_j^2) \frac{\partial^2 \phi_j}{\partial x^2} \right] \phi_i d \eta}.
 \label{eqn:gammas_linear}
  \end{alignat}\vspace{0.2cm}
Throughout the paper, $\gamma_{(j,j)}$ is simply denoted by $\gamma_{j}$ for convenience.
  
%\subsection{Nonlinear modal coefficients} 
  
\noindent The expressions for $\textnormal{NL}_{(*,j)}$ used in amplitude evolution equations \eqref{eqn:wave1}--\eqref{eqn:wave3} in \S \ref{sec:2.2.1} are provided below. Note that $\textnormal{NL}_{(*,j)}$ is used in \eqref{eqn:coupling_coefficient_j}.
\begin{align}
\textnormal{NL}_{(\Psi,1)}&=\frac{\omega_{1}}{h^4}\left[\mathcal{K}_{3}\left(\zeta_3\omega_3^2\Gamma^{(1)}_2 - \zeta_3\Gamma^{(2)}_2 - \Gamma^{(3)}_3 \right) - \mathcal{K}_{2}\left(\zeta_2\omega_2^2\Gamma^{(1)}_3 - \zeta_2\Gamma^{(2)}_3 - \Gamma^{(3)}_2\right) \right] \nonumber \\
&+\frac{\omega_{1}}{h^4}\left[\left(\mathcal{K}_{2}^2-\mathcal{K}_{3}^2\right)(\mathcal{K}_{2} \Gamma^{(1)}_3 + \mathcal{K}_{3} \Gamma^{(1)}_2) \right], \nonumber \\
\textnormal{NL}_{(\Psi,2)}&=\frac{\omega_{2}}{h^4}\left[ \mathcal{K}_{3}\left(\zeta_3\omega_3^2\Gamma^{(1)}_1 - \zeta_3\Gamma^{(2)}_1 - \Gamma^{(3)}_3\right) - \mathcal{K}_{1}\left(\zeta_1\omega_1^2\Gamma^{(1)}_3 - \zeta_1\Gamma^{(2)}_3 - \Gamma^{(3)}_1 \right)  \right] \nonumber \\
&+\frac{\omega_{2}}{h^4}\left[\left({\mathcal{K}_{1}^2}-\mathcal{K}_{3}^2\right)(\mathcal{K}_{1} \Gamma^{(1)}_3 + \mathcal{K}_{3} \Gamma^{(1)}_1) \right], \nonumber \\
\textnormal{NL}_{(\Psi,3)}&=\frac{\omega_{3}}{h^4}\left[\mathcal{K}_{1}\left(\zeta_2\omega_2^2\Gamma^{(1)}_1 - \zeta_2\Gamma^{(2)}_1-\Gamma^{(3)}_2\right) + \mathcal{K}_{2}\left(\zeta_1\omega_1^2\Gamma^{(1)}_2 - \zeta_1\Gamma^{(2)}_2 - \Gamma^{(3)}_1\right) \right] \nonumber \\
&+\frac{\omega_{3}}{h^4}\left[\left(\mathcal{K}_{2}^2-{\mathcal{K}_{1}^2}\right) \left(\mathcal{K}_{1} \Gamma^{(1)}_2 - \mathcal{K}_{2} \Gamma^{(1)}_1 \right) \right].
\label{eqn:NL_big_s2_1}
\end{align} 
\begin{align}
\textnormal{NL}_{(B,1)}&=\frac{(\mathcal{K}_{3}-\mathcal{K}_{2})}{h^4}\left[ \mathcal{K}_{2}\mathcal{K}_{3}\left(\frac{1}{\omega_2}-\frac{1}{\omega_3} \right)\Gamma^{(4)} + \left(\frac{\mathcal{K}_{3}}{\omega_3}-\frac{\mathcal{K}_{2}}{\omega_2}\right) \left(\mathcal{K}_{3} \Gamma^{(2)}_2- \mathcal{K}_{2} \Gamma^{(2)}_3 \right) \right], \nonumber \\
\textnormal{NL}_{(B,2)}&= \frac{(\mathcal{K}_{3}-\mathcal{K}_{1})}{h^4}\left[\mathcal{K}_{1}\mathcal{K}_{3}\left(\frac{1}{\omega_1}-\frac{1}{\omega_3} \right) \Gamma^{(4)} + \left(\frac{\mathcal{K}_{3}}{\omega_3}-\frac{\mathcal{K}_{1}}{\omega_1}\right) \left(\mathcal{K}_{3} \Gamma^{(2)}_1-\mathcal{K}_{1} \Gamma^{(2)}_3 \right) \right], \nonumber \\
\textnormal{NL}_{(B,3)}&=\frac{(\mathcal{K}_{1}+\mathcal{K}_{2})}{h^4}\left[ \mathcal{K}_{1}\mathcal{K}_{2}\left(\frac{1}{\omega_1}+\frac{1}{\omega_2} \right) \Gamma^{(4)} + \left(\frac{\mathcal{K}_{2}}{\omega_2}-\frac{\mathcal{K}_{1}}{\omega_1}\right) \left(\mathcal{K}_{1} \Gamma^{(2)}_2 - \mathcal{K}_{2} \Gamma^{(2)}_1\right)  \right]  
\label{eqn:NL_big_s2_2}
\end{align} 
\begin{align}
\textnormal{NL}_{(\mathcal{V},1)}&=\frac{f^2}{h^4}\left[\left(\frac{1}{\omega_3}+\frac{1}{\omega_2}\right)\left(\mathcal{K}_2+\mathcal{K}_3\right)\left(\zeta_3\omega_3^2\Gamma^{(1)}_2 - \zeta_3\Gamma^{(2)}_2 + \zeta_2\omega_2^2\Gamma^{(1)}_3 - \zeta_2\Gamma^{(2)}_3   \right)   \right] \nonumber \\
&+ \frac{f^2}{h^4}\left[\left(\frac{\mathcal{K}_2}{\omega_3}+\frac{\mathcal{K}_3}{\omega_2}\right)\left(  \Gamma^{(3)}_2  + \Gamma^{(3)}_3    \right) \right], \nonumber \\
\textnormal{NL}_{(\mathcal{V},2)}&=\frac{f^2}{h^4}\left[\left(\frac{1}{\omega_3}+\frac{1}{\omega_1}\right)\left(\mathcal{K}_1+\mathcal{K}_3\right)\left(\zeta_3\omega_3^2\Gamma^{(1)}_1 - \zeta_3\Gamma^{(2)}_1 + \zeta_1\omega_1^2\Gamma^{(1)}_3 - \zeta_1\Gamma^{(2)}_3   \right) \right] \nonumber \\
&+ \frac{f^2}{h^4}\left[\left(\frac{\mathcal{K}_1}{\omega_3}+\frac{\mathcal{K}_3}{\omega_1}\right)\left(  \Gamma^{(3)}_1  + \Gamma^{(3)}_3    \right) \right], \nonumber \\
\textnormal{NL}_{(\mathcal{V},3)}&=\frac{f^2}{h^4}\left[\left(\frac{1}{\omega_1}-\frac{1}{\omega_2}\right)\left(\mathcal{K}_1-\mathcal{K}_2\right)\left(\zeta_1\omega_1^2\Gamma^{(1)}_2 - \zeta_1\Gamma^{(2)}_2 + \zeta_2\omega_2^2\Gamma^{(1)}_1 - \zeta_2\Gamma^{(2)}_1   \right) \right] \nonumber \\
&- \frac{f^2}{h^4}\left[\left(\frac{\mathcal{K}_2}{\omega_1}+\frac{\mathcal{K}_1}{\omega_2}\right)\left(  \Gamma^{(3)}_2  + \Gamma^{(3)}_1 \right) \right].
\label{eqn:NL_big_s2_3}
\end{align} 
\noindent Moreover, $\Gamma^{(n)}_j$ are defined as follows: 
 \begin{align}
 \Gamma^{(1)}_1 & = {\int^{0}_{-1}\phi_2\phi_3\frac{\partial \phi_1}{\partial \eta} d \eta},  \hspace{1cm}  \Gamma^{(1)}_2 = {\int^{0}_{-1}\phi_1\phi_3\frac{\partial \phi_2}{\partial \eta} d \eta}, \hspace{1cm}  \Gamma^{(1)}_3 = {\int^{0}_{-1}\phi_2\phi_1\frac{\partial \phi_3}{\partial \eta} d \eta},\nonumber\\
 \Gamma^{(2)}_1 &= {\int^{0}_{-1}N^{2}\phi_2\phi_3\frac{\partial \phi_1}{\partial \eta} d \eta}, \hspace{0.6cm}  \Gamma^{(2)}_2 = {\int^{0}_{-1}N^{2}\phi_1\phi_3\frac{\partial \phi_2}{\partial \eta} d \eta}, \hspace{0.6cm}  \Gamma^{(2)}_3 = {\int^{0}_{-1}N^{2}\phi_2\phi_1\frac{\partial \phi_3}{\partial \eta} d \eta},  \nonumber\\
  \Gamma^{(3)}_1 & = {\int^{0}_{-1}\phi_2\phi_3\frac{\partial^3 \phi_1}{\partial \eta^3} d \eta},  \hspace{1cm}  \Gamma^{(3)}_2 = {\int^{0}_{-1}\phi_1\phi_3\frac{\partial^3 \phi_2}{\partial \eta^3} d \eta}, \hspace{1cm}  \Gamma^{(3)}_3 = {\int^{0}_{-1}\phi_2\phi_1\frac{\partial^3 \phi_3}{\partial \eta^3} d \eta},\nonumber\\
 \Gamma^{(4)} & = {\int^{0}_{-1} \frac{\partial N^{2}}{\partial \eta} \phi_1 \phi_2 \phi_3  d \eta}. 
 \label{eqn:gammas_nonlinear}
  \end{align}

\section{Scaling analysis for finding the relation between the small parameters}
\label{app:A}
%Scaling analysis is performed to predict the relation between various small parameters.
Here we perform a scaling analysis for all the terms appearing in \eqref{eqn:LHS_triad_after_ortho}. Equation \eqref{eqn:LHS_triad_after_ortho} is chosen here so that scaling analysis can be also done for the different terms that compose the $\beta_j$ function \eqref{eqn:WKB_cr_final}. Integrals ($\gamma_j$) in \eqref{eqn:LHS_triad_after_ortho} ($\gamma_j$ expressions are given in \eqref{eqn:gammas_linear}) cannot be analytically simplified for non-uniform stratification profiles. Hence, we adopt a numerical approach where we study how different integrals scale in an ensemble of stratification profiles that resemble the profiles used throughout the paper. Using this information, we scale the different terms. To this end, the stratification profiles are chosen such that $N_{\textnormal{max}}=(5N_b, 10N_b, 15N_b)$, $W_p=(H/100, 2H/100, 3H/100)$, and $z_c=(H/80, H/40, H/20, H/10)$; and we consider all possible ($36$) combinations.

The analysis provides a relation between the time scale of the amplitude's temporal evolution ($\epsilon_{t} t$), length scale of the amplitude function ($\epsilon_{x} x$), and the magnitude of the waves' amplitude ($\epsilon_{a} a_j$). 
Small parameters $(\epsilon_h, \epsilon_k)$ represent the bathymetry and they also influence the wave amplitude evolution. Equation \eqref{eqn:LHS_triad_after_ortho}, after some simplifications to the nonlinear term, is given below:
\begin{align}
&\hspace{-0.5cm}\frac{\partial a_j}{\partial t} + \frac{1}{\mathfrak{D}_j}\left[2\ii\gamma^{(3)}_j\left( \frac{\mathcal{K}_{j}}{h} \frac{\partial a_j}{\partial x}\right) + \frac{\gamma^{(6)}_j}{h^2}\left(\frac{d h}{dx}\right)^2  a_j \right]\nonumber\\
&\hspace{-0.5cm}+\frac{1}{\mathfrak{D}_j}\left[\frac{2\mathcal{K}_{j}}{h} \gamma^{(4)}_j +  \gamma^{(3)}_j  \frac{\partial }{\partial x}\left(\frac{\mathcal{K}_{j}}{h} \right) -\gamma^{(5)}_j\frac{2}{h}\frac{\partial h}{\partial x}\left(\frac{\mathcal{K}_{j}}{h} \right) - \frac{\gamma^{(3)}_j}{\beta_j} \frac{2\mathcal{K}_{j}}{h} \frac{d (\beta_j)}{dx} \right]a_j = \widehat{\mathfrak{N}}_j a^2,
\label{eqn:app_analysis}
\end{align}
where $\widehat{\mathfrak{N}}_j$ is defined as:
\begin{equation}
   \widehat{\mathfrak{N}}_{j} =  \frac{1}{\mathfrak{D}_j}\left[\textnormal{NL}_{(\mathcal{V},j)} + \textnormal{NL}_{(B,j)} + \textnormal{NL}_{(\Psi,j)} \right].
\end{equation}
The analysis is similar for all three waves, hence from here on all subscripts $j$ (denoting the $j-$th wave) 
are dropped for convenience. Moreover, a term containing $\gamma^{(6)}$ is also included in the above equation. It will be proved in this section that this term is an order of magnitude smaller than the other terms for the parameter regime we consider.

The time scale of wave amplitude's evolution is assumed to be at least an order of magnitude larger than the time period of the wave. Therefore ${\partial a}/{\partial t}$ will approximately scale as: ${\partial a}/{\partial t} \sim \epsilon_{t} \epsilon_{a} \omega$. The amplitude's length scale is assumed to be much larger than the wavelength of the wave. Hence ${\partial a}/{\partial x}$ will scale as ${\partial a}/{\partial x} \sim \epsilon_x \epsilon_{a} \mathcal{K}/h$. Using the above scaling, the ${\partial a}/{\partial x}$ term in \eqref{eqn:app_analysis} (including its coefficients) will scale as:
\begin{equation}
2\frac{\gamma^{(3)}}{\mathfrak{D}}\left( \frac{\mathcal{K}}{h} \frac{\partial a}{\partial x}\right) \sim \frac{1}{\omega} \frac{\gamma^{(3)} \mathcal{K}^2}{\gamma^{(1)}\mathcal{K}^2-\gamma^{(2)}}  \epsilon_x \epsilon_{a} \sim (\widehat{c}_g  \epsilon_x) \omega\epsilon_{a},
\label{eqn:CG_app}
\end{equation}
where $\widehat{c}_g$ represents the scale  
of group speed term for the packet, and is given by:
\begin{equation}
    \widehat{c}_g \equiv  \frac{(\omega^2-f^2)}{\omega^2}\left[\frac{\gamma^{(3)} }{(\omega^2-f^2)\gamma^{(1)}+\gamma^{(3)}} \right].
    \label{eqn:CG_measure}
\end{equation}
It can be noticed that as $\epsilon_x$ is reduced, the effect of group speed diminishes as expected since a decrease in $\epsilon_x$ means the length scale of the packet is increased. Here we also emphasize that for $\omega \approx N$: $\gamma^{(3)} \ll \omega^2\gamma^{(1)}$. In such kind of parameter regime, $\widehat{c}_g \ll 1$, hence ${\partial a}/{\partial x}$ term will have a reduced effect on the amplitude evolution. Moreover, for $\omega \approx f$, similar behavior is observed since $\widehat{c}_g \ll 1$.
%In such parameter regime, ${\partial a}/{\partial x}$ term will have reduced effect on the amplitude evolution. 
 
%(hereafter referred to as `$T_6$')

\begin{figure}
 \centering{\includegraphics[width=0.9\textwidth]{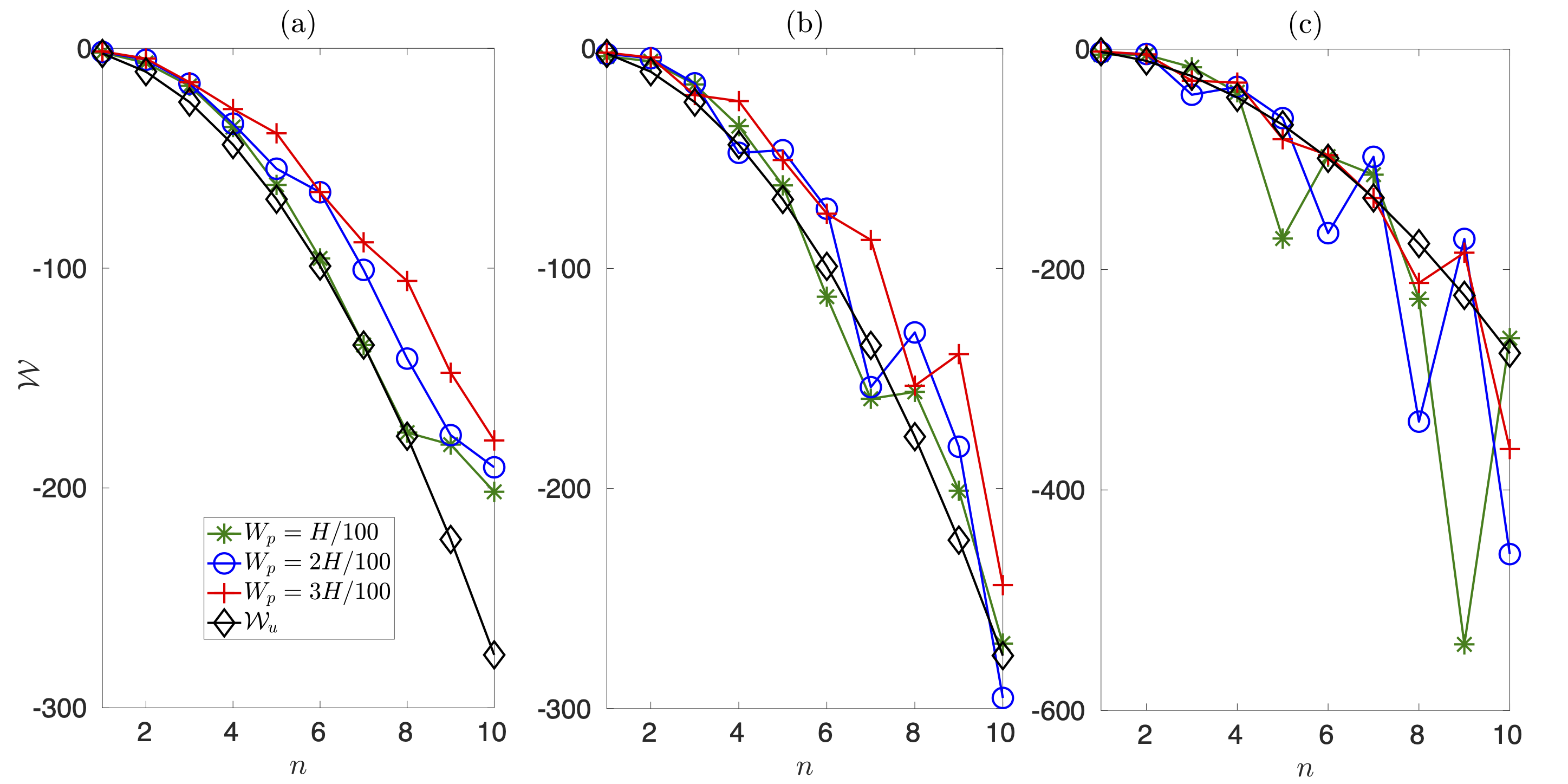}}
  \caption{The variation in $\mathcal{W}$ for modes $1$--$10$ for different stratification profiles.
  (a) $N_{\textnormal{max}} = 5N_b $ is used with $z_c = H/10$ and $W_p$ is varied.  (b) $N_{\textnormal{max}} = 10N_b $ is used with $z_c = H/10$ and $W_p$ is varied.  (c) $N_{\textnormal{max}} = 15N_b $ is used with $z_c = H/10$ and $W_p$ is varied. }
  \label{fig:W_app}
\end{figure}

\noindent Now we focus on the term containing $\gamma^{(6)}$ in  \eqref{eqn:app_analysis}, which is given below (after some simplification):
% \begin{equation}
% \textnormal{T}_6 \equiv \left(\frac{d h}{dx}\right)^2 \left[ \frac{(\omega^2-f^2)}{\mathcal{K}^2\omega^2}\frac{\gamma^{(6)} }{(\omega^2-f^2)\gamma^{(1)}+\gamma^{(3)}} \right] \frac{\omega a}{2}. 
%   \label{eqn:T6_app}
% \end{equation}
\begin{equation}
  \left(\frac{d h}{dx}\right)^2  \frac{\mathcal{W}}{\mathcal{K}^2}   \frac{\omega a}{2},
  \label{eqn:T6_app}
\end{equation} 
% The integral $\gamma^{(6)}$ is evaluated numerically to study its scaling. Furthermore to evaluate $T_6$, a non-dimensional quantity $\mathcal{W}$ is defined as follows:
%  \begin{equation}
%      \mathcal{W} = \left[ \frac{(\omega^2-f^2)}{\omega^2}\frac{\gamma^{(6)} }{(\omega^2-f^2)\gamma^{(1)}+\gamma^{(3)}} \right].
%  \end{equation}
where $\mathcal{W}$ is a non-dimensional quantity defined as:
 \begin{equation}
     \mathcal{W} =  \frac{\omega^2-f^2}{\omega^2}\frac{\gamma^{(6)} }{(\omega^2-f^2)\gamma^{(1)}+\gamma^{(3)}}.
 \end{equation}
The integral $\gamma^{(6)}$ is evaluated numerically to study its scaling. 
For uniform stratification, $\mathcal{W}$ can be evaluated analytically, which is given below:
% \begin{equation}
%          \mathcal{W}_u = -\frac{\mathcal{M}^2}{3}\left[ \frac{\omega^2-f^2}{\omega^2}\frac{N_b^2 - \omega^2}{ N_b^2 - f^2} \right] = -\frac{(n\pi)^2}{3}\left[ \frac{\omega^2-f^2}{\omega^2}\frac{N_b^2 - \omega^2}{ N_b^2 - f^2} \right],
%          \label{eqn:wu}
% \end{equation}
\begin{equation}
         \mathcal{W}_u = -{\mathcal{M}^2}\left[ \frac{\omega^2-f^2}{\omega^2}\frac{N_b^2 - \omega^2}{ N_b^2 - f^2} \right]\left(\frac{1}{3}-\frac{1}{2\mathcal{M}^2}\right)  
         \label{eqn:wu}
\end{equation}
where $\mathcal{W}_u$ is used to denote $\mathcal{W}$ in constant stratification $N_b$, and $\mathcal{M}=n\pi$ is the non-dimensionalised vertical wavenumber of the wave. Moreover, using $ (d h/d x)^2 \sim (\epsilon_h \epsilon_k)^2\mathcal{K}^2$, the term given in equation \eqref{eqn:T6_app} will scale as
\begin{equation}
    \left(\frac{(\epsilon_h \epsilon_k)^2}{2} \mathcal{W} \right) {\omega \epsilon_a}.
    \label{g6_scale}
\end{equation} 
Hence for the multiple-scale analysis to be consistent, $\mathcal{W}({(\epsilon_h \epsilon_k)^2}/{2})$ has to be a small quantity. $\mathcal{W}$ is plotted in figure \ref{fig:W_app} for nine stratification profiles, where $f=0$ and $\omega/N_b = 0.4$ were used. In all subfigures, $\mathcal{W}_u$ is also plotted for reference, where $\mathcal{W}_u$ is evaluated with constant stratification $N_b$ (hence $\mathcal{W}_u$ in all subfigures is same).  From figure \ref{fig:W_app}, it can be seen that in general for any stratification profile, $\mathcal{W}$ is almost proportional to the square of the modenumber $n$, similar to $\mathcal{W}_u$. Hence the bathymetry has to be more slowly varying ($\epsilon_k$ has to be smaller) as the modenumber increases. Other pycnocline depths ($z_c=H/20,H/40,H/80$) were also tested for different combinations of $W_p, N_{\textnormal{max}}$ used in figure \ref{fig:W_app} that provided similar results.

% Moreover, using $\mathcal{W}_u$ as a reference for non-uniform stratifications, the scaling \eqref{g6_scale} can be simplified as:
% \begin{equation}
%     \textnormal{T}_6 \sim \left({1.6(\epsilon_h \epsilon_k)^2} {n^2}  \right) {\omega \epsilon_a}
% \end{equation}
% Moreover using $\epsilon_h \sim 1$ (focusing on large amplitude topographies), the above scaling can be simplified to: 
% \begin{equation}
%     \textnormal{T}_6 \sim \left({1.6(n \epsilon_k)^2}\right) {\omega \epsilon_a}
% \end{equation}
% Hence for multiple-scale analysis to be valid, the condition ${1.6(n \epsilon_k)^2}\ll \mathcal{O}(1)$ has to be satisfied. Here in this paper, we always assume the condition that $n \epsilon_k \ll \mathcal{O}(1)$. This is because when $n \epsilon_k \ll \mathcal{O}(1)$ was satisfied, the scattering of the internal waves due to the bathymetry was found to be negligible as the wave passes through a bathymetry. 

% Under the assumption that:
% \begin{equation}
%     \left[ \frac{(\omega^2-f^2)}{\omega^2}\frac{N^2 - \omega^2}{ N^2 - f^2  } \right] \lesssim \mathcal{O}(1),
%     \label{eqn:assumption_g6}
% \end{equation}
% $\mathcal{W}_u \sim m^2/3$. This results in the requirement that $m^2[{(\epsilon_h \epsilon_k)^2}/{6}] \ll \mathcal{O}(1) $

%Moreover, different $f$ values also qualitatively provided similar results. \\

%The effect of all the LHS terms on the second line of equation \eqref{eqn:app_analysis} is captured by the $\beta$ functions. However here they are analysed to show that all the terms can be of same order of magnitude in a non-uniform stratification case. 

The term which contains the $\gamma^{(5)}$ integral is now analysed.
%The integral $\gamma_j^{(5)}$ can be rewritten as:
% \begin{equation}
%  \gamma^{(5)} = \frac{\omega^2}{2\mathcal{K}^2}\left( \frac{\partial \phi}{\partial \eta} \right)^2\bigg\rvert_{\eta = -1} - \frac{1}{2}\gamma^{(3)}.
%   \label{eqn:gamma_5}
% \end{equation}
% The first term on the RHS of \eqref{eqn:gamma_5} is a quantity which is proportional to square of horizontal velocity at the bottom surface. Moreover, the second term is a quantity which is proportional to the average of square of horizontal velocity of the wave in the fluid depth, with the same proportionality constant as the first term.  Moreover, the integral term in RHS can be rewritten as:
% \begin{equation}
%  {\int^{0}_{-1} \frac{\omega^2}{2\mathcal{K}^2}  \left( \frac{\partial \phi}{\partial \eta} \right)^2 d \eta} = \frac{1}{2} {\int^{0}_{-1}   \left({N^2-\omega^2} \right)\phi^2  d \eta} = \frac{1}{2}\gamma^{(3)}
%   \label{eqn:gamma_6}
% \end{equation}
% \textcolor{red}{gamma5 limit is upperbound.}
For all non-uniform stratification profiles used in this appendix, it was observed that
\begin{equation}
 \gamma^{(5)} \lesssim \frac{1}{2}\gamma^{(3)}.
  \label{eqn:gamma_5}
\end{equation} 
%Hence, it is assumed that $\gamma^{(5)}$ is same order of magnitude as $\gamma^{(3)}/2$.
Using \eqref{eqn:gamma_5}, the term containing $\gamma^{(5)}$ can be scaled to:
\begin{equation}
\frac{1}{\mathfrak{D}_j}\left[\gamma^{(5)}\frac{2}{h}\frac{\partial h}{\partial x}\left(\frac{\mathcal{K}}{h} \right) a \right] \sim  \left(\frac{\widehat{c}_g}{2} \epsilon_h\epsilon_k \right) \omega\epsilon_a.
\label{eqn:gamma_5_scale}
\end{equation}

% The term containing the integral $\gamma_p^{(5)}$ is given below:
% \begin{equation}
%     \textnormal{T}_4 = \frac{1}{h} \left(\frac{d^2 h}{dx^2}\right)\left[ \frac{1}{\omega_p^2}\frac{\gamma_p^{(5)}}{\gamma^{(1)}_p\mathcal{K}_p^2-\gamma^{(2)}_p}\right] \frac{\omega_p a_p}{2}  
% \end{equation}
% Using the scaling $\gamma_{(5)} \sim \gamma_{(3)}/2$ and  $(d^2 h/d x^2)/h \sim \epsilon_h (\epsilon_k)^2\mathcal{K}_p^2$, the term $\textnormal{T}_4$'s scaling can be simplified as follows:
% \begin{equation}
%     \textnormal{T}_4 \sim \epsilon_h (\epsilon_k)^2 \left[ \frac{\mathcal{K}_p^2}{\omega_p^2}\frac{\gamma_p^{(3)}}{\gamma^{(1)}_p\mathcal{K}_p^2-\gamma^{(2)}_p}\right] \frac{\omega_p a_p}{4} \sim \frac{\epsilon_h (\epsilon_k)^2}{4} \left[\frac{\gamma_p^{(2)}}{\gamma^{(1)}_p\mathcal{K}_p^2-\gamma^{(2)}_p}\right] {\omega_p a_p}  
%   \label{eqn:T4_app}
% \end{equation}
% Hence it can be seen that this term may be significant when the bathymetry's wavenumber is same order as the parent wave's wavenumber. Moreover, it also needs the condition that $\gamma_{(2)} \sim (\mathcal{K}^{2}_p\gamma_p^{(1)} - \gamma_{(2)}) $, which is not satisfied for waves which have $\omega_p \approx N$. 

Now we analyse how the wavenumber of a mode changes as $h$ changes. To this end,  \eqref{eqn:eigen2}, which provides the $n$-th eigenfunction, is differentiated in $x-$direction, yielding:
    \begin{equation}
       \left[\frac{\partial^2 }{\partial \eta^2}+ \mathcal{K}^{2}_n \chi^2 \right] \frac{\partial \phi_n}{\partial x} =  - \mathcal{K}^{2}_n \frac{\partial \chi^2}{\partial x} {\phi_n} - \frac{d \mathcal{K}_n^2}{d x} \chi^2 {\phi_n},
       \label{eq:dif_eigx}
    \end{equation}
%\begin{equation}
%      \frac{\partial^2 \phi}{\partial \eta^2} +  \mathcal{K}^2 \chi^2 \phi = 0.
%      \label{eqn:app_eigen}
% \end{equation}
% Here it can noticed it is different from equation \eqref{eqn:eigen2}, since the variation of bathymetry is considered in this equation, instead of the $x-$direction. 
% \begin{align}
% \frac{\partial}{\partial x} \left( \frac{\partial^2 \phi_p}{\partial \eta^2} \right) &+    {2}{k_p}\left(\frac{d k_p}{d x}\right) \left(\frac{N^2-\omega^{2}_p}{\omega^{2}_p}\right) h^2 \phi_p  +   2h \left(\frac{d h}{d x}\right)  k_p^2 \left(\frac{N^2-\omega^{2}_p}{\omega^{2}_p}\right) \phi_p   \nonumber \\
% &+ \frac{\partial N^2}{\partial x} k_p^2 h^2 \phi_p  + {k_{p}^{2}}{h^2} \left(\frac{N^2-\omega^{2}_p}{\omega^{2}_p}\right) \frac{\partial \phi_p}{\partial x} = 0.
%      \label{eqn:app_eigen_2}
% \end{align}
%Equation \eqref{eqn:app_eigen} 
where $\mathcal{K}_n = k_n h$. Equation \eqref{eq:dif_eigx} can have a non-trivial solution only when the RHS is orthogonal to the solution of the self adjoint operator in the LHS. Hence mutliplying RHS with $\phi_n$, and then integrating in the $\eta$--direction between the domain limits would result in:
 \begin{equation}
             \frac{1}{k_n}\frac{d k_n}{d x} + \frac{1}{h}\frac{d h}{d x} = -\frac{1}{2\gamma^{(3)}_n} \int_{-1}^{0} \frac{\partial N^2}{\partial x} \phi^2_n d \eta.
       \label{eq:Lam_var}
\end{equation} 
From  \eqref{eq:Lam_var},we notice that the dimensional wavenumber $k_n$ can change due to: (a) change in the domain height, and (b) change in the effective stratification profile. For uniform stratification, $d \mathcal{K}_n/dx=0$. 
\noindent For the lower modes ($1$--$10$) in profiles considered in this appendix, it was observed that
 \begin{equation}
 k_h \equiv  \mathcal{O}\left(\frac{1}{k_n}\frac{d k_n}{d x}\right)\bigg/ \mathcal{O}\left(\frac{1}{h}\frac{dh}{d x}\right)  \sim  \mathcal{O}(1). 
  \label{eqn:k_change}
 \end{equation} 
Moreover, in general it was observed that as the modenumber increases, $k_h$ increases. %For the lower modes ($1$--$10$), $k_h \sim \mathcal{O}(1)$ was always observed.  
\noindent Using \eqref{eqn:k_change}, the term containing the derivative of the wavenumber can be scaled as:
\begin{equation}
\frac{1}{\mathfrak{D}_j}\left[ \gamma^{(3)} \frac{d k }{d x} a \right]  \sim \left( \frac{\widehat{c}_g}{2}\epsilon_h\epsilon_k \right) \omega\epsilon_a.
\label{eqn:kx_scale}
\end{equation}

\noindent We now evaluate $\gamma^{(4)}$ for the stratification profiles considered in this appendix. For modes $1$--$10$, we find
% \begin{equation}
%     \frac{\gamma^{(4)}}{\gamma^{(3)}} \sim \mathcal{O}\left(\frac{1}{h}\frac{dh}{dx} \right) \hspace{0.3cm} \textnormal{and} \hspace{0.3cm} \frac{\gamma^{(8)}}{\gamma^{(3)}} \sim \mathcal{O}\left(\frac{n^2}{h}\frac{d^2 h}{dx^2} \right)
%     \label{eqn:gamma_4_p}
% \end{equation}
\begin{equation}
    \frac{\gamma^{(4)}}{\gamma^{(3)}} \sim \mathcal{O}\left(\frac{1}{h}\frac{dh}{dx} \right). 
    \label{eqn:gamma_4_p}
\end{equation}

% \begin{equation}
% max( \left[(N^2-\omega^2)) \mathcal{K}_m I_m^2 \right]   \sim  max( \left[(N^2-\omega^2)) \mathcal{K}_n I_n^2 \right]
%     \label{eqn:flux}
% \end{equation}
% \begin{equation}
%   \left[\int_{-1}^{0} \frac{\partial N^2}{\partial x} \phi_m \phi_n d \eta \right] \sim   \frac{\sqrt{\mathcal{K}_m}}{\sqrt{\mathcal{K}_n}}  \left[\int_{-1}^{0} \frac{\partial N^2}{\partial x} \phi_m^2 d \eta \right]
%   \label{eqn:main_scaling_assumption}
% \end{equation}
% \begin{equation}
%   max( \frac{\partial N^2}{\partial x} ) I_m I_n   \sim   \frac{\sqrt{\mathcal{K}_m}}{\sqrt{\mathcal{K}_n}} max( \frac{\partial N^2}{\partial x} ) I_m I_m  
%   \label{eqn:main_scaling_assumption}
% \end{equation}
% The vertical function can be scaled for

% \noindent Hence using \eqref{eqn:gamma_4_p}, the term containing $\gamma^{(4)}$ can be scaled to:
% \begin{equation}
%   \frac{h^2}{\mathfrak{D}_j}\left[  \frac{2\mathcal{K}}{h} \gamma^{(4)} a \right] \sim  \left({\widehat{c}_g} \frac{2h}{\mathcal{K}^2}\frac{d \mathcal{K}}{dx}\right)  \omega\epsilon_a
%     \label{eqn:gamma_4_scale}
% \end{equation} 

\noindent Using \eqref{eqn:gamma_4_p}, the term containing $\gamma^{(4)}$ can be scaled to:
\begin{equation}
  \frac{1}{\mathfrak{D}_j}\left[  \frac{2\mathcal{K}}{h} \gamma^{(4)} a \right] \sim  \left( {\widehat{c}_g}\epsilon_h\epsilon_k \right)  \omega \epsilon_a.
    \label{eqn:gamma_4_scale}
\end{equation} 

% For scenarios where $({h}/{\mathcal{K}}){d \mathcal{K}}/{dx} \sim {dh}/{dx}$, equation \eqref{eqn:gamma_4_scale} can be re-written as:
% \begin{equation}
%   {\widehat{c}_g} \frac{2h}{\mathcal{K}^2}\frac{d \mathcal{K}}{dx}\epsilon_a \sim \left( {\widehat{c}_g}\epsilon_h\epsilon_k \right)  \omega \epsilon_a
%     \label{eqn:gamma_4_scale_final}
% \end{equation}  
Using \eqref{eqn:gamma_5_scale}, \eqref{eqn:kx_scale}  and \eqref{eqn:gamma_4_scale}, we observe that for the lower modes, the three terms that compose the $\beta$ function can scale to a maximum value which is of the same order of magnitude. Hence they are all retained and are used in evaluating the $\beta$ function \eqref{eqn:WKB_cr_final}. 
Moreover, it can be seen that the topographic terms are all dependent on the magnitude of the group speed. This relation is naturally there because a wavepacket has to travel to different $h$ fast enough to feel the effect of $h$ variation. Scaling \eqref{eqn:gamma_4_scale} also holds for uniform stratification, where $\phi$ still varies in the $x-$direction. This is because of the nature of the $\phi$ normalisation, i.e.\, \eqref{eqn:nondim_energy}, used in this paper.

% Moreover, now we study the ratio of \eqref{eqn:gamma_4_scale}  with \eqref{eqn:CG_app}, we arrive at:
% \begin{equation}
%  \frac{h^2}{\mathfrak{D}_j}\left[  \frac{2\mathcal{K}}{h} \gamma^{(4)} a \right] \bigg/ 2\frac{\gamma^{(3)}}{\mathfrak{D}}\left( {\mathcal{K}}{h} \frac{\partial a}{\partial x}\right) \sim \frac{n\epsilon_k}{\epsilon_x}
% \end{equation}
% Now assuming that the length scale of the wavepacket is same order as length scale of the topographic variation, we can approximate ${n\epsilon_k}/{\epsilon_x} \sim \mathcal{O}(1)$. This simply means that the term in \eqref{eqn:gamma_4_scale} (and hence also \eqref{eqn:gamma_5_scale}, \eqref{eqn:kx_scale}) scale similar to the group speed term. This relation simply arrive because the wavepacket has to be fast 

\noindent The nonlinear coupling coefficient in the RHS cannot be further simplified, hence the nonlinear term scales as: 
\begin{equation}
    \textnormal{RHS} \sim \widehat{\mathfrak{N}}\epsilon_a^2.
\end{equation}
Hence the final scaling for \eqref{eqn:wave1}--\eqref{eqn:wave3}, using all the scaling derived, and with the inclusion of the $\gamma^{(6)}$ term, is given below after some simplification:
\begin{equation}
 \epsilon_{t} \sim \frac{{\mathfrak{N}}}{\omega } \epsilon_a  -    {\widehat{c}_g}  \epsilon_x   - \frac{(\epsilon_h \epsilon_k)^2}{2} \mathcal{W}.  \label{eqn:final_scale_1}
\end{equation}
.Here an important point to remember is that the multiple-scale analysis was derived with the assumption that internal waves do not scatter/exchange energy to different modes of the same angular frequency. Therefore the reduced order equations provide the most accurate results when the internal waves do not scatter significant amount of its energy as it passes over a bathymetry. Moreover, even when $\mathcal{O}(\epsilon_h \epsilon_k) \ll \mathcal{O}(1)$ is satisfied, there could be special circumstances when waves may still get scattered significantly. An example of such a scenario is Bragg resonance of internal waves due to small amplitude, subcritical topographies \citep{buhler_2011,li_mei_2014,couston_2017}. Scattering/energy exchange can also occur for large amplitude, slowly varying topographies. {However, it was observed that modes $1$--$8$ are scattered very little for large amplitude topographies ($\epsilon_h \approx 0.5$) with low criticality ($\lesssim 0.1$) in the presence of uniform stratification. Criticality is defined as the ratio of the maximum slope of the topography to the slope of the internal wave. Mode-8 has $\approx 8\%$ variation in its amplitude as it propagates through a Gaussian topography with $\epsilon_h = 0.5$ and criticality = $0.1$.}  Low criticality topographies for mode $n$ of any $\omega/N_b$ is obtained when the condition $n\epsilon_k \ll \mathcal{O}(1)$ is satisfied.  Moreover for the condition $n\epsilon_k \ll \mathcal{O}(1)$, the last term in \eqref{eqn:final_scale_1} becomes an $\epsilon_k^2$ term even for large amplitude topographies. This can be seen by considering $\mathcal{W}_u$ (which can also be used as a reference for non-uniform stratifications) given in \eqref{eqn:wu}. Hence this term is neglected in the governing equations \eqref{eqn:wave1}--\eqref{eqn:wave3}. 
\vspace{-0.3cm}\section{Scaling analysis for the governing equations in \S 6}
\label{app:B} 
The scaling analysis for the governing equation \eqref{eqn:s6_d_wave} derived in \S 6 is done with the help of results derived in appendix \ref{app:A}. The above-mentioned governing equations are given below:
 \begin{align}
& \left[  \left(\gamma_j^{(3)} \frac{\partial^2 \mathcal{A}_j}{\partial x^2} \right) + \mathcal{K}^2_j\left(\frac{\mathcal{A}_j}{h^2} \gamma_j^{(3)} \right) \right] \mathcal{T}_j + 2\left(\gamma_j^{(4)} - \frac{\gamma_j^{(5)}}{h}\frac{\partial h}{\partial x}\right) \frac{\partial \mathcal{A}_j}{\partial x}  \mathcal{T}_j + \gamma_j^{(8)}\mathcal{A}_j\mathcal{T}_j  \nonumber\\
+ &\left[ 
\frac{\gamma_j^{(6)}}{h^2}\left(\frac{\partial h}{\partial x}\right)^2 -  \frac{\gamma_j^{(5)} }{h}\left(\frac{\partial^2 h}{\partial x^2} \right) + \frac{2\gamma_j^{(5)}}{h^2}\left(\frac{\partial h}{\partial x}\right)^2 - 2\frac{\gamma_j^{(7)}}{h}\frac{\partial h}{\partial x} \right]\mathcal{A}_j\mathcal{T}_j    = \langle \textnormal{NL}_3 \rangle. \label{eqn:app_s6_eq}
 \end{align}
From here on the subscripts are omitted, since the analysis is similar for both the waves. The leading order terms scale as follows:
 \begin{equation}
    \left[ \frac{\partial^2 \mathcal{A}}{\partial x^2}, \mathcal{K}^2\frac{\mathcal{A}}{h^2} \right] \sim \epsilon_a \frac{\mathcal{K}^2}{h^2}.
    \label{eqn:leading_order_terms}
 \end{equation}
Using \eqref{eqn:wav_no_assumption}, the scalings derived in appendix \ref{app:A}, and the small amplitude assumption for {topography} ($\epsilon_h \ll \mathcal{O}(1)$ and $\epsilon_k \sim \mathcal{O}(1)$), we obtain: 
\begin{align}
 \hspace*{-1.5cm}   2  \frac{\gamma^{(4)}}{\gamma^{(3)}}  \frac{\partial \mathcal{A}}{\partial x} &\sim 2\epsilon_h\epsilon_a\frac{\mathcal{K}^2}{h^2}, \hspace{2.6cm} \frac{2}{h}\frac{\partial h}{\partial x}\frac{\gamma^{(5)}}{\gamma^{(3)}}  \frac{\partial \mathcal{A}}{\partial x} \sim  \epsilon_h\epsilon_a\frac{\mathcal{K}^2}{h^2}, \nonumber \\  \hspace*{-1.5cm} \left(\frac{1}{h}\frac{\partial^2 h}{\partial x^2}\right)\frac{\gamma^{(5)}}{\gamma^{(3)}}\mathcal{A} &\sim \frac{\epsilon_h}{2}\epsilon_a\frac{\mathcal{K}^2}{h^2}, \hspace{1.99cm} 
    2\left(\frac{1}{h}\frac{\partial h}{\partial x}\right)^2\frac{\gamma^{(5)}}{\gamma^{(3)}} \mathcal{A} \sim \epsilon_h^2\epsilon_a\frac{\mathcal{K}^2}{h^2}.
\end{align} 
For the profiles and the parameters used in appendix \ref{app:A} we observe that $(\omega^2-f^2)\gamma^{(1)} + \gamma^{(3)} \sim \gamma^{(3)}$. Hence the $\gamma^{(6)}$ term can be scaled as:
\begin{equation}
 \frac{\gamma^{(6)}}{\gamma^{(3)}}\frac{1}{h^2}\left(\frac{\partial h}{\partial x}\right)^2 \mathcal{A} \sim   \left(\mathcal{W}\epsilon_h^2\right){ \epsilon_a} \frac{\mathcal{K}^2}{h^2},
\end{equation}
% \begin{equation}
%  \frac{\gamma^{(6)}}{\gamma^{(3)}}\frac{1}{h^2}\left(\frac{\partial h}{\partial x}\right)^2 \mathcal{A} \sim   \left(\frac{\mathcal{W}}{h^2}\left(\frac{\partial h}{\partial x}\right)^2\right){ \epsilon_a} 
% \end{equation}
where $\mathcal{W}$ is plotted in figure \ref{fig:W_app} for various stratification profiles. Therefore similar to appendix \ref{app:A}, the term $ \mathcal{W}{\epsilon_h^2}$ has to be a small number for the multiple-scale analysis to be consistent.
Furthermore, the $\gamma_j^{(7)}$ term was observed to scale as:
% \begin{equation}
%  \frac{\gamma^{(8)}}{\gamma^{(3)}} \mathcal{A} \sim  \frac{\mathcal{K}_n^2}{\mathcal{K}_1^2}\epsilon_h\epsilon_a\frac{\mathcal{K}^2}{h^2}   , \hspace{2cm}        \left(\frac{1}{h}\frac{\partial h}{\partial x}\right)\frac{\gamma_j^{(7)}}{\gamma_j^{(3)}}\mathcal{A} \sim \frac{\mathcal{K}_n^2}{\mathcal{K}_1^2}\epsilon_h^2\epsilon_a\frac{\mathcal{K}^2}{h^2}  
% \end{equation}
% \begin{equation}
% \left(\frac{1}{h}\frac{\partial h}{\partial x}\right)\frac{\gamma^{(7)}}{\gamma^{(3)}}\mathcal{A} \sim \frac{\mathcal{K}_n^2}{\mathcal{K}_1^2}\epsilon_h^2\epsilon_a\frac{\mathcal{K}^2}{h^2},   
% \end{equation} 
\begin{equation}
\left(\frac{2}{h}\frac{\partial h}{\partial x}\right)\frac{\gamma^{(7)}}{\gamma^{(3)}}\mathcal{A} \lesssim \frac{\mathcal{K}_n^2}{\mathcal{K}_1^2}\left(\frac{1}{h}\frac{\partial h}{\partial x}\right)^2 2\epsilon_a,
\end{equation}
where $\mathcal{K}_n$ is the nondimensional wavenumber of wave-1 (or wave-3), and $n$ gives the wave's modenumber. Note that this scaling has a similar behavior as $\mathcal{W}$, which is nearly proportional to $n^2$. Now we focus on the scaling of the integral $\gamma^{(8)}$:
\begin{equation}
    \gamma^{(8)} = \left(\frac{dh}{dx}\right)^2{\int^{0}_{-1} (N^{2}-\omega^2) \phi \frac{\partial^2 \phi}{\partial h^2} d \eta} + \frac{d^2h}{dx^2}{\int^{0}_{-1} (N^{2}-\omega^2) \phi \frac{\partial \phi}{\partial h} d \eta}.
\end{equation}

For a uniform stratification, ${\partial^2 \phi}/{\partial h^2} = 0$. Moreover, for the non-uniform stratification profiles used in appendix \ref{app:A}, it was observed that:
\begin{equation}
{\int^{0}_{-1} (N^{2}-\omega^2) \phi \frac{\partial^2 \phi}{\partial h^2} d \eta} \lesssim  \frac{\mathcal{K}_n^2}{\mathcal{K}_1^2} \frac{\gamma^{(3)}}{h^2}, \hspace{1cm} {\int^{0}_{-1} (N^{2}-\omega^2) \phi \frac{\partial \phi}{\partial h} d \eta} \sim   \frac{\gamma^{(3)}}{h}.
\label{eqn:phixx_scale}
\end{equation}
Hence using \eqref{eqn:phixx_scale}, the scaling for $\gamma^{(8)}$ can be given in a simpler form which is as follows:
\begin{equation}
 \gamma^{(8)}  \sim \left[\frac{\mathcal{K}_n^2}{\mathcal{K}_1^2} \left(\frac{1}{h}\frac{dh}{dx}\right)^2 +  \frac{1}{h}\frac{d^2h}{dx^2} \right] \gamma^{(3)}.
 \label{eqn:gamma8_final}
\end{equation}
For low modes in the presence of small amplitude topographies, the second term in RHS of \eqref{eqn:gamma8_final} would be significantly higher than the first term.
For any mild-slope bathymetry, the nonlinear terms $\langle \textnormal{NL}_3 \rangle$ in  \eqref{eqn:app_s6_eq} can be scaled using the relation $d^{n}\mathcal{A}/dx^n \approx (\mathcal{K}/h)^n\mathcal{A}$, where $n\in \mathbb{Z}^+$. Using this approximation, the nonlinear term can be scaled as:
\begin{equation}
   \langle \textnormal{NL}_3 \rangle \sim \frac{1}{\gamma^{(3)}}\left[\textnormal{NL}_{(\mathcal{V},3)} + \textnormal{NL}_{(B,3)} + \textnormal{NL}_{(\Psi,3)} \right] \epsilon_a^2.
\end{equation} 

The nonlinear coupling coefficients cannot be simplified further. Moreover the 
nonlinear terms have to be at least one order of magnitude lesser than the leading order terms (given in \eqref{eqn:leading_order_terms}).

\bibliographystyle{jfm} 
\bibliography{jfm-instructions}

\end{document}